\documentclass[prd,tightenlines,nofootinbib,superscriptaddress]{revtex4}

\usepackage[utf8]{inputenc}

\usepackage{amsfonts,amssymb,amsthm,bbm}
\usepackage{comment}
\usepackage{amsmath}
\usepackage{mathrsfs}
\usepackage{mathtools}

\usepackage{hyperref}

\usepackage{xcolor}

\usepackage{subcaption}
\usepackage{color,psfrag}
\usepackage[dvips]{graphicx}
\usepackage{tkz-euclide}
\usepackage{tikz}
\usetkzobj{all}
\usetikzlibrary{calc}
\usetikzlibrary{decorations.pathmorphing}
\usetikzlibrary{shapes.geometric}
\usetikzlibrary{matrix}
\usetikzlibrary{arrows,decorations.markings}
\usetikzlibrary{shadings,intersections}

\newcommand{\red}{\textcolor{red}}
\newcommand{\blue}{\textcolor{blue}}

\newcommand{\C}{{\mathbb C}}

\newcommand{\R}{{\mathbb R}}

\newcommand{\cB}{{\mathcal B}}

\newcommand{\cF}{{\mathcal F}}
\newcommand{\cG}{{\mathcal G}}

\newcommand{\cM}{{\mathcal M}}

\newcommand{\cR}{{\mathcal R}}

\newcommand{\cC}{{\mathcal C}}

\newcommand{\cS}{{\mathcal S}}
\newcommand{\cU}{{\mathcal U}}

\newcommand{\SU}{\mathrm{SU}}

\newcommand{\ISU}{\mathrm{ISU}}

\newcommand{\SL}{\mathrm{SL}}

\newcommand{\SB}{\mathrm{SB}}
\newcommand{\su}{\mathfrak{su}}
\newcommand{\Sb}{\mathfrak{sb}}

\newcommand{\Hom}{\text{Hom}}
\newcommand{\im}{\text{Im}}

\newcommand{\be}{\begin{equation}}
\newcommand{\ee}{\end{equation}}
\newcommand{\beq}{\begin{eqnarray}}
\newcommand{\eeq}{\end{eqnarray}}
\newcommand{\bes}{\begin{eqnarray}}
\newcommand{\ees}{\end{eqnarray}}
\newcommand{\bsp}{\begin{split}}
\newcommand{\esp}{\end{split}}
\newcommand{\bt}{\begin{tabular}}
\newcommand{\et}{\end{tabular}}
\newcommand{\bc}{\begin{cases}}
\newcommand{\ec}{\end{cases}}
\newcommand{\bpm}{\begin{pmatrix}}
\newcommand{\epm}{\end{pmatrix}}
\newcommand{\ba}{\begin{array}}
\newcommand{\ea}{\end{array}}

\newcommand{\mat} [2] {\left ( \begin{array}{#1}#2\end{array} \right ) }

\renewcommand{\sl}{{\mathfrak{sl}}}

\newcommand{\isu}{{\mathfrak{isu}}}

\newcommand{\fr}{\text{FR}}

\newcommand{\tr}{{\mathrm{Tr}}}
\newcommand{\f}{\frac}

\def\nn{\nonumber}

\def\ka{\kappa}
\def\vphi{\varphi}
\def\eps{\epsilon}

\newcommand{\id}{\mathbb{I}}

\def\tx{\tilde{x}}
\def\ty{\tilde{y}}
\def\tg{\tilde{g}}

\def\vp{\vec{p}}
\def\vq{\vec{q}}
\def\vx{\vec{x}}
\def\vX{\vec{X}}
\def\vy{\vec{y}}
\def\vY{\vec{Y}}

\def\hv{\hat{v}}
\def\vsigma{\vec{\sigma}}

\def\tz{\tilde{z}}

\def\tlam{\tilde{\lambda}}
\def\vcG{\overrightarrow{\cG}}

\def\lt{\tilde{\ell}}
\def\ut{\tilde{u}}
\def\mt{\tilde{m}}
\def\vt{\tilde{v}}

\def\rD{r^\dagger}
\def\rT{r_{21}}

\def\zb{\bar{z}}
\def\ztb{\bar{\tilde{z}}}

\def\vT{\overrightarrow{T}}

\def\Tt{\tilde{T}}

\def\jt{\tilde{j}}

\def\bt{\tilde{b}}

\def\balpha{\bar{\alpha}}
\def\bbeta{\bar{\beta}}

\def\Lt{\widetilde{L}}
\def\Ut{\widetilde{U}}
\def\Mt{\widetilde{M}}
\def\Vt{\widetilde{V}}

\newcommand{\alink}[4]
{\draw[decoration={markings,mark=at position 0.6 with {\arrow[scale=1.5,>=stealth]{>}}},postaction={decorate}] (#1) -- node[#3,pos=.5]{$#4$}(#2)}
\newcommand{\link}[2]
{\draw[decoration={markings,mark=at position 0.6 with {\arrow[scale=1.5,>=stealth]{>}}},postaction={decorate}] (#1) --(#2)}

\def\centerarc[#1](#2)(#3:#4:#5)
    { \draw[#1] ($(#2)+({#5*cos(#3)},{#5*sin(#3)})$) arc (#3:#4:#5); }
    
\def\centerarcnodes[#1](#2)(#3:#4:#5)(#6,#7)
    {\coordinate(#6) at ($(#2)+({#5*cos(#3)},{#5*sin(#3)})$);
    \coordinate(#7) at ($(#2)+({#5*cos(#4)},{#5*sin(#4)})$);
    \draw[#1] ($(#2)+({#5*cos(#3)},{#5*sin(#3)})$) arc (#3:#4:#5); }

\def\angcircle(#1)(#2)(#3:#4) {\coordinate(#1) at ($(#2)+({#4*cos(#3)},{#4*sin(#3)})$); }

\begin{document}

\title{$q$-deformed 3D Loop Gravity on the Torus}

\author{{\bf Ma\"it\'e Dupuis}}
\email{mdupuis@perimeterinstitute.ca}
\affiliation{Perimeter Institute for Theoretical Physics, Waterloo, Ontario,
Canada}
\affiliation{Department of Applied Mathematics, University of Waterloo, Waterloo, Ontario, Canada}
\author{{\bf Etera R. Livine}}\email{etera.livine@ens-lyon.fr}
\affiliation{Perimeter Institute for Theoretical Physics,  Waterloo, Ontario, Canada}
\affiliation{Universit\'e  de  Lyon,  ENS  de  Lyon,  Laboratoire  de  Physique,  CNRS  UMR  5672,  F-69342  Lyon,  France}
\author{{\bf Qiaoyin Pan}}\email{qpan@perimeterinstitute.ca}
\affiliation{Perimeter Institute for Theoretical Physics,  Waterloo, Ontario, Canada}
\affiliation{Department of Applied Mathematics, University of Waterloo, Waterloo, Ontario, Canada}

\date{\today}

\begin{abstract}

The $q$-deformed loop gravity framework was introduced as a canonical formalism for the Turaev-Viro model (with $\Lambda < 0$), allowing to quantize 3D Euclidean gravity with a (negative) cosmological constant using a quantum deformation of the gauge group. We describe its application to the 2-torus, explicitly writing the $q$-deformed gauge symmetries and deriving the reduced physical phase space of Dirac observables, which leads back  to the Goldman brackets for the moduli space of flat connections. Furthermore it turns out that the $q$-deformed loop gravity can be derived through a gauge fixing from the Fock-Rosly bracket, which provides an explicit link between loop quantum gravity (for $q$ real) and the combinatorial quantization of 3d gravity as a Chern-Simons theory with non-vanishing cosmological constant $\Lambda<0$. 
A side-product is the reformulation of the loop quantum gravity phase space for vanishing cosmological constant $\Lambda=0$, based on $\SU(2)$ holonomies and $\su(2)$ fluxes, in terms of $\ISU(2)$ Poincar\'e holonomies.
Although we focus on the case of the torus as an example, our results outline the general equivalence between 3D $q$-deformed loop quantum gravity  and the combinatorial quantization of  Chern-Simons theory for arbitrary graph and topology.

\end{abstract}

\maketitle

\section*{Introduction}

Three-dimensional gravity is the ideal test bed for quantum gravity in four space-time dimensions. Indeed, 3D gravity is a topological theory, with no local degree of freedom - no gravitational wave per se, and 4D gravity can be formulated as an almost-topological theory through the Plebanski action  \cite{DePietri:1998hnx,Barrett:1997gw,Barrett:1999qw} or the McDowell-Mansouri action \cite{Smolin:2003qu,Freidel:2005ak}.
More precisely, every solution to the vacuum 3D Einstein equations has constant curvature given by the cosmological constant and, in particular, are flat in the case of a vanishing cosmological constant. Then, the theory only has global degrees of freedom reflecting the non-trivial topology of space-time, the boundary geometry and the coupling of matter fields to gravity.
This reformulation of 3D gravity as a topological field \cite{Witten:1988hc} allows for an exact quantization. We can then compare the various quantization schemes proposed for quantum gravity, test the matter-geometry coupling at the quantum level and clarify some conceptual issues arising in higher space-time dimensions, thereby gaining much insight into 4D quantum gravity.

Let us have a look at the various quantization schemes for 3D gravity. 
First, 3D gravity can  be written as a Chern-Simons theory  \cite{Witten:1988hc,Carlip:2003}. Its canonical quantization is known as the combinatorial quantization \cite{Alekseev:1994pa,Alekseev:1994au,Fock:1998nu,Buffenoir:2002tx}, whereas the path integral approach for Chern-Simons has been developed by Witten \cite{Witten:1988hc}.
From another starting point, 3D gravity, defined by the Palatini action in its first order formulation, can be quantized through a canonical quantization scheme as Loop Quantum Gravity, or directly as a path integral leading to the Ponzano-Regge state-sum for a vanishing cosmological constant \cite{PR,Freidel:2004vi,Freidel:2005bb,Barrett:2008wh,Rovelli:2007quantum} and the Turaev-Viro topological model for a non-vanishing cosmological constant \cite{Turaev:1992state}.
 For a vanishing cosmological constant, most of the approaches have been shown to be equivalent at the end of the day \cite{Freidel:2004ponzano,Meusburger:2010hilbert}.

For the Chern-Simons formulation of 3D gravity, the combinatorial quantization formalism has been developed for an arbitrary gauge group  \cite{Alekseev:1994pa, Alekseev:1994au, Buffenoir:2002tx, Meusburger:2003hc} and quantum groups symmetries arise for any choice of spacetime signature and sign of the cosmological constant. The theory of quantum groups  and Hopf algebra offers a powerful mathematical tool to describe 3D quantum gravity and the question whether such a mathematical feature is fundamental is legitimate. For a vanishing cosmological constant, it has also been shown, in the context of the Ponzano-Regge model, that a quantum group structure encodes the symmetries at the quantum level \cite{Freidel:2004ponzano,Meusburger:2010hilbert}. For a non-vanishing cosmological constant, the Turaev-Viro topological state-sum   \cite{Turaev:1992state} defined in terms of the representations of the $q$-deformed $\cU_q(\su(2))$ at $q$ root of unity describes 3D Euclidean quantum gravity with a positive cosmological constant. A relation between the path integral quantization of Chern-Simons and the Turaev-Viro model was explicitly given  in \cite{Roberts:1995skein}. 

In the loop quantum gravity formalism with a non-zero cosmological constant, the appearance of a quantum group as a symmetry group  is not straightforward, which complicates making a direct  connection to the other frameworks. A first link was made in  \cite{Noui:2011im,Pranzetti:2014xva}, where the quantum group structure was used to regularize the Hamiltonian constraint of 3D loop quantum gravity for a positive cosmological constant leading back to the  Turaev-Viro transition amplitudes. For 3D Euclidean gravity with a negative cosmological constant, it was  proposed in  \cite{Bonzom:2014wva,Bonzom:2014bua} to implement the cosmological constant via a  $q$-deformation of the standard loop gravity phase space, with the deformation paramater $q$ given in terms of the cosmological constant.  Then, the quantization of the classical $q$-deformed phase space and $q$-deformed gauge symmetry, formulated as the Heisenberg double and Drinfeld double of $\SU(2)$  naturally defines  a quantum group structure - $\cU_q(\su(2))$ with $q$ real. This framework admits an elegant interpretation in terms of (classical and quantum) discrete hyperbolic geometry \cite{Bonzom:2014wva,Dupuis:2013lka} and is related to the Turaev-Viro amplitude (generalized for $q$ real) \cite{Bonzom:2014bua}. 
It thus seems to be a promising candidate to understand 3D quantum gravity with a non-zero cosmological constant\footnotemark.
\footnotetext{
Nevertheless, the derivation of this $q$-deformed loop gravity framework from the Palatini action is still missing.
Understanding the discretization procedure to go from the Palatini action to the $q$-deformed phase space  might be the final ingredient that will allow to describe 3D loop quantum gravity in terms of a quantum  group symmetry. 
}
Another open question with this formulation is an explicit relation with the well-studied combinatorial quantization framework for the Chern-Simons theory.
One guide at the classical level  is that the phase space structures of both the $q$-deformed loop gravity phase space and the Fock and Rosly phase space \cite{Fock:1998nu} -- the classical phase space underlying the combinatorial quantization -- can be written in terms of classical $r$-matrices, which may imply a  link between the two approaches.

\medskip

The goal of this paper is to study the link between loop gravity and Chern-Simons at the discrete level. We will focus on the simplest non-trivial compact spacial hypersurface - the 2-torus.
Thus, starting from the $q$-deformed 3D loop gravity framework applied to the 2-torus, we  derive the physical phase space of Dirac observables and  show  that we recover, as expected, the Goldman brackets \cite{Goldman:1986invariant}. 
Then, the second step is the  reconstruction of the $q$-deformed loop gravity phase space on the 2-torus from the Fock-Rosly description of Chern-Simons. More precisely, we show that loop gravity for a given graph embedded in the 2-torus can be viewed as a partial gauge fixing of the Fock-Rosly phase space defined on a larger (that we will call `fatter") graph. This serves as a first step to relate the combinatorial quantization and the loop quantum gravity scheme for a non-zero cosmological constant. We expect that our construction can be directly generalized to arbitrary topologies indicating a general equivalence between q-deformed loop quantum gravity and the combinatorial quantization of Chern-Simons theory.  

This paper is organized as follows. In Section \ref{sec:flat}, we first review the loop gravity phase space construction with a zero cosmological constant and apply it on the 2-torus. Then, a key step toward the deformation of this model is the reformulation of the phase space as a Heisenberg double. In this context, the symmetries generated by the constraints (closure (or Gauss) constraint and flatness constraint) naturally  appear as Poisson-Lie groups. An interesting result obtained at this stage is the reformulation of 3D loop gravity with a zero cosmological constant as a canonical theory of a flat Poincar\'e connection.   In Section \ref{sec:deformed}, we apply the setup of \cite{Bonzom:2014wva} on the torus to build a $q$-deformed phase space with a negative cosmological constant. A set of physical observables - the Wilson loops, are checked to give the Goldman brackets. Section \ref{sec:Fock-Rosly} contains the main result of the paper. We start with the Fock-Rosly description and recover, by an asymmetric gauge fixing, the Poisson structure and constraint system of loop gravity described in Section \ref{sec:deformed}.

\section{Reviewing the flat case with vanishing cosmological constant}
\label{sec:flat}

We start by reviewing the canonical analysis for 3D loop gravity with $\Lambda=0$ on the 2-torus. This flat model will serve as the point of comparison for 3D loop gravity with a non-vanishing cosmological constant.
At the kinematical level, the phase space is defined by the holonomy-flux observables, which define 2D discrete geometries. The dynamics are then implemented through the Hamiltonian constraints, which encode the theory's gauge invariance under 3D diffeomorphisms.
Applying this framework to the 2-torus, we explicitly construct the kinematical phase space for a basic graph embedded on the torus, and solve the Hamiltonian constraints to obtain the reduced physical phase space and identify the Dirac observables.

We further show that the procedure, from the kinematical to the physical phase space, can be entirely recast in the language of the Heisenberg double algebraic structure, which will be the starting point of the generalization to the curved theory with $\Lambda\neq 0$ as described in \cite{Bonzom:2014wva}.
This allows to re-write the phase space in terms of $\ISU(2)$ holonomies and $\ISU(2)$ flatness constraints and thus reformulate 3D loop gravity at $\Lambda=0$ as the canonical theory for a flat Poincar\'e connection.

\subsection{Loop Gravity Phase Space on the Torus}

\begin{figure}[h!]

\begin{subfigure}[t]{0.3\linewidth}
\begin{tikzpicture}[scale=1.4]

\coordinate(abc) at (.5,0.35) ;
\coordinate(acd) at (.4,-0.33);
\coordinate(bce) at (.9,0.53);
\coordinate(cdj) at (.93,-0.48);
\coordinate(cej) at (1.2,0);
\coordinate(eij) at (1.63,-0.07);
\coordinate(eig) at (1.8,0.3);
\coordinate(egf) at (1.83,0.7);
\coordinate(fgh) at (2.33,0.55);
\coordinate(igh) at (2.25,0.1);

\coordinate(ab) at (-0.35,0.6);
\link{abc}{ab};
\coordinate(ad) at (-0.35,-0.5);
\link{ad}{acd};
\coordinate(dj) at (1.1,-0.9);
\link{cdj}{dj};
\coordinate(ij) at (1.9,-0.65);
\link{eij}{ij};
\coordinate(ih) at (2.6,-0.4);
\link{ih}{igh};
\coordinate(fh) at (2.9,0.75);
\link{fgh}{fh};
\coordinate(ef) at (1.5,.9);
\link{egf}{ef};
\link{bce}{ef};
\coordinate(top) at (1,1.1);
\link{top}{ef};

\alink{abc}{acd}{left}{{e}};
\link{bce}{abc};
\link{acd}{cdj};
\link{cdj}{cej};
\link{bce}{cej};
\link{cej}{eij};
\link{eij}{eig};
\link{eig}{egf};
\link{egf}{fgh};
\link{fgh}{igh};
\link{eig}{igh};

\end{tikzpicture}
\caption{Oriented graph $\Gamma$.}

\end{subfigure}
\hspace*{2mm}
\begin{subfigure}[t]{0.32\linewidth}
\begin{tikzpicture}[scale=1]

\coordinate(a) at (0,0) ;
\coordinate(b) at (2.5,0);

\alink{a}{b}{above}{g_{e}};

\draw (a)node{$\bullet$}++(0,-0.3)node{$x_{e}$};
\draw (b)node{$\bullet$}++(0.5,-0.3)node{$\tx_{e}=g_{e}\triangleright x_{e}$};

\draw (1.4,-1.3) node{$(g_{e},x_{e})\in\SU(2)\times\su(2)$};

\end{tikzpicture}
\caption{$T^*\SU(2)$ phase space on the edge $e$.}
\end{subfigure}
\hspace*{2mm}
\begin{subfigure}[t]{0.3\linewidth}
\begin{tikzpicture}[scale=1.2]

\coordinate(O) at (0,0) ;
\draw (O) node{$\bullet$};
\draw (-0.3,0) node{$v$};

\coordinate(a) at ($({cos(10)},{sin(10)})$); 
\coordinate(b) at ($({cos(75)},{sin(75)})$); 
\coordinate(c) at ($({cos(130)},{sin(130)})$); 
\coordinate(d) at ($({cos(220)},{sin(220)})$); 
\coordinate(e) at ($({cos(300)},{sin(300)})$); 

\link{O}{a};
\link{O}{b};
\link{O}{d};
\link{c}{O};
\link{e}{O};

\draw (0,-1.5) node{$\cG_{v}=\sum_{e| v=s(e)}x_{e}-\sum_{e| v=t(e)}\tx_{e}$};

\end{tikzpicture}
\caption{Closure constraint at a vertex $v$.
}
\end{subfigure}

\caption{Holonomy-Flux phase space for 3D loop gravity on a graph $\Gamma$.}
\label{fig:edgevariables}
\end{figure}
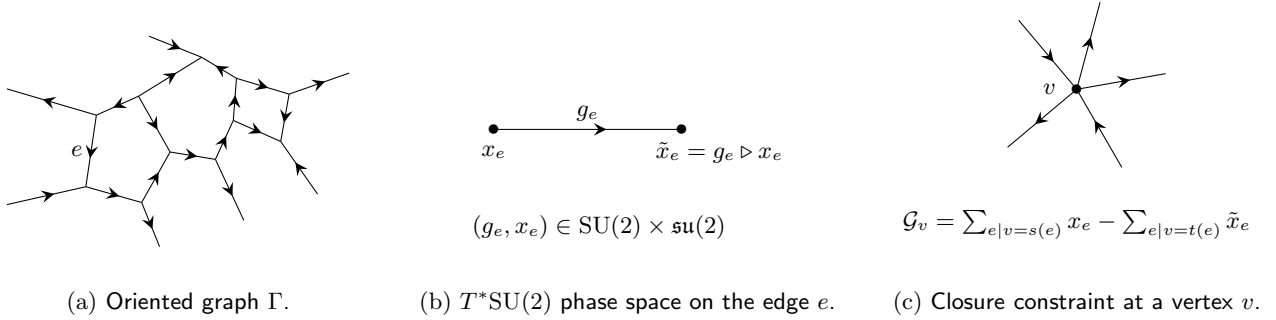

{\bf Holonomy-flux phase space for 3D loop gravity:}
The phase space for canonical 3D loop gravity encodes the basic degrees of freedom of a discretized 2D surface (see \cite{Dupuis:2017otn,Freidel:2018pbr} for a thorough and careful discretization). Considering an oriented graph $\Gamma$, the basic building block is a $T^*\SU(2)$ phase space associated to each link $e\in\Gamma$. For each oriented link, we define a group element $g_{e}\in\SU(2)$ along the link and a Lie algebra vector $x_{e}\in\su(2)\sim\R^3$ thought of as living on the source of the link, as illustrated on fig.\ref{fig:edgevariables}. The group element $g_{e}$ gives the holonomy of the $\SU(2)$ connection along the edge while the vector $x_{e}$ is the discretized geometric flux transverse to that edge, defined as the integrated triad along the edge of a 2D triangulation dual to the graph.

Decomposing the flux vector on the  Pauli matrix basis for Hermitian matrices, $x_{e}=x_{e}^a\sigma^a$, the $T^*\SU(2)$ symplectic structure is explicitly given by the Poisson brackets:
\be
\{x_{e}^a,g_{e}\}=\f i2\sigma^ag_{e}
\,,\quad
\{x_{e}^a,x_{e}^b\}=\eps^{abc}x_{e}^c
\,,\quad
\{g_{e},g_{e}\}=0
\,.
\label{eq:poisson_flat}
\ee
We also define the flux vectors at the target vertex by a parallel transport by the $\SU(2)$ holonomy, $\tx_{e}=g_{e}\triangleright x_{e}=g_{e}^{-1}x_{e}g_{e}$, satisfying flipped $\su(2)$ algebra Poisson brackets:
\be
\{x_{e}^a,\tx_{e}^b\}=0
\,,\quad
\{\tx_{e}^a,g_{e}\}=\f i2 g_{e} \sigma^a
\,,\quad
\{\tx_{e}^a,\tx_{e}^b\}=-\epsilon^{abc}\tx_{e}^c
\,.
\label{eq:poisson_flat_2}
\ee
%
It can be convenient to also project the $\SU(2)$ group element onto the Pauli basis and introduce the ``momentum'' variables:
\be
p^{0}_{e}=\f12\tr g_{e}
\,,\quad
p_{e}^{a}=\f1{2i}\tr g_{e}\sigma^a
\,,\quad
g_{e}=p_{e}^{0}\,\id+{i} p_{e}^{a}\sigma^a
\,,\qquad \text{with}\,\quad
(p_{e}^{0})^2+\vp_{e}{}^2=1
\,.
\ee
In these variables, the Poisson brackets given above in eqn.\eqref{eq:poisson_flat} now read:
\be
\{x_{e}^a,p_{e}^0\}=-\f12 p^a_{e}
\,,\quad
\{x_{e}^a,p_{e}^b\}=\f12 \delta^{ab} p_{e}^0+\f12 \eps^{{abc}}p_{e}^c
\,,\quad
\{p_{e}^\mu,p_{e}^\nu\}=0
\,,\quad
\mu,\nu=0..3
\,.
\label{eq:momentum}
\ee

Now the loop quantum gravity phase space on the graph $\Gamma$ is defined by considering the collection of the independent $T^*\SU(2)$ phase spaces living on each link $e\in\Gamma$ and coupling them at the graph nodes by a closure constraint -or Gauss law- at each vertex $v\in\Gamma$:
\be
\cG^a_{v}=\sum_{e| v=s(e)}x_{e}^a-\sum_{e| v=t(e)}\tx_{e}^a
\,.
\ee
Imposing $\vcG_{v}={\bf 0}$ amounts to requiring that the incoming flux at the vertex $v$ equals the outgoing flux. This closure constraint generates the  gauge invariance under $\SU(2)$ transformations around the vertex:
\be
\{\cG^a,\cG^b\}=\eps^{abc}\cG^c
\,,\qquad
\begin{array}{ll}
\forall e \,|\,v=s(e)\,,&
\{\cG^a,x_{e}^b\}=\eps^{abc}x_{e}^c
\,,
\vspace*{2mm}\\
\forall e \,|\,v=t(e)\,,&
\{\cG^a,\tx_{e}^b\}=\eps^{abc}\tx_{e}^c
\,.
\end{array}
\ee

The symplectic quotient of the product of the edge phase spaces $T^*\SU(2)^E$ (where $E$ counts the number of edges of the graph) by the closure constraint
defines the kinematical phase space of 3D loop quantum gravity on the graph $\Gamma$. %
In the context of loop quantum gravity in 3+1 dimensions, this holonomy-flux phase space is interpreted as defined discrete three-dimensional twisted geometries \cite{Freidel:2010aq} (see also \cite{Dupuis:2012yw,Freidel:2013bfa,Freidel:2018pvm}).
Here, in 3D space-time dimensions, these are meant to represent 2D discrete geometries. This becomes explicit once we impose the Hamiltonian constraints for 3D loop quantum gravity, implemented as flatness constraints for the $\SU(2)$ holonomies around loops of the graph, which implies that we can reconstruct a 2D geometric triangulation dual to the graph (see e.g. \cite{Bonzom:2011hm,Bonzom:2011nv,Bonzom:2013tna}).


\medskip

\begin{figure}[h!]
	\begin{tikzpicture}[scale=1]
	
	\coordinate (O) at (0,0);
	\coordinate (A) at (5,0);
	\coordinate (B) at (5,3);
	\coordinate (C) at (0,3);
	\draw (O)--(A)--(B)--(C)--(O);
	
	\coordinate (cc) at (2.5,1.5);
	\coordinate (n) at (2.5,3);
	\coordinate (s) at (2.5,0);
	\coordinate (o) at (0,1.5);
	\coordinate (e) at (5,1.5);
	
	\draw[thick,decoration={markings,mark=at position 0.55 with {\arrow[scale=1.3,>=stealth]{>}}},postaction={decorate}]  (o) -- node[pos= 0.85,below]{$\tx$}(cc); 
	\draw[thick,decoration={markings,mark=at position 0.55 with {\arrow[scale=1.3,>=stealth]{>}}},postaction={decorate}]  (cc)  -- node[pos= 0.2,below]{$x$} node[midway, above]{$g$} (e) ; 
	\draw[thick,decoration={markings,mark=at position 0.55 with {\arrow[scale=1.3,>=stealth]{>}}},postaction={decorate}]  (s) -- node[pos= 0.7,right]{$\ty$}(cc); 
	\draw[thick,decoration={markings,mark=at position 0.55 with {\arrow[scale=1.3,>=stealth]{>}}},postaction={decorate}]  (cc) -- node[pos= 0.2,right]{$y$}node[midway, left]{$h$} (n); 
	
	\end{tikzpicture}
	\caption{Twisted geometry phase on the 2-torus, parametrized by $g,h\in\SU(2)$ and $x,\tx,y,\ty\in\R^3$, as two copies of $T^*\SU(2)$ related by the closure constraint inducing the $\SU(2)$ gauge invariance.
		}
	\label{fig:basictorus}
\end{figure}
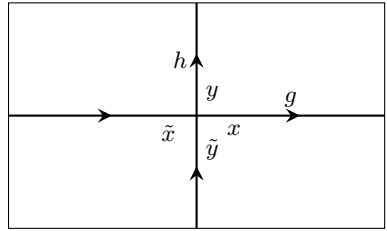

{\bf Holonomy-flux phase space on the torus:}
We apply this framework to the 2-torus, which is our main object of study. Let us thus introduce the twisted geometry phase space on the basic graph for the 2-torus, with two edges wrapping around the torus meeting at a single vertex and surrounding a single face as illustrated on fig.\ref{fig:basictorus}.

We associate each edge with a pair of holonomy-flux variables, namely $(g,x)$ on the horizontal edge, as drawn on fig.\ref{fig:basictorus}, and $(h,y)$ on the vertical edge, equipped with the Poisson structure defined in \eqref{eq:poisson_flat},
\be
\left|\begin{array}{lcl}
\{x^a,g\}&=&\tfrac i2\sigma^a g
\,,\vspace*{1mm}\\
\{x^a,x^b\}&=&\eps^{abc}x^c
\,,\vspace*{1mm}\\
\{g,g\}&=&0\,,
\end{array}
\right.
\qquad\qquad
\left|\begin{array}{lcl}
\{y^a,h\}&=&\f i2\sigma^a h
\,,\vspace*{1mm}\\
\{y^a,y^b\}&=&\eps^{abc}y^c
\,,\vspace*{1mm}\\
\{h,h\}&=&0\,.
\end{array}
\right.
\ee
The two pairs of variables are independent in the kinematical level, 
\be
\{x^a,y^b\}=\{x^a,h\}=\{y^a,g\}=\{g,h\}=0
\,.
\ee
The vectors $x$ and $y$ are sitting at the source vertex of their respective link, and we define the fluxes $\tx=g\triangleright x=g^{-1}xg$ and $\ty=h\triangleright y=h^{-1}yh$ at the target  of the links.

It is useful to introduce the vector $X^a\equiv x^a-\tx^a$, which forms a closed Lie algebra with the $\SU(2)$ holonomy $g$,
\be
\{X^a,g\}=\f i2\,\big{[}\sigma^a,g\big{]}
\,,\quad
\{X^a,X^b\}=\eps^{abc}X^c
\,,\quad
\{g,g\}=0
\,,
\label{eq:gX}
\ee
or equivalently written in terms of the momentum variables $p^\mu$ for the group element $g$:
\be
\{X^a,X^b\}=\eps^{abc}X^c
\,,\quad
\{X^{a},p^{b}\}=\eps^{abc}p^c
\,,\quad
\{p^{\mu},p^\nu\}=\{p_{0},X^a\}=0
\,.
\ee
%
This Poincar\'e algebra has two Casimirs, the mass $\vp\,{}^2=p^{a}p^{a}$ (or equivalently $p^{0}=\sqrt{1-\vp\,{}^2}$)  and the spin $\vX\cdot \vp$.
Let us point out that the vector $X=x-\tx=x-g\triangleright x$ is orthogonal to $\vp$ by definition, so that the spin $\vX\cdot \vp$ automatically vanishes.
We similarly introduce the vector $Y=y-\ty$ and the momentum variables  $q^\mu$ for the $\SU(2)$ holonomy $h$.

Then we impose the closure constraint -or Gauss law- at the vertex:
\be
\cG^a=x^a-\tx^a+y^a-\ty^a=X^a+Y^a=0\,,
\label{eq:gaussbasic}
\ee
which generates $\SU(2)$ gauge transformations, that is 3D rotations on the four vectors $x,\tx,y,\ty$ and the $\SU(2)$-action by conjugation on the two holonomies $g$ and $h$:
\be
\begin{array}{lcl}
g&\overset{G\in\SU(2)}\longmapsto& GgG^{-1}\,,
\\
h&\overset{G}\longmapsto& GgG^{-1}\,,
\end{array}
\qquad
\begin{array}{lcl}
(x,\tx)&\overset{G}\longmapsto& (GxG^{-1},G\tx G^{-1})\,,
\\
(y,\ty)&\overset{G}\longmapsto& (GyG^{-1},G\ty G^{-1})\,.
\end{array}
\ee
Taking the symplectic quotient by the Gauss law means both assuming that the variables satisfy the constraints $\cG^a=0$ and quotienting by the action it generates, i.e. considering  only $\SU(2)$-invariant observables.

\subsection{Physical Observables on the Torus}

{\bf The Hamiltonian constraint algebra on the torus:} 

The graph on the 2-torus, with a single vertex and two edges wrapping around the torus cycles as on fig.\ref{fig:basictorus}, defines a cellular decomposition for the torus with a single face. The Hamiltonian constraints consist in a flatness constraint around that face, which amounts to imposing the flatness of the $\SU(2)$ connection on the 2-torus:
\be
\cF=ghg^{-1}h^{-1}=\id
\,.
\label{eq:flatnessbasic}
\ee
While the closure constraints $\cG^a$ generates $\SU(2)$ gauge transformations at the graph vertex, the flatness constraints $\tr\,\cF\sigma^a$ generate translations of the flux vectors. 
Together they define a first class system of constraints:
\be
\{\cG^a,\cG^b\}=\eps^{abc}\cG^c
\,,\quad
\{\cG^{a},\cF\}=\f i2\,\big{[}\sigma^a,\cF\big{]}
\,,\quad
\{\cF,\cF\}=0
\,.
\ee
This constraint algebra for 3D loop gravity is identified as the Poincar\'e algebra.

\medskip

{\bf Dirac observables and Reduced phase space:}
The reduced phase space, or physical phase space, consists in the Dirac observables, which commute with both the closure  and flatness constraints. Since the closure constraint generates $\SU(2)$ transformations, amounting to 3D rotations on the vectors $\vx,\vp,\vy,\vq$, we can focus on rotation-invariant observables.
As a result, we identify four independent Dirac observables, namely $\vx\cdot\vp$, $\vy\cdot\vq$, $\vp\,{}^2$ and $\vq\,{}^2$,
\be
\{\cG,\vx\cdot\vp\}=\{\cG,\vy\cdot\vq\}=\{\cG,\vp\,{}^2\}=\{\cG,\vq\,{}^2\}=0
\,,\qquad
\{\cF,\vp\,{}^2\}=\{\cF,\vq\,{}^2\}=0
\,,
\ee
\be
\{\tr\,\sigma^a\cF,\vx\cdot\vp\}
=
\f12\tr\Big{(}
\sigma^ag(\id-\cF)
\Big{)}
\underset{\cF=\id}\sim0
\,,\qquad
\{\tr\,\sigma^a\cF,\vy\cdot\vq\}
=
\f12\tr\Big{(}
\sigma^aghg^{-1}(\id-\cF)
\Big{)}
\underset{\cF=\id}\sim0
\,,
\ee
where we have used the definition of the momentum variable for the group elements, $i\vp\cdot\vsigma=g-p^0\id$ and $i\vq\cdot\vsigma=h-q^0\id$. 

We compute the Poisson brackets between those Dirac observables, substituting  $\vp\,{}^2$ and $\vq\,{}^2$ by  $p_0$ and $q_{0}$ for the sake of simplifying the notations,
\be
\{\vx\cdot\vp,p_{0}\}
=
-\f12(1-p_{0}^2)
\,,\qquad
\{\vy\cdot\vq,q_{0}\}
=
-\f12(1-q_{0}^2)
\,.
\ee
It is fairly easy to identify Darboux coordinates on this reduced physical phase space:
\be
\Big{\{}
\f{\vx\cdot\vp}{\sqrt{\vp\,{}^2}}
\,,\,
2\arccos p_{0}
\Big{\}}
=1
\,,\qquad
\Big{\{}
\f{\vy\cdot\vq}{\sqrt{\vq\,{}^2}}
\,,\,
2\arccos q_{0}
\Big{\}}
\,,
\ee
which we interpret as pairs of physical length-angle conjugate variables.

This can be made more explicit by solving the constraints. First of all, the flatness implies that $g$ and $h$ commute, i.e. (assuming that they are not the identity) they have the same rotation axis, say the unit vector $\hv\in\cS_{2}$ on the 2-sphere:
\be
g=e^{i\f\theta2 \hv\cdot\vsigma}
\,,\quad
h=e^{i\f\vphi2 \hv\cdot\vsigma}
\,,\quad
p_{0}=\f12\tr g=\cos\f\theta2
\,,\quad
q_{0}=\f12\tr h=\cos\f\vphi2
\,.
\ee
The rotation angles $\theta$ and $\vphi$ are Dirac observables, invariant under both rotations and translations, while the rotation axis $\hat{v}$ is a gauge variable.

\begin{figure}[h]
\begin{tikzpicture}[scale=1]
	\coordinate (O) at (0,0);
	
   \begin{scope}
    \def\rx{0.71*3}
    \def\ry{0.15*3}
    \def\z{0.725*3}
	
    \coordinate (L) at (-\rx,\z);
    \coordinate (R) at (\rx,\z);
    \coordinate (g) at (0, 1.5*\z);
    \coordinate (xh) at (0,\z);
    \coordinate (L2) at (-0.2*\rx,\z);
    \coordinate (L3) at (-\rx,0);
    
    \coordinate (the) at (-0.1*\rx,0.8*\z);
    \path [name path = ellipse1]    (0,\z) ellipse ({\rx} and {\ry});
    \path [name path = angle] (0,\z) ellipse ({0.2*\rx} and {0.2*\ry});
    \path [name path = radi] (O)-- (0.8*\rx,\z);
    
    \path [name path = horizontal] (0,\z-\ry*\ry/\z)
                                -- (\rx,\z-\ry*\ry/\z);
    \path [name intersections = {of = ellipse1 and radi,by=E}];

    \draw[fill = gray!50, gray!50] (L) -- (0,0)
      -- (R) -- cycle;

    \draw[fill = gray!30,gray!30] (0,\z) ellipse ({\rx} and {\ry});
  \end{scope}

  \draw[->, thick] (O) -- (L) node [at end, left] {$\vec{x}$};
  \draw[->, thick] (O) -- (E) node [at end, right] {$\vec{\tx}$};
  \draw[->, thick] (O) -- (g) node [at end, left] {$\hv$};
  
  \draw[dashed] (L) -- (xh);
  \draw[dashed] (E) -- (xh);
  
  \tkzFindAngle (E,O,L) \tkzGetAngle{an}  \FPround\an\an{0};

  \tkzMarkAngle[arc=l,size=0.3 cm,,line width=1pt,color=red](L,xh,E)
  
  \draw[red] (the) node {$\theta$};
  
  \draw[->,line width=0.5mm,red] (O) --node[midway,right]{$\vec{x}_{\parallel}$}(xh);
  \draw[->,line width=0.5mm,blue] (O) --node[midway, below]{gauge}node[midway,above]{$\vec{x}_{\perp}$}(L3);

\end{tikzpicture}
\caption{Length-angle variables. The flux vector $\vec{\tx}$ is obtained by acting with the rotation $g$ on the vector $\vec{x}$, i.e. rotating it around $\hat{v}$ by an angle $\theta$, with the parametrization $g=e^{i\f\theta2 \hv\cdot\vsigma}\in \SU(2)$. The Dirac observables ({\it \red{in red}}) are the angle $\theta$ and the longitudinal projection of $\vec{x}$ on the direction of $g$, i.e. $\vx\cdot\hv=\vx\cdot\vp/\sqrt{\vec{p}^2}$. The gauge variables are the orthogonal component of  the flux vector, $\vec{x}_{\perp}$ ({\it \blue{in blue}}) and the direction of the rotation $\hv$. }	
\label{fig:geo}
\end{figure}
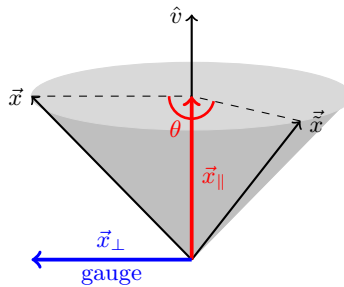

Now turning to the flux vectors, the flatness constraint does not impose any condition while the closure constraint imposes that the vector $\vX+\vY$ vanishes. The vector $\vX$ is obtained as $x-\tx$ with $\tx=g\triangleright x$ obtained by rotating $x$ around $\hv$ by an angle $\theta$. As a result, $\vX$ is always orthogonal to the rotation axis $\hat{v}$ and does not depend on the longitudinal projection $\vec{x}\cdot\hat{v}$.
Decomposing $\vec{x}=\vec{x}_{\parallel}+\vec{x}_{\perp}$ and similarly $\vec{y}=\vec{y}_{\parallel}+\vec{y}_{\perp}$ in terms of components along the direction $\hat{v}$ and orthogonal to that direction, the vectors $\vX$ and $\vY$ only depend on the transversal components, $\vec{X}=\vec{x}_{\perp} - g\triangleright \vec{x}_{\perp}$ and $\vec{Y}=\vec{y}_{\perp} - g\triangleright \vec{y}_{\perp}$. 
The closure constraint $\vX+\vY=0$ is then a condition on the transversal components of $x$ and $y$ and actually uniquely fixes $\vec{y}_{\perp}$ in terms of $\vec{x}_{\perp}$ at given $\theta$ and $\phi$. More precisely, $\vec{y}_{\perp}$ is obtained\footnotemark{} from $\vec{x}_{\perp}$ by a rotation of angle $(\theta-\vphi)/2+\pi$ and rescaling of the oriented modulus by a factor $\sin({\theta}/2)/\sin({\vphi}/2)$.
\footnotetext{
In the special case when $\theta=\vphi$, the closure constraint simply reduces to $\vec{x}_{\perp}+\vec{y}_{\perp}=0$.}
The closure constraint is thus solved by choosing the orthogonal projection for the flux along one cycle, $\vec{x}_{\perp}$, determining  $\vec{y}_{\perp}$ from $\theta$, $\vphi$ and $\vec{x}_{\perp}$, and adding to them arbitrary longitudinal components $\vec{x}_{\parallel}$ and $\vec{y}_{\parallel}$ in order to obtain  solutions for $\vx$ and $\vy$.
To summarize, as illustrated on fig.\ref{fig:geo}, the longitudinal projections $\vec{x}\cdot\hat{v}$ and $\vec{y}\cdot\hat{v}$ are invariant under translations and are Dirac observables, while the orthogonal projection $\vec{x}_{\perp}$, and thus $\vec{y}_{\perp}$, -both in direction and norm- are pure gauge.

%

\subsection{3D Loop gravity as a theory of a flat Poincar\'e connection}
\label{ISU2theory}

Here, we propose to revisit the standard loop gravity formalism, reviewed above, to reformulate the holonomy-flux phase space in terms of Poincar\'e holonomies. This is achieved in two steps.
First, we write both $\SU(2)$ holonomies and flux vectors as Poincar\'e group elements. We show that the closure constraint and the flatness constraint can be combined together into a single Poincar\'e flatness constraint. The main tool is fattening the graph $\Gamma$ into a ribbon graph, as proposed for $q$-deformed loop gravity in \cite{Bonzom:2014wva,Dupuis:2014fya}: the graph links are upgraded to ribbons and the parallel transport equation between the source flux and target flux of each link is rewritten as a flatness constraint around the corresponding ribbon.
Second, we show that the Poisson brackets between the Poincar\'e group elements are derived from a $r$-matrix, inherited from seeing $\ISU(2)$ as a Heisenberg double.
This allows to recast the 3D loop gravity phase space and constraint algebra at vanishing cosmological constant $\Lambda=0$ using the same structures as the $q$-deformed case corresponding to a non-vanishing cosmological constant $\Lambda<0$ . 

\medskip
{\bf Holonomy-flux as a Poincar\'e holonomy:}
The main idea behind reformulating the 3D loop gravity phase space in terms of Poincar\'e holonomies is to consider the flux vectors as Poincar\'e translations in $\ISU(2)$  while the $\SU(2)$ holonomies define the $\SU(2)$ rotations in $\ISU(2)$. This allows to put the flux vectors and the $\SU(2)$ holonomies on the same level and to re-package them in a single $\ISU(2)$ object.

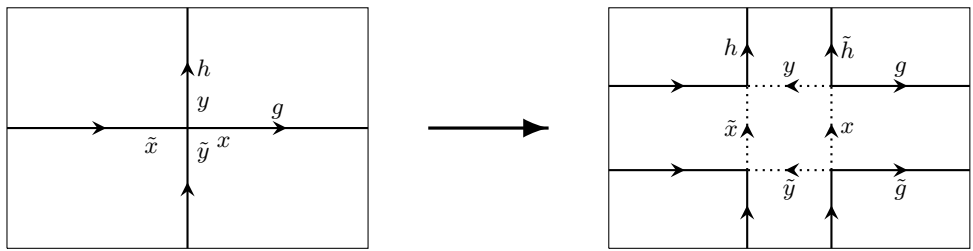
\begin{figure}[h!]

	\begin{tikzpicture}[scale=0.8]

	\coordinate (o) at (-10,0);
	\coordinate (a) at (-4,0);
	\coordinate (b) at (-4,4);
	\coordinate (c) at (-10,4);
	\draw (o)--(a)--(b)--(c)--(o);

	\coordinate (m) at (-7,2);
	\coordinate (u) at (-7,4);
	\coordinate (d) at (-7,0);
	\coordinate (r) at (-4,2);
	\coordinate (l) at (-10,2);

	\draw[thick,decoration={markings,mark=at position 0.55 with {\arrow[scale=1.3,>=stealth]{>}}},postaction={decorate}] (m)-- node[pos= 0.2,right]{$y$}node[midway, right]{$h$}(u); 
	\draw[thick,decoration={markings,mark=at position 0.55 with {\arrow[scale=1.3,>=stealth]{>}}},postaction={decorate}] (m)-- node[pos= 0.2,below]{$x$}node[midway,above]{$g$}(r); 
	\draw[thick,decoration={markings,mark=at position 0.55 with {\arrow[scale=1.3,>=stealth]{>}}},postaction={decorate}] (d)-- node[pos= 0.8,right]{$\tilde{y}$}(m); 
	\draw[thick,decoration={markings,mark=at position 0.55 with {\arrow[scale=1.3,>=stealth]{>}}},postaction={decorate}] (l)-- node[pos= 0.8,below]{$\tilde{x}$}(m); 
	
	\draw[ very thick,decoration={markings,mark=at position 1 with {\arrow[scale=1.5,>=latex]{>}}},postaction={decorate}] (-3,2)-- (-1,2); 
	
	\coordinate (O) at (0,0);
	\coordinate (A) at (6,0);
	\coordinate (B) at (6,4);
	\coordinate (C) at (0,4);
	\draw (O)--(A)--(B)--(C)--(O);
	
	\coordinate (a) at (2.3,1.3);
	\coordinate (b) at (3.7,1.3);
	\coordinate (c) at (2.3,2.7);
	\coordinate (d) at (3.7,2.7);
	
	\coordinate (ld) at (0,1.3);
    \coordinate (lu) at (0,2.7);
    \coordinate (rd) at (6,1.3);
    \coordinate (ru) at (6,2.7);
    \coordinate (dl) at (2.3,0);
    \coordinate (dr) at (3.7,0);
    \coordinate (ul) at (2.3,4);
    \coordinate (ur) at (3.7,4);
    
    \coordinate (f1) at (3,2);
    \coordinate (f2) at (1.15,0.65);
    \coordinate (f3) at (1.15,2);
    \coordinate (f4) at (3,0.65);


	\draw[thick,decoration={markings,mark=at position 0.55 with {\arrow[scale=1.3,>=stealth]{>}}},postaction={decorate}] (ld) -- (a); 
	\draw[thick,decoration={markings,mark=at position 0.55 with {\arrow[scale=1.3,>=stealth]{>}}},postaction={decorate}] (b)  --node[midway, below]{$\tilde{g}$}(rd); 
	\draw[thick,decoration={markings,mark=at position 0.55 with {\arrow[scale=1.3,>=stealth]{>}}},postaction={decorate}] (lu) --(c); 
	\draw[thick,decoration={markings,mark=at position 0.55 with {\arrow[scale=1.3,>=stealth]{>}}},postaction={decorate}] (d) --node[midway,above]{$g$}(ru);
	 
	\draw[thick,decoration={markings,mark=at position 0.55 with {\arrow[scale=1.3,>=stealth]{>}}},postaction={decorate}] (dl) -- (a); 
	\draw[thick,decoration={markings,mark=at position 0.55 with {\arrow[scale=1.3,>=stealth]{>}}},postaction={decorate}] (dr)  --(b); 
	\draw[thick,decoration={markings,mark=at position 0.55 with {\arrow[scale=1.3,>=stealth]{>}}},postaction={decorate}] (c) -- node[midway, left]{$h$}(ul); 
	\draw[thick,decoration={markings,mark=at position 0.55 with {\arrow[scale=1.3,>=stealth]{>}}},postaction={decorate}] (d)  --node[midway, right]{$\tilde{h}$}(ur) ;
	
	\draw[thick,dotted,decoration={markings,mark=at position 0.55 with {\arrow[scale=1.3,>=stealth]{>}}},postaction={decorate}] (a) --node[midway, left]{$\tx$}(c); 
	\draw[thick,,dotted,decoration={markings,mark=at position 0.55 with {\arrow[scale=1.3,>=stealth]{>}}},postaction={decorate}] (b) --node[midway, right]{$x$}(d); 
	
	\draw[thick,dotted,decoration={markings,mark=at position 0.55 with {\arrow[scale=1.3,>=stealth]{>}}},postaction={decorate}] (b) --node[midway,below]{$\ty$}(a); 
	\draw[thick,dotted,decoration={markings,mark=at position 0.55 with {\arrow[scale=1.3,>=stealth]{>}}},postaction={decorate}] (d) --node[midway,above]{$y$}(c);

\end{tikzpicture}

\caption{We upgrade the original graph $\Gamma$ by fattening it into a ribbon graph on the torus: each graph link becomes a ribbon with $\SU(2)$ holonomies running along both its sides and flux vectors along its two extremities.}
\label{fig:fatgraph}
\end{figure}

The crucial step to achieve this is to promote the original graph $\Gamma$ to the corresponding fat graph $\Gamma^{fat}$, as illustrated on fig.\ref{fig:fatgraph}. The fat graph is defined by thickening the links of the graph $\Gamma$, there by turning them into ribbons, and similarly enlarging the node into a surface patching the ribbons together. This ribbon graph now has $\SU(2)$ group elements running along the links on both side of the ribbons, $g$ and $\tg$ instead of solely $g$, and the flux vectors are attached to new lines transverse to the original link and going around the nodes, as shown on fig.\ref{fig:fatgraph}. This method was used in \cite{Bonzom:2014wva,Dupuis:2014fya} to define the phase space for $q$-deformed loop gravity.

Now we put  holonomies $g\in\SU(2)$ and flux vectors $x\in\su(2)$ in the same type of objects, embedding them in $\SU(2)\times \su(2)$ by  promoting $\SU(2)$ group elements $g$ to elements $(g,0)$ and flux vectors $x$ to elements $(\id,x)$. We endow this larger space with an associative product, turning it into the Poincar\'e group $\ISU(2)=\SU(2)\ltimes \su(2)$:
\be
(g_{1},x_{1})(g_{2},x_{2})
=
(g_{1}g_{2},x_{1}+g_{1}x_{2}g_{1}^{-1})
\,,\qquad
(g,x)^{-1}=(g^{-1},-g^{-1}xg)
\,.
\ee
This allows to put the parallel transport conditions along the links, the closure constraints at the nodes and the $\SU(2)$ flatness constraints on the same footing and write all of them as flatness of $\ISU(2)$ holonomies.

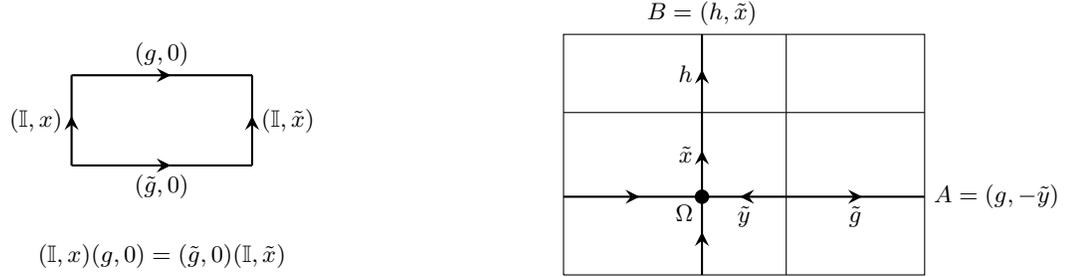
\begin{figure}[h!]
\begin{subfigure}[t]{0.45\linewidth}

	\begin{tikzpicture}[scale=0.6]

	\coordinate (O) at (0,0);
	\coordinate (A) at (4,0);
	\coordinate (B) at (4,2);
	\coordinate (C) at (0,2);	

	\draw[thick,decoration={markings,mark=at position 0.55 with {\arrow[scale=1.3,>=stealth]{>}}},postaction={decorate}] (O)  --node[midway, below]{$(\tilde{g},0)$}(A); 
	\draw[thick,decoration={markings,mark=at position 0.55 with {\arrow[scale=1.3,>=stealth]{>}}},postaction={decorate}] (O)  --node[midway, left]{$(\id,x)$}(C);
		\draw[thick,decoration={markings,mark=at position 0.55 with {\arrow[scale=1.3,>=stealth]{>}}},postaction={decorate}] (C)  --node[midway, above]{$({g},0)$}(B);
		\draw[thick,decoration={markings,mark=at position 0.55 with {\arrow[scale=1.3,>=stealth]{>}}},postaction={decorate}] (A)  --node[midway, right]{$(\id,\tilde{x})$}(B);

\draw (2,-2) node{$(\id,x)(g,0)=(\tg,0)(\id,\tx)$};
		
\end{tikzpicture}

\caption{\label{fig:ribbonflatness}
The original link dressed with the $\SU(2)$ group element $g$, the source flux $x$ and target flux $\tx$ has been lifted to a ribbon. Imposing the flatness of the Poincar\'e holonomy around the ribbon leads back to the parallel transport condition along the original link: the $\SU(2)$ holonomies along both side of the ribbon are identical, $g=\tg$, while the target flux $\tx$ is obtained by the action of $g$ on the source flux $x$.
}
\end{subfigure}
\hspace*{4mm}
\begin{subfigure}[t]{0.45\linewidth}
\begin{tikzpicture}[scale=0.8]

	\coordinate (O) at (0,0);
	\coordinate (A) at (6,0);
	\coordinate (B) at (6,4);
	\coordinate (C) at (0,4);
	\draw (O)--(A)--(B)--(C)--(O);
	
	\coordinate (a) at (2.3,1.3);
	\coordinate (b) at (3.7,1.3);
	\coordinate (c) at (2.3,2.7);
	\coordinate (d) at (3.7,2.7);
	
	\coordinate (ld) at (0,1.3);
    \coordinate (lu) at (0,2.7);
    \coordinate (rd) at (6,1.3);
    \coordinate (ru) at (6,2.7);
    \coordinate (dl) at (2.3,0);
    \coordinate (dr) at (3.7,0);
    \coordinate (ul) at (2.3,4);
    \coordinate (ur) at (3.7,4);
    
    \coordinate (f1) at (3,2);
    \coordinate (f2) at (1.15,0.65);
    \coordinate (f3) at (1.15,2);
    \coordinate (f4) at (3,0.65);

	\draw[thick,decoration={markings,mark=at position 0.55 with {\arrow[scale=1.3,>=stealth]{>}}},postaction={decorate}] (ld) -- (a); 
	\draw[thick,decoration={markings,mark=at position 0.55 with {\arrow[scale=1.3,>=stealth]{>}}},postaction={decorate}] (b)  --node[midway, below]{$\tilde{g}$}(rd)node[right]{$A=(g,-\tilde{y})$}; 
	\draw (lu) --(c); 
	\draw (d) --(ru);
	 
	\draw[thick,decoration={markings,mark=at position 0.55 with {\arrow[scale=1.3,>=stealth]{>}}},postaction={decorate}] (dl) -- (a); 
	\draw (dr)  --(b); 
	\draw[thick,decoration={markings,mark=at position 0.55 with {\arrow[scale=1.3,>=stealth]{>}}},postaction={decorate}] (c) -- node[midway, left]{$h$}(ul)node[above]{$B=(h,\tilde{x})$}; 
	\draw(d)  --(ur) ;
	
	\draw[thick,decoration={markings,mark=at position 0.55 with {\arrow[scale=1.3,>=stealth]{>}}},postaction={decorate}] (a) --node[midway, left]{$\tx$}(c); 
	\draw (b) --(d); 
	\draw[thick,decoration={markings,mark=at position 0.55 with {\arrow[scale=1.3,>=stealth]{>}}},postaction={decorate}] (b) --node[midway,below]{$\ty$}(a); 
	\draw(d) --(c);

	\fill (a) circle (0.12);
	\draw (a) node[below left]{$\Omega$};
	
\end{tikzpicture}
\caption{\label{fig:rootholo}
Choosing a root vertex $\Omega$ out of the four corners around the internal face corresponding to the original graph node, we define two $\ISU(2)$ group elements $A$, $B$, which combine flux vectors and $\SU(2)$ holonomies: these carry the same data as the holonomy-flux variables on the original graph.
}
\end{subfigure}

\caption{Ribbon flatness constraints and Poincar\'e holonomies}
\label{fig:ribbon-flat}
\end{figure}

First, considering an edge turned into a ribbon, we impose a  flatness constraint around the ribbon, requiring that the ordered product of the Poincar\'e group elements around the ribbon be equal to $\id$, as drawn on fig.\ref{fig:ribbonflatness}. This ribbon flatness constraint  becomes the parallel transport condition between source and target flux vectors:
\be \label{flatribbon}
(\id,x)(g,0)=(g,x)
\,,\quad
(\tg,0)(\id,\tx)=(\tg,\tg\tx\tg^{-1})
\,,\qquad
(\id,x)(g,0)=(\tg,0)(\id,\tx)
\Rightarrow 
\left|
\begin{array}{l}
\tg=g\\
\tx=g^{-1}xg
\end{array}
\right.
\,.
\ee
Then the flat graph consists in four faces: two ribbon faces corresponding to the upgraded graph edges, one ``internal'' face corresponding to the thickened graph node and one face corresponding to the original unique face of the graph on the torus. The flatness of the Poincar\'e holonomy around the internal face encodes the closure constraint,
\be
(\id,x)(\id,y)(\id,\tx)^{-1}(\id,-\ty)^{-1}=(\id,x-\tx+y-\ty)=(\id,0)
\,,
\ee
while the flatness around the original face still encodes the $\SU(2)$ flatness constraint defining the Hamiltonian constraints of 3D loop gravity,
\be
(g,0)(h,0)(g,0)^{-1}(h,0)^{-1}=(ghg^{-1}h^{-1},0)=(\id,0)
\,,
\ee
where we used $g=\tg$, $h=\tilde{h}$ coming from the ribbon flatness constraints \eqref{flatribbon}.

Assuming the ribbon flatness, we can actually go further and gather both closure and flatness constraints in a single $\ISU(2)$ flatness constraint. More precisely, we choose a root vertex $\Omega$ around the internal face, i.e. one of the four corners of the face corresponding to the original graph node as shown on fig.\ref{fig:rootholo}. Then, we define the Poincar\'e holonomies rooted at that point $\Omega$:
\be
A=(\id,\ty)^{-1}(g,0)=(g,-\ty)
\,,\qquad
B=(\id,\tx)(h,0)=(h,\tx)
\,.
\ee
These two $\ISU(2)$ holonomies, $A$ and $B$, wrapping around the two torus cycles  contain the exact same information as the original holonomy-flux variables $g,h,x,y$. We keep the graph drawn by these two holonomies and put aside the remaining structure of the ribbon graph: this reduces the fat graph back to the original graph, as we can see on fig.\ref{fig:rootholo}.
We introduce the Poincar\'e flatness constraint around the single face:
\be
AB=BA
\Leftrightarrow
(gh,-\ty+x)
=
(hg,\tx-y)
\Leftrightarrow
\left|
\begin{array}{l}
gh=hg \\
x-\tx+y-\ty=0
\end{array}
\right.
\,\,,
\ee
showing that $\SU(2)$ flatness and closure are indeed repackaged into a single $\ISU(2)$ flatness constraint.

This shows how to reformulate the loop gravity holonomy-flux variables into Poincar\'e holonomies, by promoting the graph to a ribbon graph then by choosing a root vertex on the ribbon graph around each graph node in order to define $\ISU(2)$ holonomies running along the original graph. This procedure is a priori straightforwardly generalizable to arbitrary graphs.

The resulting Poincar\'e group elements, $A$ and $B$, inherit Poisson brackets from the 3D loop  gravity Poisson brackets described earlier. We show below that these can be re-derived entirely from the perspective of the Poincar\'e group $\ISU(2)$ in terms of the $r$-matrix associated to $\ISU(2)$ interpreted as a Heisenberg double. This would conclude the entire reformulation of 3D loop gravity in terms of a flat $\ISU(2)$ connection.

\medskip
{\bf Poisson bracket between $\ISU(2)$ holonomies:}
The Poisson brackets for the Poincar\'e holonomies $A$ and $B$ follows from loop gravity brackets:
\be
\label{AA}
A=(g,-\ty)
\,,\qquad
\{g,g\}=0
\,,\quad
\{\ty^a,\ty^b\}=-\eps^{abc}\ty^c
\,,\quad
\{g,\ty\}=0
\,,
\ee
\be
\label{BB}
B=(h,\tx)
\,,\qquad
\{h,h\}=0
\,,\quad
\{\tx^a,\tx^b\}=-\eps^{abc}\tx^c
\,,\quad
\{h,\tx\}=0
\,,
\ee
as well as the Poisson brackets $\{A,B\}$ between the $\ISU(2)$ holonomies:
\be
\label{AB}
\{g,h\}=0
\,,\quad
\{\ty,\tx\}=0
\,,\quad
\{\ty^a,h\}=\f i2 h\sigma^a
\,,\quad
\{g,\tx^a\}=-\f i2 g\sigma^a
\,.
\ee
Taking into account the previous analysis of the  algebra formed by the closure and flatness constraints in the loop gravity phase space, it is  clear that the equivalent $\ISU(2)$ flatness constraint $ABA^{-1}B^{-1}=\id$ define a constraint system of first class.

\smallskip

Using the fat graph tool, we have started from the graph $\Gamma$ decorated with $\SU(2)$ holonomies $g$ and $h$ and flux vectors $x,y$, together with closure and $\SU(2)$ flatness constraints, and reformulated the 3D loop gravity phase space again on the graph $\Gamma$ but decorated with $\ISU(2)$ holonomies $A$ and $B$, constrained by a single $\ISU(2)$ flatness constraint.

Let us underline that $A$ and $B$ mix the flux vectors and $\SU(2)$ holonomies belonging to different links of the graph according to the combinatorial structure of the fat graph and the chosen root vertex $\Omega$ on it. This means that $A$ and $B$ are {\it not} the $\ISU(2)$ group elements following the ribbons:
\be
A,\,B \quad \ne \quad (g,x),\,(h,y)
\,.
\ee
In particular the Poisson brackets $\{A,A\}$ has a different structure than the Poisson bracket $\{(g,x),(g,x)\}$. Indeed, in the case of $A$, its $\SU(2)$ component commutes with the translation, while in the case of the Poincar\'e holonomy  on the ribbon $(g,x)$ the $\SU(2)$ and $\su(2)$ components have non-trivial Poisson brackets  inherited from the initial $T^*\SU(2)$ structure of the loop gravity phase space.

Moreover, there is a trade-off: the non-vanishing Poisson brackets $\{g,x\}$ around a ribbon leads to the non-vanishing of the Poisson bracket $\{A,B\}\ne 0$. This is a key point of this reformulation in terms of Poincar\'e holonomies: the two $\ISU(2)$ do not commute with each other, $\{A,B\}\ne 0$, while the original $\SU(2)$ holonomies along the same links did commute, $\{g,h\}=0$. The fact that holonomies living on different graph edges nevertheless meeting at a same node commute in the loop gravity phase space while their Fock-Rosly brackets does not vanish is the main point of discord between the loop quantum gravity scheme and the combinatorial quantization procedure for 3D gravity as a Chern-Simons theory. Here the shift of perspective from $\{g,h\}=0$ to  $\{A,B\}\ne 0$ promises a reconciliation between the loop gravity  and the Fock-Rosly phase spaces and hints towards a way to bridge between 3D loop quantum gravity  and the combinatorial quantization.
We will show in the next section \ref{sec:deformed} that this is indeed the case in general for a non-vanishing cosmological constant, with an explicit map between the $q$-deformed loop gravity phase space and the Fock-Rosly brackets. This is the main result of this paper.

\medskip
{\bf Poincar\'e group as Heisenberg double and the $\ISU(2)$ Poisson bracket:}
To anticipate the methods used for a non-vanishing cosmological constant, both in the $q$-deformed loop gravity framework or in the combinatorial quantization framework using the Fock-Rosly bracket, it is crucial to realize that the brackets between the $\ISU(2)$ holonomies, given above in eqn. \eqref{AA}, \eqref{BB} and \eqref{AB}, can be written in terms of a $r$-matrix.
Indeed, the Poincar\'e group can be considered as a Heisenberg double $\ISU(2)=\SU(2)\ltimes \su(2)$. Then, the Poisson structure of $\ISU(2)$ is encoded in the $r$-matrix given as the tensor product of the generators of $\SU(2)$ and $\su(2)$ (see e.g. \cite{Ahluwalia:1993rq,Kosmann:1997lie}).

We start by representing Poincar\'e group elements $(g,x)\in\SU(2)\times\su(2)$ in terms of 4$\times$ 4 matrices written as 2$\times$2 block matrices\footnotemark:
\be
(g,x)=\mat{c|c}{g & ixg \\ \hline 0 & g}
\,,
\ee
where $g\in\SU(2)$ is represented in its fundamental representations as 2$\times$2 unitary matrices and $x=\vx\cdot\vsigma\in\su(2)$ is represented as a 2$\times$2 traceless Hermitian matrix. For $x\ne 0$, the expression $xg$ is the polar decomposition of an arbitrary 2$\times$2  invertible complex matrix.
\footnotetext{
Another parametrization of Poincar\'e group elements was used in  \cite{Bonzom:2014wva}, the spin-1 representation for the $\SU(2)$ group elements instead of the spin-$\tfrac 12$ used here. It still led to 4$\times$4 matrices, but the $\SU(2)$ group elements were encoded in a 3$\times$3 block while the $\su(2)$ vectors were written as 3-vectors:
\be
(g,x)=\mat{c|c}{D^1(g) & \vec{x} \\\hline 0 & 1}\,,
\ee
with $D^1(g)$ the 3$\times$3 Wigner matrix representing the $\SU(2)$ group element as a 3D rotation acting on 3-vectors.
}
These provide a representation of the Poincar\'e group multiplication:
\be
(g_{1},x_{1})(g_{2},x_{2})
=
\mat{c|c}{g_{1} & ix_{1}g_{1} \\ \hline 0 & g_{1}}
\mat{c|c}{g_{2} & ix_{2}g_{2} \\ \hline 0 & g_{2}}
=
\mat{c|c}{g_{1}g_{2}  & i(x_{1}+g_{1}x_{2}g_{1}^{-1})g_{1}g_{2}  \\ \hline 0 & g_{1}g_{2} }
=
(g_{1}g_{2},x_{1}+g_{1}x_{2}g_{1}^{-1})
\,.
\ee
Each Poincar\'e group element admits a unique Iwasawa decomposition as the product of a translation and a rotation:
\be
(g,x)
=
\mat{c|c}{g & ixg \\ \hline 0 & g}
=
\mat{c|c}{\id & ix \\ \hline 0 & \id}
\mat{c|c}{g & 0 \\ \hline 0 & g}
=
(\id,x)(g,0)
=
\ell u
\qquad\textrm{with}\quad
\ell=(\id,x)
\quad\textrm{and}\quad
u=(g,0)
\,.
\ee
The generators of the Poincar\'e Lie algebra are the $J^a$'s for the $\SU(2)$ subgroup and $E^a$ for the $\su(2)$ subgroup :
\be
J^a=\f12\mat{c|c}{\sigma^a & 0 \\ \hline 0 & \sigma^a}
\,,\qquad
u=e^{iv^aJ^a} = \mat{c|c}{g & 0 \\ \hline 0 & g}
\quad\textrm{with}\quad
g=e^{\f i2 v^a \sigma^a}
=\cos\f{|v|}2\id_{2}+i\f1{|v|}\sin\f{|v|}2\,v^a\sigma^a
\,,
\ee
\be
E^a=\mat{c|c}{0 & \sigma^a \\ \hline0  & 0}
\,,\quad
E^aE^b=0\,,\,\,\forall a,b
\,,\qquad
\ell=e^{ix^aE^a}
=
\id+ix^aE^a
=
\mat{c|c}{\id & ix \\ \hline 0 & \id}
\quad\textrm{with}\quad
x=x^a\sigma^a
\,,
\ee
which satisfies the Poincar\'e algebra commutators:
\be
[J^a,J^b]=i\eps^{abc}J^c
\,,\qquad
[J^a,E^b]=i\eps^{abc}E^c
\,,\qquad
[E^a,E^b]=0
\,.
\ee

Provided with the bilinear form on the Lie algebra $\isu(2)$ spanned by the $E^a$ and $J^a$,
\be
\cB(M,N):= \tr \Bigg{[}MN\mat{c|c}{0 & \id \\ \hline\id & 0}\Bigg{]}\,,\quad \forall M,N\in\isu(2)\,,
\ee
which defines a pairing between the rotation and translation generators, $\cB(E^a,J^b)=\delta^{ab}$, $\cB(E^a,E^b)=\cB(J^a,J^b)=0$, 
the Heisenberg double structure defines a $r$-matrix:
\be
r=\sum_{a}E^a\otimes J^a\,,\qquad 
\rT=\sum_{a}J^a\otimes E^a\,,
\ee
which naturally satisfies the classical Yang-Baxter equation\footnotemark{} \cite{Ahluwalia:1993rq,Kosmann:1997lie}.
\footnotetext{
It is straightforward to check that the postulated $r$-matrix indeed satisfies the classical Yang-Baxter equation:
\be
[r_{12},r_{13}]+[r_{12},r_{23}]+[r_{13},r_{23}]
=
\sum_{a,b}
E_{a}\otimes [J_{a},E_{b}]\otimes J_{b}
+
\sum_{a,b}
E_{a}\otimes E_{b}\otimes  [J_{a},J_{b}]
=
i\left(\sum_{a,b}
\epsilon^{abc}E_{a}\otimes E_{c}\otimes J_{b}
+
\epsilon^{abc}E_{a}\otimes E_{b}\otimes J_{c}
\right)=
0
\,,
\nn
\ee
where we used that $E_{a}E_{b}$ vanishes for all indices $a$ and $b$. Moreover the symmetric part of the $r$-matrix, $r^s=(r+r_{21})/2$ defines a Casimir for the Lie group:
\be
r^s=\tfrac12(r+r_{21})=\tfrac12\sum_{a}E^a\otimes J^a+J^a\otimes E^a
\,,\qquad
\big{[}(g,x)\otimes(g,x),r^s\big{]}
=
0
\,,\,\,\forall (g,x)\in\ISU(2)
\,.
\nn
\ee
}
$\rT$ denotes the $r$-matrix after swapping the two components of the tensor product, with the implicit convention that $r=r_{12}$.
Using the standard notation for tensor products, $M_{1}=M\otimes \id$ and $M_{2}=\id\otimes M$,  this $r$-matrix defines a Poisson bracket on the Poincar\'e group $\ISU(2)$ endowing it with a phase space structure:
\be
\label{Poincarebracket}
\{\ell_1,\ell_2\}=-[r,\ell_1\ell_2]
\,,\qquad
\{u_1,u_2\}=-[\rT,u_1u_2]
\,,\qquad
\{\ell_1,u_2\}=-\ell_1ru_2
\,,\qquad
\{u_1,\ell_2\}=\ell_2\rT u_1
\,,
\ee
which can be directly written in a compact form as a Poisson bracket for an arbitrary Poincar\'e group element :
\be
D\equiv (g,x)=\ell u\,,\qquad
\{D_1,D_2\}=-rD_1D_2+D_1D_2\rT.
\ee
Explicitly computing the Poisson brackets between $\ell$ and $u$ leads back to the $T^*\SU(2)$ brackets of the 3D loop gravity phase space for the variables on an edge:
\be
\{g,g\}=0
\,,\qquad
\{x^a,g\}=\f i2 \sigma^a g
\,,\qquad
\{x^a,x^b\}=\eps^{abc}x^c
\,.
\ee
In particular, it is interesting that the Poisson bracket $\{u_{1},u_{2}\}$ vanishes because $u\otimes u$ commutes with $r$ for $u\in\SU(2)$. This Poisson bracket will become non-trivial in the deformed case accounting for a non-vanishing cosmological constant.

In order to get the bracket for the target flux $\tx$, assuming the parallel transport equation along the edge or equivalently the ribbon flatness constraint amounts to switching from the right to left Iwasawa decomposition:
\be
D=(g,x)=(\id,x)(g,0)=(g,0)(\id,\tx).
\ee
Then decomposing the Poincar\'e group element $D=\tilde{u}\tilde{\ell}$ with actually $\tilde{u}=u$ gives similar Poisson brackets:
\be
\{\tilde{\ell}_1,\tilde{\ell}_2\}=[r,\tilde{\ell}_1\tilde{\ell}_2]
\,,\qquad
\{\tilde{\ell}_1,\ut_2\}=-\ut_2 r\tilde{\ell}_1
\,,\qquad
\{\ut_1,\tilde{\ell}_2\}=\ut_1 \rT \tilde{\ell}_2
\,,\qquad
 \{\ut_1,\ut_2\}=[\rT,\ut_1\ut_2]
\,,
\ee
which leads to the switched $T^*\SU(2)$ brackets for $\tilde{g}=g$ and $\tx$:
\be
\{g,g\}=0
\,,\qquad
\{\tx^a,g\}=\f i2  g\sigma^a
\,,\qquad
\{\tx^a,\tx^b\}=-\eps^{abc}\tx^c
\,.
\ee

\smallskip

This allows to reformulate the Poisson bracket for the $\ISU(2)$ holonomy $D=(g,x)$ going along the ribbon in terms of the $r$-matrix for the Poincar\'e group seen as a Heisenberg double. It turns out that we can also write the Poisson brackets for the  $\ISU(2)$ holonomies, $A=(g,-\ty)$ and $B=(h,\tx)$, wrapping around the two torus cycles in terms of the same $r$-matrix but with slightly different formulas. It is straightforward to show that\footnotemark:
\be
\label{PoissonABflat}
\left|\begin{array}{lcl}
\{A_{1},A_{2}\}&=&-[r,A_{1}A_{2}]
\vspace*{1mm}\\
\{B_{1},B_{2}\}&=&[r,B_{1}B_{2}]
\vspace*{1mm}\\
\{A_{1},B_{2}\}&=&B_{2}rA_{1}+A_{1}\rT B_{2}
\,.
\end{array}\right.
\ee
\footnotetext{
Writing $\ell=(\id,x)$ and $u=(g,0)$ for the first ribbon and $m=(\id,y)$ and $v=(h,0)$ for the second ribbon, the pairs of variables $\ell,u$ and $m,v$ commute with each other and independently satisfy the Poisson brackets \eqref{Poincarebracket}. Identifying the two Poincar\'e holonomies as $A=(g,-\ty)=\tilde{m}^{-1}u$ and $B=(h,\tx)=\tilde{\ell} v$, it is simple to compute their brackets in terms of the $r$-matrix.
}
This concludes the reformulation of 3D loop gravity kinematics and dynamics as the canonical theory of a flat Poincar\'e connection. The logic and method seem a priori straightforward to be generalized to arbitrary graphs beyond the mere torus.

\section{$q$-deformed Loop Gravity on the Torus}
\label{sec:deformed}

We now turn to the 3D loop quantum gravity with a non-vanishing cosmological constant, which is the heart of the present paper. It is based on the phase space with quantum-deformed braided gauge symmetries developed in \cite{Bonzom:2014wva,Dupuis:2014fya}.
It deforms the $T^*\SU(2)$ phase space for the holonomy-flux variables on each link of the standard loop gravity into the $\SL(2,\C)$ Heisenberg double. This defines the 3D loop gravity phase space for a negative cosmological constant $\Lambda<0$ (and Euclidean signature) and was shown to provide the canonical framework - quantum states and Hamiltonian- for the Turaev-Viro topological spinfoam path integral for 3D quantum gravity \cite{Dupuis:2013quantum,Dupuis:2013lka,Bonzom:2014bua}.

We start by reviewing the $\SL(2,\C)$ phase space on each graph link formulated as ribbons. While the holonomies along the edges still live in $\SU(2)$, the flux do not live anymore in the additive abelian group defined by the $\su(2)$ Lie algebra but now live in the non-abelian group $\SB(2,\C)$.
 This extends the $\ISU(2)$ phase space construction for standard 3D loop quantum gravity  described in the previous section with a deformation parameter 
$\ka$ entering the Poisson bracket through the $r$-matrix on $\SL(2,\C)$. We then  show how to  recover the flat theory when the deformation parameter $\ka$  is sent to 0.

Gluing the edges together at the graph nodes with deformed closure constraints generating a braided non-linear $\SU(2)$ gauge symmetry, this defines the deformed holonomy-flux phase space for the holonomy-flux observables on an arbitrary graph for 3D loop gravity at $\Lambda\ne 0$. Applying this formalism to the torus, we write down the explicit action of the deformed gauge symmetry on the ribbon graph. This allows us to identify simple $\SL(2,\C)$ holonomies wrapping about the torus' cycles, whose traces give the Dirac observables and physical phase space endowed with the expected Goldman  bracket for the moduli space of flat connections on the torus.
This is the main result of this paper and achieves an explicit relation between the 3D loop quantum gravity framework and the combinatorial quantization for 3D gravity as a Chern-Simons theory.

\subsection{The $q$-deformed holonomy-flux phase space}

\medskip

\begin{figure}[h]
	\begin{tikzpicture}[scale=1]
	
	\coordinate (X) at (-6,0.7);
	\coordinate (gl) at (-5.5,0.55);
	\coordinate (gr) at (-3.5,0.55);
	\coordinate (Xt) at (-3,0.7);
	
	\coordinate (L) at (-2.3,0.5);
	\coordinate (R) at (-1.3,0.5);
	
	\coordinate (O) at (0,0);
	\coordinate (A) at (3,0);
	\coordinate (B) at (3,1);
	\coordinate (C) at (0,1);
	
	\draw[thick,decoration={markings,mark=at position 0.55 with {\arrow[scale=1.3,>=stealth]{>}}},postaction={decorate}] (gl) node[below]{$\ell$}--node[midway,above]{$u$}	(gr)node[below]{$\lt$};
	
	\draw[->](L)--(R);

	\draw[thick,dashed,decoration={markings,mark=at position 0.55 with {\arrow[scale=1.3,>=stealth]{>}}},postaction={decorate}] (O) -- node[midway, left]{$\ell$}(C); 
	\draw[thick,dashed,decoration={markings,mark=at position 0.55 with {\arrow[scale=1.3,>=stealth]{>}}},postaction={decorate}] (A) -- node[midway, right]{$\lt$}(B);
	\draw[thick,decoration={markings,mark=at position 0.55 with {\arrow[scale=1.3,>=stealth]{>}}},postaction={decorate}] (C) -- node[midway, above]{$u$}(B);
	\draw[thick,decoration={markings,mark=at position 0.55 with {\arrow[scale=1.3,>=stealth]{>}}},postaction={decorate}] (O) -- node[midway, below]{$\ut$}(A);
	
	\end{tikzpicture}
\caption{
A graph link in $q$-deformed loop gravity carries a $\SL(2,\C)$ group element $D$. Using the Iwasawa decomposition of the $\SL(2,\C)$ group element $D=\ell u$ as the product of a $\SB(2,\C)$ element  times a $\SU(2)$ group element, it defines both the $\SU(2)$ holonomy along the edge $u\in\SU(2)$ and a deformed notion of the flux living at the source node of the link $\ell\in\SB(2,\C)$. These deformed holonomy-flux variables are endowed with a symplectic structure inherited from the classical $r$-matrix for $\sl_{2}$, which defines the phase space on the edge.
The flux at the target node $\lt$ is defined through the opposite Iwasawa decomposition, $D=\ell u = \ut\lt$, where $\ut$ defines the $\SU(2)$ holonomy along the link with flipped orientation. This transport condition $\lt=\ut^{-1}\ell u$ is best reformulated as a $\SL(2,\C)$ flatness condition around a ribbon $\ell u\lt^{-1}\ut^{-1}=\id$. Thickening all the links of the graph into ribbons defines a ribbon graph, or fat graph, where the nodes are  turned into polygons connecting the ribbons.}
\label{fig:box}
\end{figure}
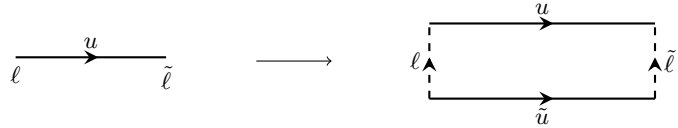

{\bf $\SL(2,\C)$ as the deformed holonomy-flux phase space on an edge:} 
A non-vanishing (negative) cosmological constant deforms the phase space. As argued in \cite{Dupuis:2013quantum,Dupuis:2013lka,Bonzom:2014bua,Bonzom:2014wva,Dupuis:2014fya}, the phase space of holonomy-flux variables on a graph link gets deformed from the standard $T^*\SU(2)$ phase space, equivalent to the $\ISU(2)$ Heisenberg double, to $\SL(2,\C)$. Every group element in $\SL(2,\C)$, written as a 2$\times$2 matrix, admits a unique (left) Iwasawa decomposition as the product of a lower triangular matrix and a $\SU(2)$ matrix:
\be
D\in\SL(2,\C) \quad\Longrightarrow\quad
\exists \,! \,\,(\ell,u)\in\SB(2,\C)\times\SU(2)
\quad\textrm{such that}\quad
D=\ell u
\,,
\ee
with the two subgroups defined as:
\be
\ell=\mat{cc}{\lambda & 0 \\ z & \lambda^{-1}}
\in\SB(2,\C)
\,,\quad
(\lambda,z)\in\R^+\times \C
\qquad
\textrm{and}
\qquad
u=\mat{cc}{\alpha & \beta \\ -\bbeta & \balpha}\in\SU(2)
\,,\quad
|\alpha|^2+|\beta|^2=1
\,.
\ee
The $\SU(2)$ group element $u$ defines the $\SU(2)$ holonomy along the graph link, while the $\SB(2,\C)$ group element $\ell$ is interpreted as a deformed version of the flux vector (at the link's source), as drawn on fig.\ref{fig:box}.

The Iwasawa decomposition translates to a decomposition of the Lie algebra $\sl(2,\C)$ as the direct sum of the Lie algebras  $\su(2)$ and $\Sb(2,\C)$. A basis of generators are the Pauli matrices  $\sigma^a$ and the matrices   $\tau^a=i\kappa(\sigma^a-\f12[\sigma^3,\sigma^a])=\kappa(i\sigma^a+\epsilon^{3ab}\sigma^b)$, with an arbitrary parameter $\ka\in\R$, explicitly:
\be
\sigma^{3}=\mat{cc}{1 & 0 \\ 0 & -1}
\,,\quad
\sigma^{1}=\mat{cc}{0 & 1 \\ 1 & 0}
\,,\quad
\sigma^{2}=\mat{cc}{0 & -i \\ i & 0}
\,,\qquad
[\sigma^a,\sigma^b]=2i\eps^{abc}\sigma^c
\,,
\ee
\be
\tau^3=i\ka \sigma^3
\,,\quad
\tau^1
=2i\ka\sigma^{-}
=2i\ka\mat{cc}{0&0 \\ 1 &0}
\,,\quad
\tau^2=-2\ka\sigma^{-}
\,,\qquad
\left|\begin{array}{l}
{[}\tau^{3},\tau^1{]}=-2i\ka \tau^1
\\
{[}\tau^{3},\tau^2{]}=-2i\ka \tau^2
\\
{[}\tau^{1},\tau^{2}{]}=0 
\end{array}
\right.
\,
\ee
\be
\left|\begin{array}{l}
{[}\sigma^3,\tau^{3}{]}=0
\,,\\
{[}\sigma^3,\tau^{1}{]}=2i\tau_{2}
\,,\\
{[}\sigma^3,\tau^{2}{]}=-2i\tau_{1}
\,,
\end{array}
\right.
\qquad
\left|\begin{array}{l}
{[}\sigma^1,\tau^{3}{]}=-2i(\tau_{2}+\ka\sigma_{1})
\,,\\
{[}\sigma^1,\tau^{1}{]}=2i\ka\sigma_{3}
\,,\\
{[}\sigma^1,\tau^{2}{]}=2i\tau_{3}
\,,
\end{array}
\right.
\qquad
\left|\begin{array}{l}
{[}\sigma^2,\tau^{3}{]}=2i(\tau_{1}-\ka\sigma_{2})
\,,\\
{[}\sigma^2,\tau^{1}{]}=-2i\tau_{3}
\,,\\
{[}\sigma^2,\tau^{2}{]}=2i\ka\sigma_{3}
\,.
\end{array}
\right.
\ee
We can see, from the last sets of commutation relations between the $\sigma$'s and the $\tau$'s, that the $\Sb(2,\C)$ generators $\tau^a$ do not transform as a 3-vector under the $\SU(2)$ action, as for the $\ISU(2)$ group in the flat 3D loop gravity phase space presented in the previous section. In fact, the commutators ${[}\sigma^1,\tau^{3}{]}$ and ${[}\sigma^2,\tau^{3}{]}$ involve both $\sigma$'s and $\tau$'s, which means that there is a feedback action of the translations on the rotations. This is the technical point which distinguishes the deformed phase space from the standard loop gravity phase space for $\Lambda=0$.

The key step is to recognize this double action of the rotations on the translations and vice-versa as identifying $\SL(2,\C)$ as the Heisenberg double $\SL(2,\C)=\SU(2) \bowtie\SB(2,\C)$ (see \cite{Bonzom:2014wva} for more details). The Heisenberg double is provided with a bilinear form on the the Lie algebra $\sl(2,\C)$ that pairs together the rotations $\su(2)$ generators with the deformed translation generators:
\be
\cB(M,N):= \frac{1}{2\kappa}\im \Big{(}\tr \,MN\Big{)}\,,\quad 
M,N\in\sl(2,\C)
\,.
\label{eq:bilinear}
\ee
This bilinear form is such that  $\cB(\tau^a,\sigma_b)=\delta^{ab}$ and $\cB(\tau^a,\tau^b)=\cB(\sigma^a,\sigma^b)=0$. In this sense, $\tau^a$ and $\sigma^a$ are  dual to each other. This pairing leads to the $r$-matrix for $\SL(2,\C)$:
\be
r
=
\f14 \sum_{a}\tau^a\otimes \sigma^a
=
\f{i\ka}4 \sum_{a}\sigma^a\otimes \sigma^a
-\f\ka4\sigma^1\otimes\sigma^2+\f\ka4\sigma^2\otimes\sigma^1
=
\f{i\ka}4\big{(}
\sigma^3\otimes\sigma^3
+
4\,\sigma^-\otimes\sigma^+
\big{)}
\,.
\label{eq:r_matrix_def}
\ee
The swapped $r$-matrix is then equal to its transpose,
\be
\rT=\f14\sum_{a}\sigma^a\otimes \tau^a
\,,\qquad
r^t=\rT
\,,\qquad
r^\dagger=-\rT
\,,
\ee
and its  symmetric part gives a $\SL(2,\C)$ Casimir:
\be
r^s=\f12(r+\rT)=\f{i\ka}4\sum_{a}\sigma^a\otimes \sigma^a
\,,\qquad
(D\otimes D)\,r^s=r^s\,(D\otimes D)
,\quad\forall D\in\SL(2,\C)
\,.
\ee
This $r$-matrix naturally satisfies the classical Yang-Baxter equation and defines the symplectic structure of the Heisenberg double as:
\be
\{\ell_1,\ell_2\}=-[r,\ell_1\ell_2]
\,,\quad
\{\ell_1,u_2\}=-\ell_1ru_2
\,,\quad
\{u_1,\ell_2\}=\ell_2\rT u_1
\,,\quad
\{u_1,u_2\}=-[\rT,u_1u_2]
\,,
\label{eq:poisson_left}
\ee
which can be written in a more compact form in the Poisson bracket for the $\SL(2,\C)$ group element $D=\ell u$:
\be
\{D_1,D_2\}=-rD_1D_2+D_1D_2\rT
\,.
\label{eq:bivector1}
\ee

{\bf Ribbon flatness and parallel transport along the link:} 
The $\SL(2,\C)$ Poisson-Lie group structure described above encodes the holonomy-flux variables along a graph link, more precisely the holonomy $u\in\SU(2)$ flowing along the link and the flux variable at the link's source $\ell\in\SB(2,\C)$. We would like to also define the target flux.
Drawing the link as a ribbon, as on fig.\ref{fig:box}, the target flux $\lt$ is the $\SB(2,\C)$ group element living on the edge at the other extremity of the ribbon. The relation between source and target flux is ensured  by  a ribbon flatness constraint around the ribbon:
\be
\cR =\ell u\lt^{-1}\ut^{-1}=\id\,.
\label{eq:ribbon_constraint}
\ee

The ribbon flatness constraint simply means that composing the source flux with the $\SU(2)$ holonomy is equivalent to composing the $\SU(2)$ holonomy running along the other side of the ribbon with the target flux, thus yielding a unique $\SL(2,\C)$ group element along the ribbon or graph link:
\be
D=\ell u = \ut\lt\,.
\ee
Therefore while the source flux $\ell$ is given by the left Iwasawa decomposition of the $\SL(2,\C)$ group element $D$, the target flux $\lt$ is given by its  right Iwasawa decomposition. As shown in \cite{Bonzom:2014wva},  the Heisenberg double structure gives the Poisson brackets  between $\lt$ and $\ut$, as above in eqn.\eqref{eq:poisson_left} for $\ell$ and $u$:
\be
\{\lt_1,\lt_2\}=[r,\lt_1\lt_2]
\,,\quad
\{\lt_1,\ut_2\}=-\ut_2 r\lt_1
\,,\quad
\{\ut_1,\lt_2\}=\ut_1 \rT \lt_2
\,,\quad
\{\ut_1,\ut_2\}=[\rT,\ut_1\ut_2]
\,.
\label{eq:poisson_right}
\ee
Taking into account that the symmetric part of the $r$-matrix $r^s=\tfrac12(r+\rT)$ is a $\SL(2,\C)$ Casimir, this gives exactly as wanted the same Poisson bracket for the $\SL(2,\C)$ group element:
\be
\{D_1,D_2\}
=
\rT D_1D_2-D_1D_2 r
=
-r D_1D_2+D_1D_2 \rT
\,.
\label{eq:bivector2}
\ee
Moreover, since both left and right Iwasawa decompositions are unique, we can view $\lt$ and $\ut$ as functions of $\ell$ and $u$ defined by $\ell u=\ut\lt$. As shown in  \cite{Bonzom:2014wva}, a little algebra allows to derive the Poisson brackets between the original sector and the tilded sector:
\be
\left|\begin{array}{l}
\{\lt_1,\ell_2\}=0\,,
\\
\{\ut_1,u_2\}=0\,,
\end{array}
\right.
\qquad
\left|\begin{array}{l}
\{\ell_1,\ut_2\}=-r\ell_1\ut_2\,,
\\
\{\ut_1,\ell_2\}=\rT\ut_1\ell_2\,,
\end{array}
\right.
\qquad
\left|\begin{array}{l}
\{\lt_1,u_2\}=-\lt_1u_2r\,,
\\
\{u_1,\lt_2\}=\lt_2u_1\rT\,.
\end{array}
\right.
\label{eq:poisson_mix}
\ee
We can check that these Poisson brackets properly implement the ribbon flatness constraint\footnotemark:
\be
\{\cR_{1},\ell_{2}\}=r_{21}\cR_{1}\ell_{2}-\cR_{1}r_{21}\ell_{2}
\underset{\cR=\id}{\sim}0
\,,\qquad
\{\cR_{1},\ut_{2}\}=\cR_{1}r\ut_{2}-r\cR_{1}\ut_{2}\sim0
\,,\qquad
\{\cR_{1},u_{2}\}=\{\cR_{1},\lt_{2}\}=0\,,
\ee
so that the ribbon flatness constraint $\cR=\id$ is hardcoded in the symplectic structure of the ribbon phase space parametrized by the $\SU(2)$ holonomies $u,\ut$ and non-abelian fluxes $\ell,\lt$.
\footnotetext{
If we start with decoupled variables $\ell,u$ on the one hand and $\lt,\ut$ on the other hand, the ribbon flatness constraint $\cR=\ell u \lt^{-1}\ut^{-1}$ define a 2nd class system of constraints, which fully determine the tilded variables from the original ones. Computing the resulting Poisson brackets for $\lt(\ell,u)$ and $\ut(\ell,u)$ as functions of $\ell$ and $u$ gives eqn.\eqref{eq:poisson_mix}. Another method would be to compute the Dirac brackets. A lengthy but straightforward calculation allows to check that it leads to the same Poisson brackets.
}

\medskip

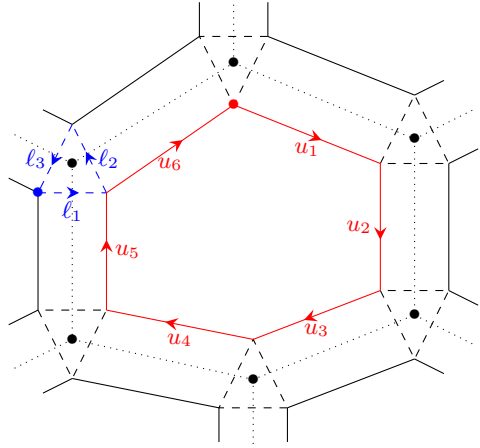
\begin{figure}[h]
	\begin{tikzpicture}[scale=1.3]

\coordinate (a1) at (0,0);
\coordinate (a2) at (0.35,0.7);
\coordinate (a3) at (0.7,0);
\coordinate (a) at (0.35,0.3);

\coordinate (b1) at (2.35,1.6);
\coordinate (b2) at (1.65,1.6);
\coordinate (b3) at (2,0.9);
\coordinate (b) at (2,1.33);
\draw[dashed] (b1)--(b2)--(b3)--(b1);				

\coordinate (c1) at (3.5,0.3);
\coordinate (c2) at (3.85,1);
\coordinate (c3) at (4.2,0.3);
\coordinate (c) at (3.85,0.55);
\draw[dashed] (c1)--(c2)--(c3)--(c1);				

\coordinate (d1) at (0,-1.2);
\coordinate (d2) at (0.35,-1.9);
\coordinate (d3) at (0.7,-1.2);
\coordinate (d) at (0.35,-1.5);
\draw[dashed] (d1)--(d2)--(d3)--(d1);		

\coordinate (f1) at (3.5,-1);
\coordinate (f2) at (3.85,-1.7);
\coordinate (f3) at (4.2,-1);
\coordinate (f) at (3.85,-1.25);
\draw[dashed] (f1)--(f2)--(f3)--(f1);				

\coordinate (e1) at (2.55,-2.2);
\coordinate (e2) at (1.85,-2.2);
\coordinate (e3) at (2.2,-1.5);
\coordinate (e) at (2.2,-1.91);
\draw[dashed] (e1)--(e2)--(e3)--(e1);				

\draw (a2)--(b2);
\draw (b1)--(c2);
\draw (c3)--(f3);
\draw (f2)--(e1);
\draw (e2)--(d2);
\draw (d1)--(a1);
\draw[dotted] (a)--(b)--(c)--(f)--(e)--(d)--(a);

\draw (a) node{$\bullet$};
\draw (b) node{$\bullet$};
\draw (c) node{$\bullet$};
\draw (d) node{$\bullet$};
\draw (e) node{$\bullet$};
\draw (f) node{$\bullet$};

\draw[dotted] (a)--++(-0.6,0.3);
\draw[dotted] (b)--++(0,0.7);
\draw[dotted] (c)--++(0.6,0.3);
\draw[dotted] (d)--++(-0.6,-0.3);
\draw[dotted] (e)--++(0,-0.7);
\draw[dotted] (f)--++(0.6,-0.3);

\draw (a1)--++(-0.3,0.15);
\draw (a2)--++(-0.3,0.15);
\draw (b1)--++(0,0.35);
\draw (b2)--++(0,0.35);
\draw (c2)--++(0.3,0.15);
\draw (c3)--++(0.3,0.15);
\draw (d1)--++(-0.3,-0.15);
\draw (d2)--++(-0.3,-0.15);
\draw (e1)--++(0,-0.35);
\draw (e2)--++(0,-0.35);
\draw (f2)--++(0.3,-0.15);
\draw (f3)--++(0.3,-0.15);

\draw[color=blue] (a1) node {$\bullet$};
\draw[dashed,color=blue,decoration={markings,mark=at position 0.6 with {\arrow[scale=1.5,>=stealth]{>}}},postaction={decorate}] (a1) -- node[below,pos=.5]{$\ell_{1}$}(a3);
\draw[dashed,color=blue,decoration={markings,mark=at position 0.6 with {\arrow[scale=1.5,>=stealth]{>}}},postaction={decorate}] (a3) -- node[right,pos=.5]{$\ell_{2}$}(a2);
\draw[dashed,color=blue,decoration={markings,mark=at position 0.6 with {\arrow[scale=1.5,>=stealth]{>}}},postaction={decorate}] (a2) -- node[left,pos=.5]{$\ell_{3}$}(a1);
				
\draw[color=red] (b3) node {$\bullet$};
\draw[,color=red,decoration={markings,mark=at position 0.6 with {\arrow[scale=1.5,>=stealth]{>}}},postaction={decorate}] (b3) -- node[below,pos=.5]{$u_{1}$}(c1);
\draw[,color=red,decoration={markings,mark=at position 0.6 with {\arrow[scale=1.5,>=stealth]{>}}},postaction={decorate}] (c1) -- node[left,pos=.5]{$u_{2}$}(f1);
\draw[,color=red,decoration={markings,mark=at position 0.6 with {\arrow[scale=1.5,>=stealth]{>}}},postaction={decorate}] (f1) -- node[below,pos=.5]{$u_{3}$}(e3);
\draw[,color=red,decoration={markings,mark=at position 0.6 with {\arrow[scale=1.5,>=stealth]{>}}},postaction={decorate}] (e3) -- node[below,pos=.5]{$u_{4}$}(d3);
\draw[,color=red,decoration={markings,mark=at position 0.6 with {\arrow[scale=1.5,>=stealth]{>}}},postaction={decorate}] (d3) -- node[right,pos=.5]{$u_{5}$}(a3);
\draw[,color=red,decoration={markings,mark=at position 0.6 with {\arrow[scale=1.5,>=stealth]{>}}},postaction={decorate}] (a3) -- node[below,pos=.5]{$u_{6}$}(b3);
				
	\end{tikzpicture}
\caption{
A ribbon graph as the thickening or fat version of its skeleton graph in dotted line. The links become ribbons with the plain edges carrying $\SU(2)$ holonomies and the dashed edges carrying the $\SB(2,\C)$ fluxes. The nodes (black bullets) become little polygons in dashed lines. The flatness constraint around each ribbon translates the parallel transport condition from the source to the target of the corresponding link. The $\SB(2,\C)$ flatness constraint, for example  {\color{blue} $\ell_{1}\ell_{2}\ell_{3}=\id$}   from the root vertex in {\color{blue} blue}, around each node gives the $\SB(2,\C)$ closure constraint at that node and generates $\SU(2)$ gauge transformations for all group elements around that node.
On the other hand, the $\SU(2)$ flatness constraints around large loops, for example {\color{red} $u_{1}..u_{6}=\id$}   from the root vertex in {\color{red} red}, still enforce the flatness constraints of the $\SU(2)$ connection as on the original skeleton graph and generate $\SB(2,\C)$ gauge translations.}
\label{fig:ribbongraph}
\end{figure}

{\bf  $q$-Deformed phase space on a ribbon graph:} 
We now work with ribbon graphs. A ribbon graph is drawn on the canonical surface and trivially embedded, i.e. such that the graph defines a cellular decomposition of the surface, with every elementary cycle of the graph defining  a region of the surface topologically isomorphic to a disk - a face.
We distinguish three types of loops, as illustrated on fig.\ref{fig:ribbongraph}: the ribbons themselves corresponding to the graph links, the vertex loops running around the graph nodes, and the graph loops going around the ``large'' faces. The ribbon graph consists in ribbons glued together at the graph nodes.

Having a $\SL(2,\C)$ phase space on each ribbon, the $q$-deformed phase space for 3D loop gravity on a ribbon graph is defined by the  product of those phase spaces together with $\SB(2,\C)$ closure constraints around every graph node and $\SU(2)$ flatness constraints around graph loop. 

More precisely, every ribbon edge is dressed either with a $\SB(2,\C)$ group element -a flux- or with a $\SU(2)$ group element -a holonomy.  Indeed, the ribbons consist in alternating $\SB(2,\C)$ and $\SU(2)$ group elements, $\ell,u,\lt,\ut$. The ribbon flatness constraint around each ribbon, $\cR=\ell u \lt^{-1}\ut^{-1}$, defines  a set of 2nd class constraints, directly taken into account in the symplectic structure on $\SL(2,\C)$  (by assuming that the left and right Iwasawa decomposition of a $\SL(2,\C)$ group element lead to the same Poisson brackets).
Consequently, the group elements belonging to the same ribbon are endowed with the Poisson brackets given above, while group elements belonging to different ribbons commute with each other. 

Then, as described in details in \cite{Bonzom:2014wva,Dupuis:2014fya}, the vertex loops define $\SB(2,\C)$ closure constraints and the graph loops define $\SU(2)$ flatness constraints. Indeed, a vertex loop consists entirely in $\SB(2,\C)$ fluxes, $\ell_{1},\ell_{2}, ...$, going from corner to corner around the vertex. The flatness around the vertex loop imposes that their oriented product is the identity, $\ell_{1}\ell_{2}\dots=\id$ (for a detail discussion on sign and orientation conventions, we refer the reader to \cite{Bonzom:2014wva,Dupuis:2014fya}). These $\SB(2,\C)$ closure constraints are first class and generate the gauge invariance under local $\SU(2)$ transformations. The $\SU(2)$ action is braided by the fluxes: starting from an assigned root corner for the vertex loop as drawn on fig.\ref{fig:ribbongraph}, if the gauge parameter is the group element $h\in\SU(2)$ at the source of $\ell_{1}$, then the gauge parameter at the next corner will be $h^{(\ell_{1})}$ defined by the braiding relation $h\ell_{1}=\ell_{1}^{(h)}h^{(\ell_{1})}$ with $\ell_{1}^{(h)}\in\SB(2,\C)$ and $h^{(\ell_{1})}\in\SU(2)$. And so on around the vertex. These gauge transformations can be entirely derived from the Hamiltonian flow of the vertex loop flatness constraint defined by the Poisson bracket with the properly oriented $\SB(2,\C)$ product $\ell_{1}\ell_{2}\dots$.

Finally, the graph loops consists entirely in $\SU(2)$ holonomies, $u_{1},u_{2},u_{3},...$ going around a ``large'' face. The flatness around the graph loop imposes that their oriented product is the identity, $u_{1}u_{2}u_{3}\dots=\id$. These $\SU(2)$ flatness constraints are first class and generate a gauge invariance under deformed translations \cite{Bonzom:2014wva}, which implement the gauge invariance of the theory under (space-time) diffeomorphisms. These are the Hamiltonian constraints of 3D loop gravity. The  flow of the  $\SU(2)$ flatness constraints and their braiding is described in details in \cite{Bonzom:2014wva}.

In the present work, we will not review the details of the general structure of those three types of $\SL(2,\C)$ flatness constraints on a generic ribbon graph, but we will focus on the specific example of the 2-petal flower graph on the torus, presented below in section \ref{sec:qtorus}. This will allow to present the formalism in the simplest non-trivial graph and show explicitly the inner-working of the rotation and translation gauge transformations and their braiding. Having identified the gauge transformations will then allow us to define Dirac observables in terms of $\SL(2,\C)$ holonomies and describe the physical Poisson bracket on the reduced phase space, finally recovering the Goldman bracket for the moduli space of flat discrete $\SL(2,\C)$ connections.

\medskip

{\bf Recovering the standard holonomy-flux in the flat limit :} 
It was shown in \cite{Bonzom:2014wva} that, upon imposing $\SU(2)$ flatness constraints around loops of the ribbon graph, the $\SL(2,\C)$ phase space describes discrete 2D hyperbolic triangulations, and it was further proved  in \cite{Bonzom:2014bua} that quantizing the $\SL(2,\C)$ phase space leads to $q$-deformed spin networks for 3D loop quantum gravity. These previous works shows that this framework allows to take into account a (negative) cosmological constant through the deformation of the $T^*\SU(2)$ phase space of standard loop gravity to the $\SL(2,\C)$ ribbon phase space and give the identification of the deformation parameter $\kappa\equiv c^{-1}{G\sqrt{-\Lambda}}$ in terms of the cosmological constant $\Lambda$, where $G$ is the gravitational coupling constant and $c$ the speed of light. 

Indeed, as on the one hand we have shown in the previous section \ref{ISU2theory} that it is possible to reformulate the $T^*\SU(2)$ holonomy-flux phase space in terms of the $r$-matrix of the Poincar\'e gorup $\ISU(2)$ considered as Heisenberg double, we can on the other hand reformulate the $\SL(2,\C)$ phase space in terms of a $\SU(2)$ holonomy and a flux vector and show that we do recover the standard loop gravity phase space in the flat limit $\ka\rightarrow 0$.

\smallskip

The flux vectors are defined from projecting the $\SB(2,\C)$ group elements onto the Pauli matrices:
\be
\ell=\bpm \lambda & 0 \\ z & \lambda^{-1} \epm
\,,\qquad 
T\equiv\ell\ell^\dagger =\bpm \lambda^2 & \lambda \zb \\ \lambda z & \lambda^{-2}+{|z|}^2 \epm
\,,\qquad
T^0\equiv\frac{1}{2 \kappa}\tr \,T\,,\quad
T^a\equiv\frac{1}{2\kappa}\tr\,T{\sigma}^a\,,
\ee
where the 4-vector $T^\mu$ lives on the space-like 3-hyperboloid in the 3+1 Minkowski space,  $T^\mu T_\mu =T_0^2-\overrightarrow{T}^2=\ka^{-2}$.  The deformation parameter $\ka$ clearly plays the role of the curvature. The 3D component of $T^\mu$ defines the flux vector $\vT$ at the ribbon source.

We similarly define the target flux vector by projecting the $\SB(2,\C)$ group element $\lt$ on the Pauli matrices,
\be
\lt=\bpm \tlam & 0 \\ \tz & \tilde{\lambda}^{-1} \epm
\,,\qquad
\Tt\equiv\lt\lt^\dagger =\bpm \tlam^2 & \tlam \ztb \\ \tlam \tz & \tlam^{-2}+{|\tz|}^2 \epm
\,,\qquad
\Tt^0\equiv\frac{1}{2 \kappa}\tr\, \Tt
\,,\quad
\Tt^a\equiv\frac{1}{2\kappa}\tr\,\Tt{\sigma}^a
\,.
\ee
Since $\ell u=\ut\lt$ and thus $T=\ut \Tt\ut^{-1}$, the $\SU(2)$ holonomy $\ut$ transports the source flux vector $T^a$ to the target flux vector $\Tt^a$ as for the flat holonomy-flux variables.

\smallskip

One can write\footnotemark{} the Poisson brackets \eqref{eq:poisson_left}, \eqref{eq:poisson_right} and \eqref{eq:poisson_mix} explicitly in terms of
the flux vectors $T^\mu$ and $\Tt^\mu$, as done in \cite{Bonzom:2014wva,Dupuis:2014fya}:
\be\ba{lll}
\{T^0,T^a\}=\{\Tt^0,\Tt^a\}=\{T^a,\Tt^b\}=0\,,\quad 
&\{T^a,T^b\}=-\epsilon_{abc}\kappa(T^0+T^3)T^c\,,\quad
&\{\Tt^a,\Tt^b\}=\epsilon_{abc}\kappa(\Tt^0+\Tt^3)\Tt^c\,,
\vspace*{1.5mm}\\
\{T^1,\ut\}=-\frac{i}{2}\kappa(T^0+T^3)\sigma^1 \ut\,,\quad
&\{T^2,\ut\}=-\frac{i}{2}\kappa(T^0+T^3)\sigma^2 \ut\,,\quad
&\{T^3+T^0,\ut\}=-\frac{i}{2}\kappa(T^0+T^3)\sigma^3 \ut\,,
\vspace*{1.5mm}\\
\{\Tt^1,\ut\}=-\frac{i}{2}\kappa(\Tt^0+\Tt^3) \ut\sigma^1\,,\quad
&\{\Tt^2,\ut\}=-\frac{i}{2}\kappa(\Tt^0+\Tt^3) \ut\sigma^2\,,\quad
&\{\Tt^3+\Tt^0,\ut\}=-\frac{i}{2}\kappa(\Tt^0+\Tt^3) \ut\sigma^3
\,.
\ea
\label{eq:q_Poisson}
\ee
\footnotetext{
In order to compute the Poisson bracket with the flux vectors $T$ and $\Tt$, we actually need the Poisson brackets with the complex conjugate of the $\SB(2,\C)$ fluxes $\ell^\dagger$ and $\lt^\dagger$.
The method used in \cite{Bonzom:2014wva} to derive the Poisson brackets with $\ell^\dagger$ and $\lt^\dagger$ is to notice that the 
the $\SB(2,\C)$ group structure and the $\Sb(2,\C)$ Lie algebra are preserved under the map: $\ell \mapsto \left(\ell^\dagger\right)^{-1},\tilde{\ell} \mapsto \left(\tilde{\ell}^\dagger\right)^{-1}, \tau^a \mapsto (\tau^a)^\dagger$. Thus switching $r$-matrix $r \mapsto \rD=-\rT$
in  the Poisson brackets \eqref{eq:poisson_left}, \eqref{eq:poisson_right} and \eqref{eq:poisson_mix} gives:
\be\ba{llll}
\left|\ba{l}
\{\ell^\dagger_1,\ell_2 \}=-\ell^\dagger_1 \rT \ell_2+\ell_2 \rT \ell^\dagger_1\,,\\
\{\ell_1,\ell^\dagger_2 \}= -\ell_1 r \ell^\dagger_2 + \ell^\dagger_2 r \ell_1 \,,
\ea\right.
\qquad
&\left|\ba{l}
\{\ell^\dagger_1,\ell^\dagger_2\}={[}r,\ell^\dagger_1 \ell^\dagger_2{]}\,,\\
\{\ell^\dagger_1,\lt_2\}=0\,,
\ea\right.
\qquad
&\left|\ba{l}
\{\ell^\dagger_1,u_2\}=-\rT \ell^\dagger_1 u_2\,,\\
\{u_1, \ell^\dagger_2\} =  r u_1 \ell^\dagger_2\,,
\ea\right.
\qquad
&\left|\ba{l}
\{\ell^\dagger_1,\ut_2\}=-\ell^\dagger_1 \rT \ut_2\,,\\
\{\ut_1,\ell^\dagger_2\}=\ell^\dagger_2 r \ut_1\,,
\ea\right.
\ea\nn
\ee
\be
\ba{llll}
&\left|\ba{l}
\{\tilde{\ell}^\dagger,\tilde{\ell}_2\}=\tilde{\ell}^\dagger_1 \rT \tilde{\ell}_2 - \tilde{\ell}_2 \rT \tilde{\ell}^\dagger_1\,,\\
\{\tilde{\ell}_1,\tilde{\ell}^\dagger_2\}=\tilde{\ell}_1 r \tilde{\ell}^\dagger_2- \tilde{\ell}^\dagger_2 r \tilde{\ell}_1\,,
\ea\right.
\qquad
\left|\ba{l}
\{\tilde{\ell}^\dagger_1,\ell_2\}=0\,,\\
\{\tilde{\ell}^\dagger_1,\tilde{\ell}^\dagger_2\}=-{[}r,\tilde{\ell}^\dagger_1 \tilde{\ell}^\dagger_2{]} \,,
\ea\right.
\qquad
&\left|\ba{l}
\{\tilde{\ell}^\dagger_1,u_2\}= -u_2 \rT \tilde{\ell}^\dagger_1\,,\\
\{u_1,\tilde{\ell}^\dagger_2\}=u_1 r \tilde{\ell}^\dagger_2\,,
\ea\right.
\qquad
&\left|\ba{l}
\{\tilde{\ell}^\dagger_1,\ut_2\}=-\tilde{\ell}^\dagger_1 \ut_2 \rT\,,\\
\{\ut_1,\tilde{\ell}^\dagger_2\}=\ut_1 \tilde{\ell}^\dagger_2 r\,.
\ea\right.
\ea
\nn
\ee
}
The difference with the standard flat holonomy-flux brackets of the $T^*\SU(2)$ phase space, given in  \eqref{eq:poisson_flat} and \eqref{eq:poisson_flat_2}, are the rescaling factors $\lambda^{2}=\kappa(T^0+T^3)$ and $\tlam^{2}=\kappa(\Tt^0+\Tt^3)$.
The flat limit $\ka\rightarrow 0$ is an inhomogeneous scaling of the flux 4-vector. As $\ka$ is sent to 0, the hyperboloid  $T_0^2-\overrightarrow{T}^2=\ka^{-2}$ becomes the flat $\R^3$ space, with the inhomogeneous limit $\ka T^0\rightarrow 1$, with $T^0\sim\ka^{-1}$ sent to $+\infty$, while the 3D flux vector $T^a$ remains finite. 
Up to a global sign switch for the $T^a$ and $\Tt^a$, this actually sends the $\SL(2,\C)$ Poisson  brackets written above to the  $T^*\SU(2)$ Poisson brackets of standard ``flat'' loop gravity, given in  \eqref{eq:poisson_flat} and \eqref{eq:poisson_flat_2}.

To avoid the subtleties of the inhomogeneous re-scaling limit, we can re-parametrize the $\SB(2,\C)$ fluxes in terms of the $\Sb(2,\C)$ generators:
\be
\ell=e^{i\,j^a\tau^a}
\,,\qquad
\lambda=e^{-\frac{\kappa}{2}j^3}\,,\quad
z=-\kappa e^{\frac{\kappa}{2}j^3}(j^1+ij^2)=-e^{\frac{\kappa}{2}j^3}\kappa j^{+}\,,\quad
\zb=-e^{\frac{\kappa}{2}j^3}\kappa j^{-}\,,
\label{eq:lamda_z}
\ee
where $j^a\in \R^3$ is an arbitrary 3-vector.
The flat limit $\ka \rightarrow 0$ is taken keeping the 3-vector $j^a$ finite:
\be
\kappa T^0= \cosh\kappa j^3+\frac{\kappa^2}{2}e^{\kappa j^3}j^{+}j^{-}\,\xrightarrow{\kappa\rightarrow 0}\,1
\,,\qquad
\left|\begin{array}{l}
T^1=-j^1\,,
\vspace*{1mm}\\
T^2=-j^2\,,
\end{array}\right.
\qquad
T^3=\frac{-\sinh\kappa j^3}{\kappa}-\frac{\kappa}{2}e^{\kappa j^3}j^{+}j^{-}\,\xrightarrow{\kappa\rightarrow 0}\,-j^3
\,.
\label{eq:j_limit}
\ee
Similarly defining the 3-vector $\jt^a$ for the target flux $\lt$, we recover the $T^*\SU(2)$ Poisson brackets  \eqref{eq:poisson_flat} and \eqref{eq:poisson_flat_2} of flat loop gravity in the limit $\ka\rightarrow0$:
\be\ba{lll}
\{j^a,j^b\}\rightarrow \epsilon_{abc}j^c\,,\qquad\qquad
&\{j^a,\ut\}\rightarrow \frac{i}{2}\sigma^a \ut\,,\qquad\qquad
&\{\ut,\ut\}\rightarrow 0\,,
\vspace*{1.5mm}\\
\{\jt^a,\jt^b\}\rightarrow -\epsilon_{abc}\jt^c\,,\qquad\qquad
& \{\jt^a,\ut\}\rightarrow \frac{i}{2}\ut\sigma^a \,,\qquad\qquad
&\{j^a,\jt^b\}=0\,.
\ea\ee
Reciprocally, the $\SL(2,\C)$ Poisson brackets \eqref{eq:q_Poisson} define the curved deformation of the $T^*\SU(2)$ Poisson brackets extending them to take into account a non-vanishing cosmological constant.

\subsection{Deformed Holonomy-Flux on the Torus}
\label{sec:qtorus}

{\bf Ribbon graph on the Torus:} 
We apply the $q$-deformed loop gravity framework described above, with the deformed  holonomy-flux phase space provided with the $\SL(2,\C)$ Poisson brackets, to the simple case of the torus as reviewed in section \ref{sec:flat} for  flat loop gravity with the $T^*\SU(2)$ symplectic structure.

\begin{figure}[h!]
	\begin{tikzpicture}[scale=1]
	
	\coordinate (O) at (0,0);
	\coordinate (A) at (6,0);
	\coordinate (B) at (6,4);
	\coordinate (C) at (0,4);
	\draw (O)--(A)--(B)--(C)--(O);
	
	\coordinate (a) at (2.3,1.3);
	\coordinate (b) at (3.7,1.3);
	\coordinate (c) at (2.3,2.7);
	\coordinate (d) at (3.7,2.7);
	
	\coordinate (ld) at (0,1.3);
    \coordinate (lu) at (0,2.7);
    \coordinate (rd) at (6,1.3);
    \coordinate (ru) at (6,2.7);
    \coordinate (dl) at (2.3,0);
    \coordinate (dr) at (3.7,0);
    \coordinate (ul) at (2.3,4);
    \coordinate (ur) at (3.7,4);
    
    \coordinate (f1) at (3,2);
    \coordinate (f2) at (1.15,0.65);
    \coordinate (f3) at (1.15,2);
    \coordinate (f4) at (3,0.65);
	
	\draw[thick,decoration={markings,mark=at position 0.55 with {\arrow[scale=1.3,>=stealth]{>}}},postaction={decorate}] (ld) -- node[midway, below]{$\ut$}(a); 
	\draw[thick,decoration={markings,mark=at position 0.55 with {\arrow[scale=1.3,>=stealth]{>}}},postaction={decorate}] (b)  --node[midway, below]{$\ut$}(rd); 
	\draw[thick,decoration={markings,mark=at position 0.55 with {\arrow[scale=1.3,>=stealth]{>}}},postaction={decorate}] (lu) --node[midway,above]{$u$}(c); 
	\draw[thick,decoration={markings,mark=at position 0.55 with {\arrow[scale=1.3,>=stealth]{>}}},postaction={decorate}] (d) --node[midway,above]{$u$}(ru);
	 
	\draw[thick,decoration={markings,mark=at position 0.55 with {\arrow[scale=1.3,>=stealth]{>}}},postaction={decorate}] (dl) -- node[midway, left]{$v$}(a); 
	\draw[thick,decoration={markings,mark=at position 0.55 with {\arrow[scale=1.3,>=stealth]{>}}},postaction={decorate}] (dr)  --node[midway, right]{$\vt$}(b); 
	\draw[thick,decoration={markings,mark=at position 0.55 with {\arrow[scale=1.3,>=stealth]{>}}},postaction={decorate}] (c) -- node[midway, left]{$v$}(ul); 
	\draw[thick,decoration={markings,mark=at position 0.55 with {\arrow[scale=1.3,>=stealth]{>}}},postaction={decorate}] (d)  --node[midway, right]{$\vt$}(ur) ;
	
	\draw[thick,dashed,decoration={markings,mark=at position 0.55 with {\arrow[scale=1.3,>=stealth]{>}}},postaction={decorate}] (a) --node[midway, right]{$\lt$}(c); 
	\draw[thick,,dashed,decoration={markings,mark=at position 0.55 with {\arrow[scale=1.3,>=stealth]{>}}},postaction={decorate}] (b) --node[midway, left]{$\ell$}(d); 
	
	\draw[thick,,dashed,decoration={markings,mark=at position 0.55 with {\arrow[scale=1.3,>=stealth]{>}}},postaction={decorate}] (b) --node[midway,above]{$\mt$}(a); 
	\draw[thick,,dashed,decoration={markings,mark=at position 0.55 with {\arrow[scale=1.3,>=stealth]{>}}},postaction={decorate}] (d) --node[midway,below]{$m$}(c);

	\draw (b) node {$\bullet$};
	
	\draw (f1) node {$f_1$};
	\draw (f2) node {$f_2$};
	\draw (f3) node {$f_3$};
	\draw (f4) node {$f_4$};
	
	\draw[fill=gray!50,opacity=.4](b) -- (d) -- (ru) -- (rd) -- cycle ;
	\draw[fill=gray!50,opacity=.4](a) -- (c) -- (lu) -- (ld) -- cycle ;
	\draw[fill=gray!50,opacity=.4](a) -- (b) -- (dr) -- (dl) -- cycle ;
	\draw[fill=gray!50,opacity=.4](c) -- (d) -- (ur) -- (ul) -- cycle ;
	
\end{tikzpicture}
\caption{Ribbon graph on the torus, parametrized by $\ell,\lt,m,\mt\in \SB(2,\C)$ and $u,\ut,v,\vt\in\SU(2)$. The ribbons, shaded in grey, are the thickened graph links.
The ribbon flatness constraints around the faces $f_{3}$ and $f_{4}$ give the relations between the $\SB(2,\C)$ fluxes at the source and target of the ribbons:  $\cR^{(\ell)} =\ell u\lt^{-1}\ut^{-1}=\id$ and  $\cR^{(m)} =mv\mt^{-1}\vt^{-1}=\id$. 
The Gauss constraint $\cG=\ell m\lt^{-1}\mt^{-1} =\id$ is a $\SB(2,\C)$ flatness around the face $f_1$ while the holonomy flatness constraint $\cF= \ut v^{-1}u^{-1}\vt =\id$ is the $\SU(2)$ flatness around the face $f_2$.
}
\label{fig:ribbon}
\end{figure}
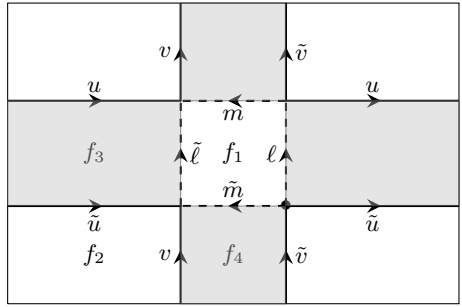

We draw a ribbon graph on the torus, as depicted on fig.\ref{fig:ribbon}, dressed with $\SB(2,\C)$ fluxes and $\SU(2)$ holonomies.
The ribbon graph defines four faces, noted $f_{1,..,4}$. The face $f_{1}$ is bounded by the vertex loop running around the graph node. The face $f_{2}$ is the large face defined by the graph.
The two shaded faces, $f_3$ and $f_4$, are the two ribbons. The horizontal ribbon is decorated by the variables $u, \ut \in \SU(2)$ along the long edges and $\ell, \lt \in \SB(2,\C)$  on the short edges. The vertical ribbon is decorated by the variables $v, \vt \in \SU(2)$ along the long edges and $m, \mt \in \SB(2,\C)$  on the short edges.
The variables associated to the two ribbons, $(\ell,u,\lt,\ut)$ and $(m,v,\mt,\vt)$, Poisson-commute with each other. And each set of ribbon variables is provided with the Poisson brackets, \eqref{eq:poisson_left}, \eqref{eq:poisson_right} and \eqref{eq:poisson_mix},  defined above from the symplectic structure on $\SL(2,\C)$ as the Heisenberg double $\SU(2)\bowtie\SB(2,\C)$.

Flatness constraints around the ribbons amount to the equivalence of the two Iwasawa decompositions $\ell u=\ut \lt$, $mv=\vt \mt$ on both ribbons:
\beq
\cR^{(\ell)} &=&\ell u\lt^{-1}\ut^{-1}=\id\,,
\label{ribbon-constraint}
\\
\cR^{(m)} &=&mv\mt^{-1}\vt^{-1}=\id\,.
\nn
\eeq
These ribbon flatness constraints are second class and already taken into account in the Poisson brackets \eqref{eq:poisson_mix}.

\medskip

{\bf The constraint algebra:} 
The two loops around the faces $f_1$ and $f_2$ define the closure constraint and the $\SU(2)$ flatness constraint, which we both root at the same corner around the central node of the graph, as drawn on fig.\ref{fig:ribbon}:
\beq
\cG&=\ell m\lt^{-1}\mt^{-1} =\id\,,\label{eq:gauss_def} \\
\cF&= \ut v^{-1}u^{-1}\vt =\id\,. \label{eq:flatness_def}
\eeq
Imposing the ribbon flatness constraints, the closure and the $\SU(2)$ flatness constraints, amounts to imposing the $\SL(2,\C)$ flatness around the four faces of the ribbon graph. As a consequence, the ordered oriented product of $\SU(2)$ and $\SB(2,\C)$ group elements along any path on the ribbon graph does not depend on the path itself but simply on where it starts and ends, as for a flat connection theory.

The Poisson brackets of the closure and flatness constraints form a closed algebra and therefore define a system of first class constraints: 
\be
\{\cF_1,\cG_2\}=\cG_2\rT\cF_1-\cF_1\rT\cG_2\,,\quad
 \{\cG_1,\cG_2\}=-[r,\cG_1\cG_2]\,,\quad
\{\cF_1,\cF_2\}=-[\rT,\cF_1\cF_2]\,.
\label{eq:constraint_alg}
\ee
As shown in \cite{Bonzom:2014wva,Dupuis:2014fya}, $\cG$ generate $\SU(2)$ gauge transformations (which implement the Gauss law of loop quantum gravity) while $\cF$ generate the translational gauge transformations (implementing the action of the space-time diffeomorphisms). Finite gauge transformations are described below.
The goal will then be to identify Dirac observables, that Poisson-commute with both closure and flatness constraints.

\medskip

{\bf Deformed braided  gauge symmetries:}
The $\SB(2,\C)$ closure constraint $\cG$ generates $\SU(2)$ action acting at the graph node. Since the node has been fattened, the explicit $\SU(2)$ action  gets braided by the $\SB(2,\C)$ fluxes and is slightly different at each corner around the node.
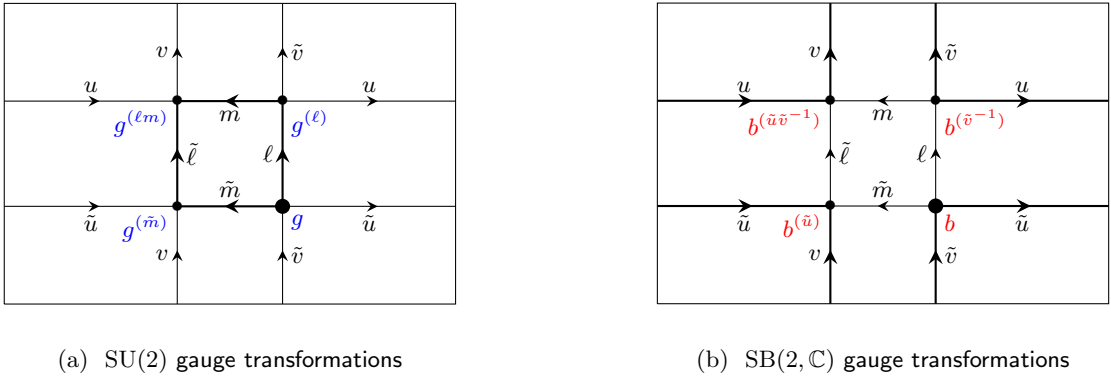
\begin{figure}[h!]

\begin{subfigure}[t]{0.45\linewidth}
	\begin{tikzpicture}[scale=1]
	
	\coordinate (O) at (0,0);
	\coordinate (A) at (6,0);
	\coordinate (B) at (6,4);
	\coordinate (C) at (0,4);
	\draw (O)--(A)--(B)--(C)--(O);
	
	\coordinate (a) at (2.3,1.3);
	\coordinate (b) at (3.7,1.3);
	\coordinate (c) at (2.3,2.7);
	\coordinate (d) at (3.7,2.7);

\draw (b) node[blue,below right]{$g$};
\draw (d) node[blue,below right]{$g^{(\ell)}$};
\draw (c) node[blue,below left]{$g^{(\ell m)}$};
\draw (a) node[blue,below left]{$g^{(\mt)}$};
	
	\coordinate (ld) at (0,1.3);
    \coordinate (lu) at (0,2.7);
    \coordinate (rd) at (6,1.3);
    \coordinate (ru) at (6,2.7);
    \coordinate (dl) at (2.3,0);
    \coordinate (dr) at (3.7,0);
    \coordinate (ul) at (2.3,4);
    \coordinate (ur) at (3.7,4);
    
    \coordinate (f1) at (3,2);
    \coordinate (f2) at (1.15,0.65);
    \coordinate (f3) at (1.15,2);
    \coordinate (f4) at (3,0.65);
	
	\draw[,decoration={markings,mark=at position 0.55 with {\arrow[scale=1.3,>=stealth]{>}}},postaction={decorate}] (ld) -- node[midway, below]{$\ut$}(a); 
	\draw[,decoration={markings,mark=at position 0.55 with {\arrow[scale=1.3,>=stealth]{>}}},postaction={decorate}] (b)  --node[midway, below]{$\ut$}(rd); 
	\draw[,decoration={markings,mark=at position 0.55 with {\arrow[scale=1.3,>=stealth]{>}}},postaction={decorate}] (lu) --node[midway,above]{$u$}(c); 
	\draw[,decoration={markings,mark=at position 0.55 with {\arrow[scale=1.3,>=stealth]{>}}},postaction={decorate}] (d) --node[midway,above]{$u$}(ru);
	 
	\draw[,decoration={markings,mark=at position 0.55 with {\arrow[scale=1.3,>=stealth]{>}}},postaction={decorate}] (dl) -- node[midway, left]{$v$}(a); 
	\draw[,decoration={markings,mark=at position 0.55 with {\arrow[scale=1.3,>=stealth]{>}}},postaction={decorate}] (dr)  --node[midway, right]{$\vt$}(b); 
	\draw[,decoration={markings,mark=at position 0.55 with {\arrow[scale=1.3,>=stealth]{>}}},postaction={decorate}] (c) -- node[midway, left]{$v$}(ul); 
	\draw[,decoration={markings,mark=at position 0.55 with {\arrow[scale=1.3,>=stealth]{>}}},postaction={decorate}] (d)  --node[midway, right]{$\vt$}(ur) ;
	
	\draw[thick,,decoration={markings,mark=at position 0.55 with {\arrow[scale=1.3,>=stealth]{>}}},postaction={decorate}] (a) --node[midway, right]{$\lt$}(c); 
	\draw[thick,decoration={markings,mark=at position 0.55 with {\arrow[scale=1.3,>=stealth]{>}}},postaction={decorate}] (b) --node[midway, left]{$\ell$}(d); 
	
	\draw[thick,decoration={markings,mark=at position 0.55 with {\arrow[scale=1.3,>=stealth]{>}}},postaction={decorate}] (b) --node[midway,above]{$\mt$}(a); 
	\draw[thick,decoration={markings,mark=at position 0.55 with {\arrow[scale=1.3,>=stealth]{>}}},postaction={decorate}] (d) --node[midway,below]{$m$}(c);

	\fill (b) circle (0.1);
	\draw (a) node {$\bullet$};
	\draw (c) node {$\bullet$};
	\draw (d) node {$\bullet$};
	
\end{tikzpicture}

\caption{\label{fig:SU2action}
$\SU(2)$ gauge transformations
}
\end{subfigure}
\hspace*{4mm}
\begin{subfigure}[t]{0.45\linewidth}
	\begin{tikzpicture}[scale=1]
	\coordinate (O) at (0,0);
	\coordinate (A) at (6,0);
	\coordinate (B) at (6,4);
	\coordinate (C) at (0,4);
	\draw (O)--(A)--(B)--(C)--(O);
	
	\coordinate (a) at (2.3,1.3);
	\coordinate (b) at (3.7,1.3);
	\coordinate (c) at (2.3,2.7);
	\coordinate (d) at (3.7,2.7);

\draw (b) node[red,below right]{$b$};
\draw (d) node[red,below right]{$b^{(\vt^{-1})}$};
\draw (c) node[red,below left]{$b^{(\ut\vt^{-1})}$};
\draw (a) node[red,below left]{$b^{(\ut)}$};
	
	\coordinate (ld) at (0,1.3);
    \coordinate (lu) at (0,2.7);
    \coordinate (rd) at (6,1.3);
    \coordinate (ru) at (6,2.7);
    \coordinate (dl) at (2.3,0);
    \coordinate (dr) at (3.7,0);
    \coordinate (ul) at (2.3,4);
    \coordinate (ur) at (3.7,4);
    
    \coordinate (f1) at (3,2);
    \coordinate (f2) at (1.15,0.65);
    \coordinate (f3) at (1.15,2);
    \coordinate (f4) at (3,0.65);
	
	\draw[thick,decoration={markings,mark=at position 0.55 with {\arrow[scale=1.3,>=stealth]{>}}},postaction={decorate}] (ld) -- node[midway, below]{$\ut$}(a); 
	\draw[thick,decoration={markings,mark=at position 0.55 with {\arrow[scale=1.3,>=stealth]{>}}},postaction={decorate}] (b)  --node[midway, below]{$\ut$}(rd); 
	\draw[thick,decoration={markings,mark=at position 0.55 with {\arrow[scale=1.3,>=stealth]{>}}},postaction={decorate}] (lu) --node[midway,above]{$u$}(c); 
	\draw[thick,decoration={markings,mark=at position 0.55 with {\arrow[scale=1.3,>=stealth]{>}}},postaction={decorate}] (d) --node[midway,above]{$u$}(ru);
	 
	\draw[thick,decoration={markings,mark=at position 0.55 with {\arrow[scale=1.3,>=stealth]{>}}},postaction={decorate}] (dl) -- node[midway, left]{$v$}(a); 
	\draw[thick,decoration={markings,mark=at position 0.55 with {\arrow[scale=1.3,>=stealth]{>}}},postaction={decorate}] (dr)  --node[midway, right]{$\vt$}(b); 
	\draw[thick,decoration={markings,mark=at position 0.55 with {\arrow[scale=1.3,>=stealth]{>}}},postaction={decorate}] (c) -- node[midway, left]{$v$}(ul); 
	\draw[thick,decoration={markings,mark=at position 0.55 with {\arrow[scale=1.3,>=stealth]{>}}},postaction={decorate}] (d)  --node[midway, right]{$\vt$}(ur) ;
	
	\draw[,,decoration={markings,mark=at position 0.55 with {\arrow[scale=1.3,>=stealth]{>}}},postaction={decorate}] (a) --node[midway, right]{$\lt$}(c); 
	\draw[,decoration={markings,mark=at position 0.55 with {\arrow[scale=1.3,>=stealth]{>}}},postaction={decorate}] (b) --node[midway, left]{$\ell$}(d); 
	
	\draw[,decoration={markings,mark=at position 0.55 with {\arrow[scale=1.3,>=stealth]{>}}},postaction={decorate}] (b) --node[midway,above]{$\mt$}(a); 
	\draw[,decoration={markings,mark=at position 0.55 with {\arrow[scale=1.3,>=stealth]{>}}},postaction={decorate}] (d) --node[midway,below]{$m$}(c);

	\fill (b) circle (0.1);
	\draw (a) node {$\bullet$};
	\draw (c) node {$\bullet$};
	\draw (d) node {$\bullet$};

\end{tikzpicture}

\caption{\label{fig:SB2action}
$\SB(2,\C)$ gauge transformations
}
\end{subfigure}

\caption{
The $\SB(2,\C)$ closure constraints $\cG$ and the $\SU(2)$ flatness constraints respectively generate $\SU(2)$ gauge transformations and $\SB(2,\C)$ gauge transformations. Since we started with a graph with a single node and a single face, then thickened into a ribbon graph, the $\SU(2)$ and $\SB(2,\C)$ gauge transformations respectively act at that node and around that face. With the node thickened into a polygon (a square here), the $\SU(2)$ gauge transformation can be defined as acting by a group element $g$ at a root vertex chosen around that polygon  and simply transforming the group elements on the edges attached to that vertex by the linear group action. This gauge transformation is then braided by the $\SB(2,\C)$ group elements all around the polygon, in order to obtain the gauge transformations of the group elements attached to the other vertices. Similarly, the $\SB(2,\C)$ gauge transformations, generated by the $\SU(2)$ flatness constraint, is braided by $\SU(2)$ group elements all around the face.
}

\end{figure}

Let us root the $\SU(2)$ transformation at the corner between $\ell$ and $\mt$ as on fig.\ref{fig:SU2action} and let us call $g\in\SU(2)$ the group element of the transformation. This transformation will act simply on $\ell$ and $\mt$. But, in order to get its action on $m$, we need to parallel transport the action along the ribbon edge from the source corner of $\ell$ to the source corner of $m$. This leads to a braiding by the $\SB(2,\C)$ flux $\ell$ living on that ribbon edge. As a result, the $\SU(2)$ transformation acting on $m$ is not simply given by the group element $g\in\SU(2)$ but by a group element $g^{(\ell)}$ that depends both on the original transformation parameter $g$ and on the flux $\ell$. Doing this consistently around the loop vertex gives the following $\SU(2)$ gauge transformation:
\be
\label{SU2gauge}
\left|
\ba{lll}
\ell &\rightarrow & \ell^{(g)}=g\ell {g^{(\ell)}}^{-1} \\
\lt &\rightarrow & \lt^{(g)}=g^{(\mt)}\lt{g^{(\ell m)}}^{-1}\\
m &\rightarrow & m^{(g)}=g^{(\ell)}m{g^{(\ell m)}}^{-1} \\
\mt &\rightarrow &\mt^{(g)}=g\mt{g^{(\mt)}}^{-1}
\ea\right. \,,\quad
\left|
\ba{lll}
u &\rightarrow & u^{(g)}=g^{(\ell)}u{g^{(\ell m)}}^{-1} \\
\ut &\rightarrow & \ut^{(g)}=g\ut{g^{(\mt)}}^{-1}\\
v &\rightarrow & v^{(g)}=g^{(\ell m)}v{g^{(\mt)}}^{-1} \\
\vt &\rightarrow & \vt^{(g)}=g^{(\ell)}\vt g^{-1}
\ea\right. \,,
\ee
where the gauge transformed flux $\ell^{(g)}\in\SB(2,\C)$ and the parallelly transported gauge transformation $g^{(\ell)}\in\SU(2)$ are uniquely determined in terms of $g$ and $\ell$ from the left Iwasawa decomposition of $g\ell=\ell^{(g)}g^{(\ell)}$. And similarly $g\mt=\mt^{(g)}{g^{(\mt)}}$ and $m^{(g)}{g^{(\ell m)}}=g^{(\ell)}m$.

It is fairly direct to check that the closure and flatness constraint are transformed by conjugation  to $\cG^{(g)}=g\cG g^{-1}$ and $\cF^{(g)}=g\cF g^{-1}$, so that the conditions $\cG=\cF=\id$ are  invariant under $\SU(2)$ gauge transformations as expected.

Similarly, the $\SU(2)$ flatness constraint $\cF$ generates a $\SB(2,\C)$ action. Rooting again the transformation at the corner between $\ell$ and $\mt$, which is also the corner between $\ut$ and $\vt$, the gauge parameter $b\in\SB(2,\C)$ at the root corner gets braided by the $\SU(2)$ holonomies and needs to be parallelly transported from one corner  to the next, as depicted on fig.\ref{fig:SB2action}. Doing so consistently around the large face
gives the $\SB(2,\C)$ gauge transformations as:
\be
\label{SB2gauge}
\left|
\ba{lll}
	\ell &\rightarrow &\ell^{(b)}=b\ell {b^{(\vt^{-1})}}^{-1} \\
	\lt &\rightarrow &\lt^{(b)}=b^{(\ut)}\lt{b^{(\ut \vt^{-1})}}^{-1}\\
	m &\rightarrow &m^{(b)}=b^{(\vt^{-1})}m{b^{(\ut\vt^{-1})}}^{-1} \\
	\mt &\rightarrow &\mt^{(b)}=b\mt{b^{(\ut)}}^{-1}
\ea\right. \,,\quad
\left|
\ba{lll}
	u &\rightarrow &u^{(b)}=b^{(\vt^{-1})}u{b^{(\ut \vt^{-1})}}^{-1} \\
	\ut &\rightarrow &\ut^{(b)}=b\ut{b^{(\ut)}}^{-1}\\
	v &\rightarrow &v^{(b)}=b^{(\ut \vt^{-1})}v{b^{(\ut)}}^{-1} \\
	\vt &\rightarrow &\vt^{(b)}=b^{(\vt^{-1})}\vt b^{-1}
\ea\right. \,,
\ee
where the gauge transformed holonomy $\ut^{(b)}\in\SU(2)$ and the  transported gauge transformation $b^{(\ut)}\in\SB(2,\C)$ are uniquely determined in terms of $b$ and $\ut$ from the right Iwasawa decomposition of $b\ut=\ut^{(b)}{b^{(\ut)}}$. And similarly $\vt b^{-1}=(b^{(\vt^{-1})})^{-1}\vt^{(b)}$ and $v(b^{(\ut)})^{-1}=v^{(b)}(b^{(\ut \vt^{-1})})^{-1}$.
These define the deformed translation gauge transformations.

It is also direct to check that the closure and flatness constraint are once again simply transformed by conjugation  to $\cG^{(b)}=b\cG b^{-1}$ and $\cF^{(b)}=b\cF b^{-1}$, leaving the conditions $\cG=\cF=\id$ invariant under finite $\SB(2,\C)$ gauge transformations as expected.

\medskip

\begin{figure}[h!]
	\begin{tikzpicture}[scale=1]
	
	\coordinate (O) at (0,0);
	\coordinate (A) at (6,0);
	\coordinate (B) at (6,4);
	\coordinate (C) at (0,4);
	\draw (O)--(A)--(B)--(C)--(O);
	
	\coordinate (a) at (2.3,1.3);
	\coordinate (b) at (3.7,1.3);
	\coordinate (c) at (2.3,2.7);
	\coordinate (d) at (3.7,2.7);

	\fill (b) circle (0.1);
\draw (b) node[below right]{$\Omega$};
	
	\coordinate (ld) at (0,1.3);
    \coordinate (lu) at (0,2.7);
    \coordinate (rd) at (6,1.3);
    \coordinate (ru) at (6,2.7);
    \coordinate (dl) at (2.3,0);
    \coordinate (dr) at (3.7,0);
    \coordinate (ul) at (2.3,4);
    \coordinate (ur) at (3.7,4);

\draw (ur) node[above]{$A=\ell \vt$};
\draw (rd) node[right]{$B=\mt \ut^{-1}$};
	
	\draw[thick,decoration={markings,mark=at position 0.55 with {\arrow[scale=1.3,>=stealth]{>}}},postaction={decorate}] (ld) -- node[midway, below]{$\ut$}(a); 
	\draw[thick,decoration={markings,mark=at position 0.55 with {\arrow[scale=1.3,>=stealth]{>}}},postaction={decorate}] (b)  --node[midway, below]{$\ut$}(rd); 
	\draw[,decoration={markings,mark=at position 0.55 with {\arrow[scale=1.3,>=stealth]{>}}},postaction={decorate}] (lu) --node[midway,above]{$u$}(c); 
	\draw[,decoration={markings,mark=at position 0.55 with {\arrow[scale=1.3,>=stealth]{>}}},postaction={decorate}] (d) --node[midway,above]{$u$}(ru);
	 
	\draw[,decoration={markings,mark=at position 0.55 with {\arrow[scale=1.3,>=stealth]{>}}},postaction={decorate}] (dl) -- node[midway, left]{$v$}(a); 
	\draw[thick,decoration={markings,mark=at position 0.55 with {\arrow[scale=1.3,>=stealth]{>}}},postaction={decorate}] (dr)  --node[midway, right]{$\vt$}(b); 
	\draw[,decoration={markings,mark=at position 0.55 with {\arrow[scale=1.3,>=stealth]{>}}},postaction={decorate}] (c) -- node[midway, left]{$v$}(ul); 
	\draw[thick,decoration={markings,mark=at position 0.55 with {\arrow[scale=1.3,>=stealth]{>}}},postaction={decorate}] (d)  --node[midway, right]{$\vt$}(ur) ;
	
	\draw[decoration={markings,mark=at position 0.55 with {\arrow[scale=1.3,>=stealth]{>}}},postaction={decorate}] (a) --node[midway, right]{$\lt$}(c); 
	\draw[thick,decoration={markings,mark=at position 0.55 with {\arrow[scale=1.3,>=stealth]{>}}},postaction={decorate}] (b) --node[midway, left]{$\ell$}(d); 
	
	\draw[thick,decoration={markings,mark=at position 0.55 with {\arrow[scale=1.3,>=stealth]{>}}},postaction={decorate}] (b) --node[midway,above]{$\mt$}(a); 
	\draw[decoration={markings,mark=at position 0.55 with {\arrow[scale=1.3,>=stealth]{>}}},postaction={decorate}] (d) --node[midway,below]{$m$}(c);
	
\end{tikzpicture}

\caption{\label{fig:SL2Cholo}
$\SL(2,\C)$ holonomies rooted at the corner $\Omega$. Having chosen the root corner around the central polygon and the face for both $\SU(2)$ and $\SB(2,\C)$ gauge transformations, these two sets of gauge transformations are combined into a single set of $\SL(2,\C)$ gauge transformations. Then $\SL(2,\C)$ holonomies going around loops starting and ending at that root corner simply transform under the $\SL(2,\C)$ action  by conjugation and allow to define simple Wilson loop observables.
}

\end{figure}
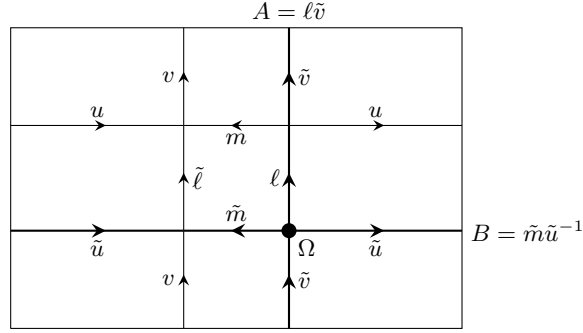

{\bf $\SL(2,\C)$ Holonomies and physical observables:}
From the structure of the ribbon graph, it is clear that the $\SL(2,\C)$ group elements running along the ribbons, and used to define the symplectic structure, $D=\ell u$ and $D'=mv$ do not start and end at the same point. It seems more natural to introduce $\SL(2,\C)$ holonomies that wrap around the cycles of the torus and come back to their initial point. Similarly to the definition of Poincar\'e holonomies for standard flat 3D loop gravity introduced earlier in section \ref{ISU2theory}, we introduce $\SL(2,\C)$ holonomies rooted at a corner\footnotemark{} around the graph node, as drawn on fig.\ref{fig:SL2Cholo}:
\be
A=\ell \vt\,,\qquad
B=\mt\ut^{-1}\,,\qquad
A,B\in\SL(2,\C)\,.
\ee
\footnotetext{
We could consider the $\SL(2,\C)$ holonomies rooted at any corner. This would not change anything, as long as the two $\SL(2,\C)$ holonomies are rooted at the point. 
}

First, $A$ and $B$ contain all the information about the holonomy-flux variables around the ribbons (once we assume the ribbon flatness constraints). There are not the $\SL(2,\C)$ group elements, $D=\ell u$ and $D'=m v$, around the ribbons but mix the $\SB(2,\C)$ fluxes and $\SU(2)$ holonomies of the different ribbons.

Second, their Poisson brackets form a closed algebra,
\be
\left|\begin{array}{lcl}
\{A_1,A_2\}&=&[r,A_1A_2]\,,
\vspace*{1.5mm}\\
\{B_1,B_2\}&=&-[r,B_1B_2]\,,
\vspace*{1.5mm}\\
\{A_1,B_2\}&=&A_1rB_2+B_2\rT A_1\,.
\end{array}\right.
\label{eq:Poisson_AB}
\ee
expressed in terms of the $r$-matrix similarly as for standard flat loop gravity formulated in terms of Poincar\'e holonomies in \eqref{PoissonABflat}.

Third, the $\SB(2,\C)$ closure constraint $\cG$ and the $\SU(2)$  flatness constraint $\cF$ can be repackaged in a single $\SL(2,\C)$ flatness constraint, implying that the two $\SL(2,\C)$ holonomies $A$ and $B$ commute:
\be
\cG=\cF=\id
\quad\Longleftrightarrow\quad
\cC:=ABA^{-1}B^{-1}=\id
\,.
\label{eq:flat_AB}
\ee
This is exactly the $\SL(2,\C)$ constraint for a flat $\SL(2,\C)$ connection on the torus, as arising  in the Chern-Simons phase space and its combinatorial quantization.
However, due to the non-trivial braiding between $\SB(2,\C)$ and $\SU(2)$ group elements which is reflected in the non-trivial commutator between the constraints $\cG$ and $\cF$, the Poisson brackets of the $\SL(2,\C)$ constraint $\cC$ have a more complicated form than the closed algebra \eqref{eq:constraint_alg} formed by  $\cG$ and $\cF$:
\be
\{\cC_1,\cC_2\}={[}r,\cC_1\cC_2{]}+A_2B_2{[}\rT,\cC_1{]}A_2^{-1}B_2^{-1}-A_1B_1{[}r,\cC_2{]}A_1^{-1}B_1^{-1} \sim 0\,.
\ee
Finally, the two holonomies $A$ and $B$ are rooted at the same vertex. Their gauge transformations are straightforward. Under both the $\SU(2)$ gauge transformations given above in \eqref{SU2gauge} and the $\SB(2,\C)$ gauge transformations given in \eqref{SB2gauge}, the $\SL(2,\C)$ holonomies $A$ and $B$ transform under the action by conjugation:
\be
A,B\quad\xmapsto{g\in\SU(2)}\quad gAg^{-1},\,gBg^{-1}
\,,\qquad\qquad
A,B\quad\xmapsto{b\in\SB(2,\C)}\quad bAb^{-1},\,bBb^{-1}
\,.
\ee
Moreover, it is straightforward to check that the $\SL(2,\C)$ flatness constraint $ABA^{-1}B^{-1}=\id$, together with the Poisson brackets \eqref{eq:Poisson_AB}, generates these gauge transformations\footnotemark.
\footnotetext{
To check that the $\SL(2,\C)$ flatness constraint $\cC$ does generate  $\SL(2,\C)$ gauge transformations, we compute the Poisson flow of its projection onto the $\sl(2,\C)$ generators $t^k\in\{\sigma^a,\tau^b\}$ with $k=1,..,6$ distinguishing the $\su(2)$ generators from the $\mathfrak{sb}(2,\C)$ generators. For example, computing the Poisson bracket  of $\tr \cC t^k$ with the $\SL(2,\C)$ holonomy $A$ leads to a variation $\delta A$ corresponding to an  infinitesimal $\SL(2,\C)$ gauge transformation on $A$:
\beq
\delta_{\eps}A\equiv
\sum_{k}\{\epsilon_k \tr\, \cC t^k,A\}
&=&
\epsilon_k \tr_1 (\{A_1B_1A_1^{-1}B_1^{-1}t_1^k,A_2\})
=
\epsilon_k{[}A,\tr_1(A_1B_1rA_1^{-1}B_1^{-1}t_1^k){]}+\epsilon_k A\tr_1(\cC_1\rT t_1^k)-\epsilon_k\tr_1(\rT \cC_1 t_1^k)A
\nn\\
&\sim&
\frac{\epsilon_k}{4} \tr(AB\tau^aA^{-1}B^{-1}t^k){[}A,\sigma^a{]}+\epsilon_k\tr(\sigma^a t^k){[}A,\tau^a{]}
\eeq
We see that the resulting $\SU(2)$ gauge transformations depend non-linearly on the group element $A$.
}
This means that Dirac observables, invariant under both rotation and translation gauge transformations, are simply the components of $A$ and $B$ invariant under conjugation by $\SL(2,\C)$. This leaves us with the two complex Dirac observables given by the two Wilson loops, $\tr A$ and $\tr B$. We can compute their Poisson bracket using the bracket $\{A_1,B_2\}$:
\beq
\{\tr A,\tr B\}
&=&
\tr \left(A_1B_2(r+\rT)\right)
=\f{i\ka}2\,(\tr A \sigma^a)\,(\tr B\sigma^a)
\nn\\
&=&
i\kappa\left(\tr(AB)-\f12\tr A \, \tr B\right)
\,,
\label{eq:Poisson_AB_tr}
\eeq
We recognize the Goldman bracket for gauge-invariant functions on the moduli space $\Hom(\pi,\SL(2,\C))/\SL(2,\C)$, $\pi$ being the fundamental group of the surface, with the bilinear map identified with \eqref{eq:bilinear} up to a constant: $\cB(M,N)=\kappa\, \im (\tr\,MN)$ for $M,N\in\sl(2,\C)$ \cite{Goldman:1986invariant}.
This is the desired result since upon imposing constraints on all the faces $A$ and $B$ indeed live on the space of flat connections, thus the Wilson loops $\tr A$ and $\tr B$ are the projection on the corresponding moduli space. 

\smallskip

Another useful way to formulate the Goldman bracket above is to write it in terms of the eigenvalues of $A$ and $B$. Indeed, $A$ and $B$ commute, so they are simultaneously diagonalizable (except in degenerate cases of measure zero in $\SL(2,\C)^{\times 2}$). The logarithm of their eigenvalues are canonically conjugate, i.e. they provide Darboux coordinates for the Goldman bracket:
\be
A=U\mat{cc}{e^\alpha & \\ & e^{-\alpha}}U^{-1}\,,\qquad
B=U\mat{cc}{e^\beta & \\ & e^{-\beta}}U^{-1}\,,
\nn
\ee
\be
\{\alpha,\beta\}=\f{i\ka}2
\quad\Longleftrightarrow\quad
\{e^{\alpha}+e^{-\alpha},e^{\beta}+e^{-\beta}\}
=
\f{i\ka}2\big{(}
e^{\alpha+\beta}+e^{-\alpha-\beta}-e^{\alpha-\beta}-e^{-\alpha+\beta}
\big{)}
\,.
\ee

\smallskip

At the end of the day, we have shown that the $q$-deformed loop gravity phase space, provided with non-abelian $\SB(2,\C)$ closure constraints and $\SU(2)$ flatness constraints, can be reformulated as a phase space of discrete $\SL(2,\C)$ connections with a simple Poisson bracket \eqref{eq:Poisson_AB} and $\SL(2,\C)$ flatness constraints \eqref{eq:flat_AB}. The physical phase space is then the moduli space of flat $\SL(2,\C)$ connection with the expected Goldman bracket.
Although we have focused on the example of the torus, we see no obstacle in extending the method and the results to arbitrary ribbon graph  and arbitrary orientable surface topology.

This shows that the the $q$-deformed approach to 3D loop gravity (for $q\in\R$), which is the canonical framework for the Turaev-Viro model, is consistent with the combinatorial quantization of 3D gravity (with a negative cosmological constant $\Lambda<0$) as a Chern-Simons theory.

\section{3d Loop Gravity from Fock-Rosly construction}
\label{sec:Fock-Rosly}

Now that we have reformulated the phase space of 3d (Riemannian) loop gravity with (negative) cosmological constant in terms of $\SL(2,\C)$ group elements, it is natural to compare it with the Fock-Rosly Poisson brackets introduced in \cite{Fock:1998nu}. Indeed, 3d gravity with $\Lambda<0$ can be reformulated as a $\SL(2,\C)$ Chern-Simons theory, then the Fock-Rosly brackets define a symplectic structure of the moduli space of flat $\SL(2,\C)$ graph connections up to $\SL(2,\C)$ gauge transformations. Quantizing this bracket and promoting it to operator commutators leads to the combinatorial quantization of Chern-Simons theory \cite{Alekseev:1994pa,Alekseev:1994au,Buffenoir:2002tx}. It is thus a recurring theme of the loop quantum gravity framework in three space-time dimensions to reconcile it with the Fock-Rosly phase space and thereby with quantum Chern-Simons theory.

At a very simple level, the main mismatch in the symplectic structure is that, on the one hand in loop gravity,  the $\SU(2)$ holonomies on different links Poisson-commute with each other, as explained in \ref{sec:deformed}, while on the other hand in the Fock-Rosly framework, the $\SL(2,\C)$ holonomies on links sharing a node have non-vanishing Poisson brackets. At a first glance, this problem seems to be remedied by our definition of $\SL(2,\C)$ group elements, combining the $\SU(2)$ holonomies with the $\SB(2,\C)$ fluxes, as introduced in \eqref{eq:Poisson_AB} in the previous section. These $\SL(2,\C)$ group elements do not commute and their Poisson brackets is simply expressed in terms of the $r$-matrix. However, a careful analysis shows that these brackets do not match the Fock-Rosly brackets.

Thus, the present section aims to underline the differences between the loop gravity phase space and the Fock-Rosly approach, explaining why it is natural that the resulting Poisson brackets for $\SL(2,\C)$ holonomies are not the same, but further show that the loop gravity phase space can nevertheless be obtained from the Fock-Rosly phase space from a partial asymmetric gauge fixing, distinguishing half of the $\SL(2,\C)$ holonomies as $\SU(2)$ holonomies and the other half as $\SB(2,\C)$ fluxes.

\medskip

{\bf The Fock-Rosly bracket for $\SL(2,\C)$ graph connections:}
Let us first quickly review the Fock-Rosly construction introduced in \cite{Fock:1998nu}. 
We work on a compact, oriented Riemann surface. We consider a cellular decomposition of that surface and focus on the graph defined by its 1-skeleton. For simplicity, we  consider a Riemann surface without boundaries. We define a graph connection, as in lattice gauge theory, by assigning a $\SL(2,\C)$ group element, or holonomy, to each oriented link of the graph. The Fock-Rosly construction provides the space of flat graph connections with a symplectic structure compatible with gauge transformations. More precisely, we impose the flatness of the $\SL(2,\C)$ connection around every face of the graph and consider equivalence classes under $\SL(2,\C)$ gauge transformations at every node of the graph.

To this purpose, we introduce another combinatorial structure to the graph, a linear order of the links around each node.
%
This is visually realized by adding a cilium at each node, signalling the first  link  from which the links are ordered counterclockwise around the node. On these {\it ciliated  graphs}, we further assign a $r$-matrix $r(n)$ to each node $n$. Each $r$-matrix is decomposed as $r(n)=r_s+r_a(n)$ in terms of its  symmetric part $2r_s= r(n)+r_{21}(n)$ and its antisymmetric part $2r_a=r(n)-r_{21}(n)$. As the notation suggests, we require that the symmetric parts $r_{s}$ of the  $r$-matrices are all the same and do not depend on the node, while their antisymmetric parts $r_{a}(n)$ are left free.
This allows to define a Poisson structure on the space of graph connections. 

Let us call $G^e\in\SL(2,\C)$ the holonomy along the edge $e$. We call $s(e)$ and $t(e)$ respectively the source and target vertices of the edge $e$. We also call $e_{(s)}$ the part of the edge attached to its source vertex $s(e)$ and $e_{(t)}$ the part of the edge attached to its target vertex $t(e)$.
Then the Fock-Rosly bracket $\{G^e_1,G^{e'}_2\}_\fr$ between the holonomies along two edges $e$ and $e'$ is defined by distinguishing the various configurations (the list below is not exhaustive but  representative):
\begin{itemize}
\item For a single edge $e$ with distinct source and target, i.e. $s(e)\neq t(e)$:
\be
\{G^e_1,G^e_2\}_\fr= r_a(s(e))G^e_1G^e_2+G^e_1G^e_2r_a(t(e))\,.
\label{eq:FR_1}
\ee
	\item For a single closed curve $e$ oriented counterclockwise, i.e. $s(e) =  t(e)$ and $e_{(s)} <  e_{(t)}$:
\be
\{G^e_1,G^e_2\}_\fr= [r_a,G^e_1G^e_2]_{+}+G^e_2 \rT G^e_1-G^e_1rG^e_2\,.
\label{eq:FR_2}
\ee
	\item For two edges $e$, $e'$ with the same source but different targets distinct from the source, i.e. $s(e)=s(e')$ but $s(e)$, $t(e)$, $t(e')$ all three distinct, and $e_{(s)}<e'_{(s)}$:
\be
\{G^e_1,G^{e'}_2\}_\fr=rG^e_1G^{e'}_2\,.
\label{eq:FR_3}
\ee
	\item For two closed loops $e$, $e'$ intersecting at a single vertex, i.e. $s(e)=s(e')=t(e)= t(e')$, with the order $e_{(s)}<e'_{(s)}<e_{(t)}<e'_{(t)}$:
\be
\{G^e_1,G^{e'}_2\}_\fr={[}r,G^e_1G^{e'}_2{]}_{+} + G^{e'}_2 \rT G^e_1-G^e_1rG^{e'}_2\,.
\label{eq:FR_4}
\ee
\end{itemize}
All the Poisson brackets between two non-intersecting edges vanish.
Then gauge transformations at the vertices, $G^e\mapsto H_{s(e)}G^e H_{t(e)}^{-1}$ for $H_{v}\in\SL(2,\C)$, is a Poisson map leaving the Fock-Rosly bracket invariant \cite{Fock:1998nu}. This means that the Fock-Rosly bracket provides the moduli space of flat graph connection up to gauge transformations with a symplectic structure. One further shows the definition of the Fock-Rosly bracket is stable under contraction and deletion of edges and leads back to the Goldman bracket \cite{Fock:1998nu}.

\medskip

Let us compare the Fock-Rosly and loop quantum gravity approaches.
The Fock-Rosly phase space is defined on the (ciliated) embedded graph $\Gamma$, in the sense that the definition of $\Gamma$ also contains faces (on top of vertices and edges) identified as loops on the graph. It defines a discrete version of the Chern-Simons phase space for a discrete $\SL(2,\C)$ connection. It assigns $\SL(2,\C)$ group elements  to the edges of $\Gamma$, imposes flatness conditions for those group elements around every face of $\Gamma$ and quotients by the $\SL(2,\C)$ action at every vertex of the graph. It is a priori not defined as a symplectic quotient\footnotemark{}. 
\footnotetext{
A special case where the Fock-Rosly phase space is nevertheless directly defined as a symplectic quotient is the {\it flower graph} on a closed oriented connected surface. The flower graph has a single vertex and edges wrapping around every non-contractible cycle of the surface, thus forming a single face. Calling $(A_{i},B_{i})$ the pairs of $\SL(2,\C)$ group elements living on pairs of conjugate cycles, with $i$ running from 1 to the surface genus $g$, the flatness condition $\prod_{i}^g A_{i}B_{i}A_{i}^{-1}B_{i}^{-1}=\id$ also generates $\SL(2,\C)$ gauge transformations at the vertex acting by conjugation simultaneously on all the group elements. In the specific case of the 2-torus with genus $g=1$, the Fock-Rosly framework leads to a symplectic structure defined on $\SL(2,\C)^{\times 2}//\textrm{Ad}\,\SL(2,\C)$ described in details in the appendix of their original paper \cite{Fock:1998nu}.
}

On the other hand, the loop quantum gravity aims to quantize holonomy-flux variables. On the same graph $\Gamma$, it assigns $\SU(2)$ and $\SB(2,\C)$ group elements to the edges of the graph, with both  flatness  constraints around the faces and closure  constraints at the vertices. The loop gravity phase space is defined as a symplectic quotient, with the flatness constraints generating $\SB(2,\C)$ translations for each face and the closure constraints generating $\SU(2)$ rotation at each vertex. 

From this perspective, the loop gravity phase space looks rather different from the Fock-Rosly phase space. We have nevertheless seen in the previous section that one can fatten graph $\Gamma$ into a ribbon graph $\Gamma^{fat}$ by turning every edge of $\Gamma$ into a ribbon and every vertices of $\Gamma$ into polygons to which the ribbons are attached. Now dressing the edges of the ribbon graph $\Gamma^{fat}$ with $\SU(2)$ and $\SB(2,\C)$ group elements, the closure constraints and the flatness constraints, as well as ribbon constraints encoding the transport of the $\SB(2,\C)$ group elements by the $\SU(2)$ group elements across the ribbons, can be all be formulated as $\SL(2,\C)$ flatness conditions. This allows to reconstruct $\SL(2,\C)$ group elements on the initial slim graph $\Gamma$ from the $\SU(2)$ and $\SB(2,\C)$ group elements on the ribbon graph $\Gamma^{fat}$, which satisfy the same $\SL(2,\C)$ flatness constraint as in the Fock-Rosly framework.

However, the resulting  Poisson brackets on the $\SL(2,\C)$ group elements does not match the Fock-Rosly bracket on the $\SL(2,\C)$ holonomies. They nevertheless lead to the same Goldman bracket on the moduli space of flat $\SL(2,\C)$ graph connections on $\Gamma$. In order to reconcile the Fock-Rosly approach with the loop gravity phase space, it seems natural to work on the fat graph $\Gamma^{fat}$ and compare the Fock-Rosly bracket for $\SL(2,\C)$ holonomies on $\Gamma^{fat}$ with the loop gravity phase space. In fact, we explain below that we need to go one step further and introduce a ``{\it fatter graph}'' $\Gamma^{fatter}$ and that the loop gravity on $\Gamma^{fat}$ turns out to result from the Fock-Rosly structure on $\Gamma^{fatter}$ by a partial gauge-fixing.

\medskip

{\bf Gauge-fixing Fock-Rosly on the fatter graph to recover loop gravity on the fat graph:}

\begin{figure}[t!]
	\begin{tikzpicture}[scale=1]
	
	\coordinate (O) at (0,0);
	\coordinate (A) at (9,0);
	\coordinate (B) at (9,6);
	\coordinate (C) at (0,6);
	\draw (O)--(A)--(B)--(C)--(O);
	
	\coordinate (a1) at (3.8,1.3);
	\coordinate (b1) at (5.2,1.3);
	\coordinate (c1) at (3.8,4.7);
	\coordinate (d1) at (5.2,4.7);
	
	\coordinate (a2) at (2.5,2.3);
	\coordinate (b2) at (6.5,2.3);
	\coordinate (c2) at (2.5,3.7);
	\coordinate (d2) at (6.5,3.7);
	
	\coordinate (ld) at (0,2.3);
    \coordinate (lu) at (0,3.7);
    \coordinate (rd) at (9,2.3);
    \coordinate (ru) at (9,3.7);
    \coordinate (dl) at (3.8,0);
    \coordinate (dr) at (5.2,0);
    \coordinate (ul) at (3.8,6);
    \coordinate (ur) at (5.2,6);

    \coordinate (aa1) at (5.1,1.7);
    \coordinate (aa2) at (6.1,2.5);
    \coordinate (bb1) at (6.1,3.5);
    \coordinate (bb2) at (5.1,4.3);
    \coordinate (cc1) at (3.9,4.3);
    \coordinate (cc2) at (2.9,3.5);
    \coordinate (dd1) at (2.9,2.5);
    \coordinate (dd2) at (3.9,1.7);
    
    \coordinate (f2) at (2,1.1);
    \coordinate (f1) at (4.5,3);
    \coordinate (f3) at (1.2,3);
    \coordinate (f4) at (4.5,0.5);
	
	\draw[thick,decoration={markings,mark=at position 0.55 with {\arrow[scale=1.3,>=stealth]{>}}},postaction={decorate}] (ld) -- node[midway, below]{$\Ut$}(a2); 
	\draw[thick,decoration={markings,mark=at position 0.55 with {\arrow[scale=1.3,>=stealth]{>}}},postaction={decorate}] (b2)  --node[midway, below]{$\Ut$}(rd); 
	\draw[thick,decoration={markings,mark=at position 0.55 with {\arrow[scale=1.3,>=stealth]{>}}},postaction={decorate}] (lu) --node[midway,above]{$U$}(c2); 
	\draw[thick,decoration={markings,mark=at position 0.55 with {\arrow[scale=1.3,>=stealth]{>}}},postaction={decorate}] (d2) --node[midway,above]{$U$}(ru);
	 
	\draw[thick,decoration={markings,mark=at position 0.55 with {\arrow[scale=1.3,>=stealth]{>}}},postaction={decorate}] (dl) -- node[midway, left]{$V$}(a1); 
	\draw[thick,decoration={markings,mark=at position 0.55 with {\arrow[scale=1.3,>=stealth]{>}}},postaction={decorate}] (dr)  --node[midway, right]{$\Vt$}(b1); 
	\draw[thick,decoration={markings,mark=at position 0.55 with {\arrow[scale=1.3,>=stealth]{>}}},postaction={decorate}] (c1) -- node[midway, left]{$V$}(ul); 
	\draw[thick,decoration={markings,mark=at position 0.55 with {\arrow[scale=1.3,>=stealth]{>}}},postaction={decorate}] (d1)  --node[midway, right]{$\Vt$}(ur) ;
	
	\draw[thick,dashed,decoration={markings,mark=at position 0.55 with {\arrow[scale=1.3,>=stealth]{>}}},postaction={decorate}] (a2) --node[midway, left]{$\Lt$}(c2); 
	\draw[thick,dashed,decoration={markings,mark=at position 0.55 with {\arrow[scale=1.3,>=stealth]{>}}},postaction={decorate}] (b2) --node[midway, right]{$L$}(d2); 
	\draw[thick,dashed,decoration={markings,mark=at position 0.55 with {\arrow[scale=1.3,>=stealth]{>}}},postaction={decorate}] (b1) --node[midway,below]{$\Mt$}(a1); 
	\draw[thick,dashed,decoration={markings,mark=at position 0.55 with {\arrow[scale=1.3,>=stealth]{>}}},postaction={decorate}] (d1) --node[midway,above]{$M$}(c1);
	
	\draw[thick,dotted,decoration={markings,mark=at position 0.55 with {\arrow[scale=1.3,>=stealth]{>}}},postaction={decorate}] (d2) --node[midway,above]{$P$}(d1); 
	\draw[thick,dotted,decoration={markings,mark=at position 0.55 with {\arrow[scale=1.3,>=stealth]{>}}},postaction={decorate}] (c1) --node[midway,above]{$Q$}(c2); 
	\draw[thick,dotted,decoration={markings,mark=at position 0.55 with {\arrow[scale=1.3,>=stealth]{>}}},postaction={decorate}] (b1) --node[midway,below]{$T$}(b2); 
	\draw[thick,dotted,decoration={markings,mark=at position 0.55 with {\arrow[scale=1.3,>=stealth]{>}}},postaction={decorate}] (a2) --node[midway,below]{$S$}(a1);

    \draw (f1) node {$f_1$};
    \draw (f2) node {$f_2$};
    \draw (f3) node {$f_3$};
    \draw (f4) node {$f_4$};
	
	\draw[color=blue] (a1) node {$\bullet$};
	\draw[color=red] (b1) node {$\bullet$};
	\draw[color=red] (c1) node {$\bullet$};
	\draw[color=blue] (d1) node {$\bullet$};
	\draw[color=red] (a2) node {$\bullet$};
	\draw[color=blue] (b2) node {$\bullet$};
	\draw[color=blue] (c2) node {$\bullet$};
	\draw[color=red] (d2) node {$\bullet$};
	
	\draw[thick] (b1)--(aa1); 
	\draw[thick] (b2)--(aa2); 
	\draw[thick] (d2)--(bb1); 
	\draw[thick] (d1)--(bb2); 
	\draw[thick] (c1)--(cc1); 
	\draw[thick] (c2)--(cc2); 
	\draw[thick] (a2)--(dd1); 
	\draw[thick] (a1)--(dd2); 
	
	\draw[fill=gray!50,opacity=.4](b2) -- (d2) -- (ru) -- (rd) -- cycle ;
	\draw[fill=gray!50,opacity=.4](a2) -- (c2) -- (lu) -- (ld) -- cycle ;
	\draw[fill=gray!50,opacity=.4](a1) -- (b1) -- (dr) -- (dl) -- cycle ;
	\draw[fill=gray!50,opacity=.4](c1) -- (d1) -- (ur) -- (ul) -- cycle ;
	
\end{tikzpicture}
\caption{Ciliated fat ribbon graph, or fatter graph for short, on the torus. Vertices $s(P)$, $s(Q)$, $s(S)$ and $s(T)$ ({\it in red}) are assigned the $r$-matrix $r$, while vertices $t(P)$, $t(Q)$, $t(S)$ and $t(T)$ ({\it in blue}) are assigned $\rT$. At each vertex, a cilium is introduced to fix the linear order of the links around this vertex. The cilia are chosen to all look into the face $f_1$.}
\label{fig:ribbon_fat}
\end{figure}
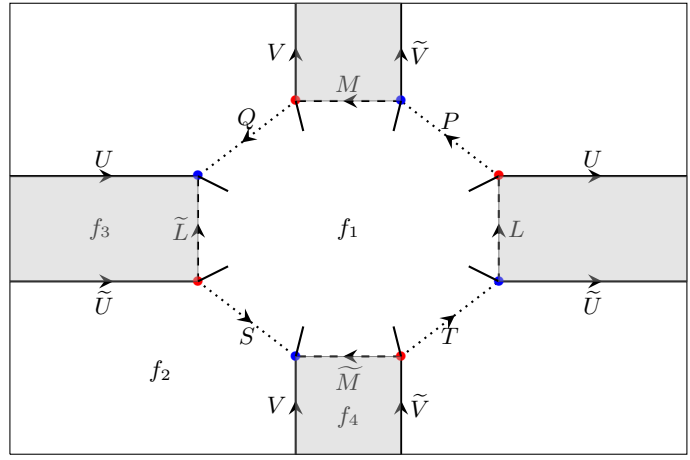

We consider te Fock-Rosly phase space on the torus on a fatter graph, illustrated on fig.\ref{fig:ribbon_fat}, where we have unfolden the four 4-valent nodes of the ribbon graph into pairs of 3-valent vertices by adding an intermediate edge. This leads to a graph with twelve edges, each decorated with  $\SL(2,\C)$ holonomies. We note  $L,U,\Lt,\Ut,M,V,\Mt,\Vt$ the eight group elements along the ribbon edges, and $P,Q,S,T$ the four group elements on the new intermediate edges. To simplify the notations, we refer to the edge through the $\SL(2,C)$ group element it carries.

In order to recover the loop gravity Poisson structure, we assign the $r$-matrix $r$ to the source vertices of the intermediate edges $s(P)$, $s(Q)$, $s(S)$ and $s(T)$, while we  assign the $r$-matrix $\rT$ to their target vertices $t(P)$, $t(Q)$, $t(S)$ and $t(T)$. We further choose  all the cilia looking inwards to the face $f_1$ to fix the convention. A different choice of cilia would still lead to the loop gravity phase space. 

The Fock-Rosly brackets can be read directly from their definition \eqref{eq:FR_1}-\eqref{eq:FR_4} applied to the ciliated fat ribbon graph, or fatter graph for short, shown in fig.\ref{fig:ribbon_fat}.  All the non-vanishing Fock-Rosly brackets related to $P$ are
\be\ba{llll}
\{P_1,L_2\}_\fr=L_2\rT P_1\,,&
\{L_1,P_2\}_\fr=-L_1r A_2\,,&
\{P_1,U_2\}_\fr=-\rT P_1U_2\,,&
\{U_1,P_2\}_\fr =rU_1P_2\,,\\
\{P_1,M_2\}_\fr=P_1\rT M_2\,,&
\{M_1,P_2\}_\fr=-P_2r M_1\,,&
\{P_1,\Vt_2\}_\fr =P_1\rT\Vt_2\,,&
\{\Vt_1,P_2\}_\fr=-P_2r\Vt_1\,,\\
\{P_1,P_2\}_\fr=[r,P_1P_2]\,.
\ea
\label{eq:Fock-Rosly_P}
\ee

Closed holonomies $A$, $B$ can be reconstructed based on the vertex $s(P)$,  as $A\equiv UQ^{-1}M^{-1}P^{-1}$ and $B\equiv P\Vt TL$. A direct computation shows the Fock-Rosly brackets between $A$ and $B$ are indeed the ones given in \eqref{eq:FR_2} and \eqref{eq:FR_4}:
\begin{align}
&\{A_1,A_2\}_{\fr}=r_a A_1A_2+A_1A_2r_a +A_2\rT A_1-A_1rA_2\,,\label{eq:Fock-Rosly_AB_1}\\
&\{B_1,B_2\}_{\fr}=r_a B_1B_2+B_1B_2r_a +B_2\rT B_1-B_1rB_2\,,\label{eq:Fock-Rosly_AB_2}\\
&\{A_1,B_2\}_{\fr}=r A_1B_2+A_1B_2r +B_2\rT A_1-A_1rB_2\,.\label{eq:Fock-Rosly_AB_3}
\end{align}
One notices that although these brackets define a different Poisson structure from \eqref{eq:Poisson_AB}, they lead to the same Goldman brackets on gauge invariant observables as  the loop gravity phase space: $\tr\left(\{A_1,B_2\}_\fr\right)=\tr\left(A_1B_2(r+\rT)\right)$. 

The Fock-Rosly phase space is defined on top of the Poisson brackets given above by imposing the flatness of the $\SL(2,\C)$ connection around the four faces of the fatter graph:
\begin{align}
&\cC^{f_1}=TLPMQ\Lt^{-1}S\Mt^{-1}\,,\\
&\cC^{f_2}=UQ^{-1}VS^{-1}\Ut^{-1}T^{-1}\Vt^{-1}P^{-1}\,,\\
&\cC^{f_3}=LU\Lt^{-1}\Ut^{-1}\,,\label{eq:Cf3}\\
&\cC^{f_4}=MV\Mt^{-1}\Vt^{-1}\,.\label{eq:Cf4}
\,,
\end{align}
and quotienting by the $\SL(2,\C)$ group action at the eight nodes of the graph.

\medskip

If we look at the Fock-Rosly brackets for the eight group elements $L,U,\Lt,\Ut,M,V,\Mt,\Vt$, they are the same Poisson brackets as for the loop gravity phase space parametrized by the group elements $\ell,u,\tilde{\ell},\ut,m,v,\mt,\vt$. The only difference is that the Fock-Rosly brackets involve $\SL(2,\C)$ group elements, while the loop gravity group elements live alternatively in the subgroups $\SU(2)$ and $\SB(2,\C)$.
Thus, in order to recover the loop gravity phase space, we perform a partial gauge fixing of these $\SL(2,\C)$ group elements. We start at the vertex $t(T)=s(L)=s(\Ut)$ and will go around the face $f_{1}$. We use the $\SL(2,\C)$ gauge invariance at that vertex to fix $T=\id$. Then moving to the following node $t(L)=s(U)=s(P)$, we use the Iwasawa decompositions $L=\ell u^L$ and $U=\ell^U u$ and perform a gauge transformation:
\be
u^{L}\ell^{U}=\ell' u'
\,,\quad
G=(\ell')^{{-1}}u^{L}
\,,\qquad
\left|\begin{array}{lclll}
L&\mapsto& LG^{-1}&=\ell \ell'&\in\SB(2,\C)
\,,\vspace*{1mm}\\
U&\mapsto& G U&=u' u&\in\SU(2)
\,,\vspace*{1mm}\\
P&\mapsto& G P&= (\ell')^{{-1}}u^{L} P
\,,&
\end{array} \right.
\ee
thus gauge fixing $L$ to live in $\SB(2,\C)$  and $U$ to live in $\SU(2)$.

Next, at the following node, $t(P)=s(M)=s(\Vt)$, we perform a $\SL(2,\C)$ gauge transformation to gauge fix to $P=\id$. Then we repeat this pair of gauge fixings all around the central face $f_{1}$. So at the node $t(M)=s(V)=s(Q)$, we gauge fix to $M\in\SB(2,\C)$ and $V\sin\SU(2)$. At the node $t(Q)=t(U)=t(\Lt)$ we gauge fix to $Q=\id$. At the node $s(S)=s(\Lt)=t(\Ut)$, we gauge fix to $\Lt\in\SB(2,\C)$ and $\Ut\in\SU(2)$. At the node $t(S)=t(V)=t(\Mt)$, we gauge fix to $S=\id$.

Finally at the last node $s(T)=s(\Mt)=t(\Vt)$, we do not do anything. The flatness condition around the face $f_{1}$ automatically implies that the group element $\Mt$ lives in $\SB(2,\C)$ while the flatness condition around the face $f_{2}$ automatically implies that the group element $\Vt$ lives in the $\SU(2)$ subgroup.
Since we do not gauge-fix the group action at the last vertex of the graph, we are left with a single $\SL(2,\C)$ gauge invariance of our partially gauge-fixed group variables.

This reproduces exactly the setting of the $q$-deformed loop quantum gravity variables on the ribbon graph, with the group elements $\ell,\tilde{\ell},m,\mt\,\in\SB(2,\C)$ and $u,\ut,v,\vt\,\in\SU(2)$ satisfying the flatness constraints around the 4 faces of the graph:
\begin{align}
&\cC^{f_1}=uv\ut^{-1}\vt^{-1}\,,\\
&\cC^{f_2}=\ell m\lt^{-1}\mt^{-1}\,,\\
&\cC^{f_3}=\ell u\lt^{-1}\ut^{-1}\,,\\
&\cC^{f_4}=mv\mt^{-1}\vt^{-1}\,.
\end{align}
Since the $\SL(2,\C)$ group action at the nodes is a Poisson map for the Fock-Rosly symplectic structure \cite{Fock:1998nu}, it is straightforward to check that the Fock-Rosly brackets on the original $\SL(2,\C)$ group elements $L,U,\Lt,\Ut,M,V,\Mt,\Vt$ directly descend to Poisson brackets on the gauge-fixed group variables $\ell,u,\tilde{\ell},\ut,m,v,\mt,\vt$:
\be\ba{llll}
\{\ell_1,\ell_2\}_\fr=-[r,\ell_1\ell_2]\,,&
\{u_1,u_2\}_\fr=-[\rT,u_1u_2]\,,&
\{\lt_1,\lt_2\}_\fr=[r,\lt_1\lt_2]\,,&
\{\ut_1,\ut_2\}_\fr=[\rT,\ut_1\ut_2]\,,\\
\{m_1,m_2\}_\fr=-[r,m_1m_2]\,,&
\{v_1,v_2\}_\fr=-[\rT,v_1v_2]\,,&
\{\mt_1,\mt_2\}_\fr=[r,\mt_1\mt_2]\,,&
\{\vt_1,\vt_2\}_\fr=[\rT,\vt_1\vt_2]\,,\\
\{\ell_1,u_2\}_\fr=-\ell_1ru_2\,,&
\{\lt_1,\ut_2\}_\fr=-\ut_2r\lt_1\,,&
\{m_1,v_2\}_\fr=-m_1rv_2\,,&
\{\mt_1,\vt_2\}_\fr=-\vt_2r\mt_1\,.
\ea\label{eq:lu_fr}
\ee
These are precisely the (flat or $q$-deformed) loop gravity Poisson brackets. $\cC^{f_1}$ and $\cC^{f_2}$ are still first class constraints generating respectively the gauge invariance under $\SB(2,\C)$ translation and $\SU(2)$ rotations, while $\cC^{f_3}$ and $\cC^{f_4}$ are second class constraints directly hardcoded in the Poisson brackets.
This explicitly shows that the loop gravity phase space can be reconstructed from the Fock-Rosly description by a specific gauge fixing. 

\medskip

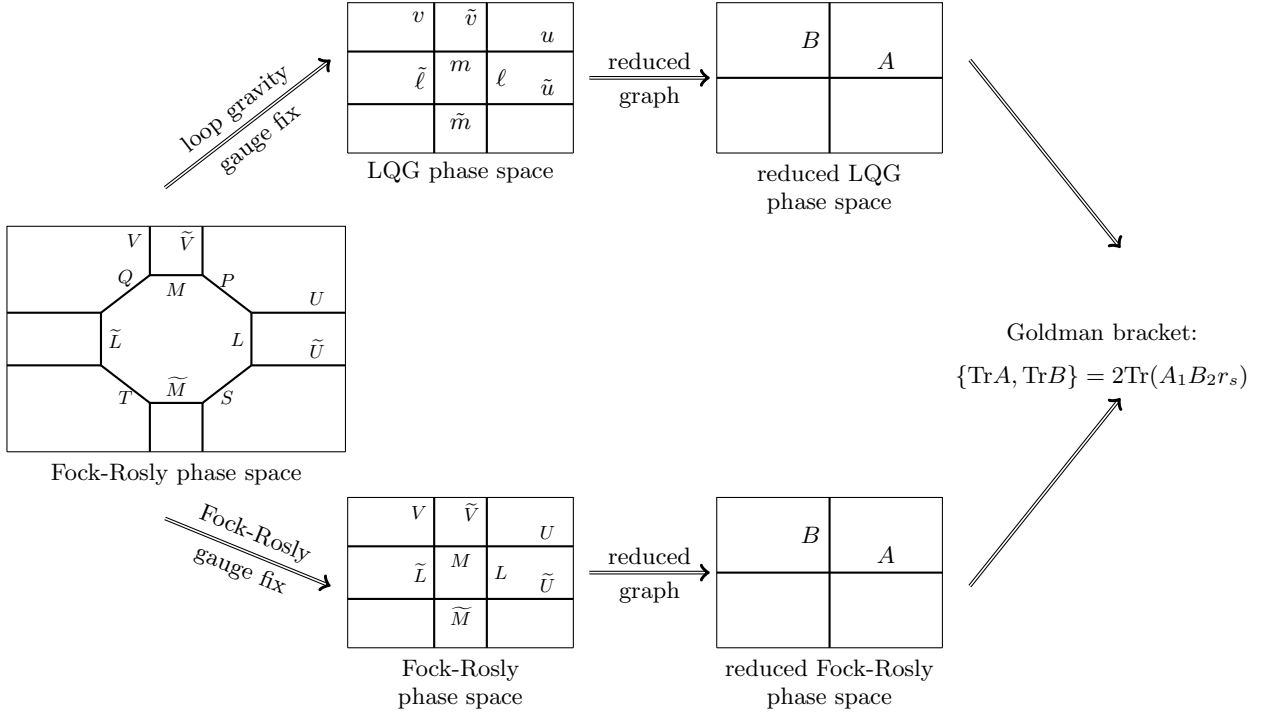
\begin{figure}[h!]
 
	\begin{tikzpicture}[scale=1]
 
 \def\sca{0.5};
 	
 	\matrix(m)
 	{
 	{1}&
 	 {
 	\coordinate (O) at (0,0);
	\coordinate (A) at (3,0);
	\coordinate (B) at (3,2);
	\coordinate (C) at (0,2);
	\draw (A)--(B)--(C)--(O);
	\draw (O)--node[midway,below]{$\text{LQG phase space}$}(A);
	
	\coordinate (a) at (2.3/2,1.3/2);
	\coordinate (b) at (3.7/2,1.3/2);
	\coordinate (c) at (2.3/2,2.7/2);
	\coordinate (d) at (3.7/2,2.7/2);
	
	\coordinate (ld) at (0,1.3/2);
    \coordinate (lu) at (0,2.7/2);
    \coordinate (rd) at (6/2,1.3/2);
    \coordinate (ru) at (6/2,2.7/2);
    \coordinate (dl) at (2.3/2,0);
    \coordinate (dr) at (3.7/2,0);
    \coordinate (ul) at (2.3/2,4/2);
    \coordinate (ur) at (3.7/2,4/2);
	
	\draw[thick] (ld) --(a); 
	\draw[thick] (b) --node[pos=0.7,above]{$\ut$}(rd); 
	\draw[thick] (lu) --(c); 
	\draw[thick] (d) --node[pos=0.7,above]{$u$}(ru);
	 
	\draw[thick] (dl) -- (a); 
	\draw[thick] (dr)  --(b); 
	\draw[thick] (c) --node[pos=0.7,left]{$v$} (ul); 
	\draw[thick] (d)  --node[pos=0.7,left]{$\vt$}(ur) ;
	
	\draw[thick] (a) --node[midway,left]{$\lt$}(c); 
	\draw[thick] (b) --node[midway,right]{$\ell$}(d); 

	\draw[thick] (b) --node[midway,below]{$\mt$}(a); 
	\draw[thick] (d) --node[midway,below]{$m$}(c);
	}&
	{
 	\coordinate (O) at (0,0);
	\coordinate (A) at (2,0);
	\coordinate (B) at (2,2);
	\coordinate (C) at (0,2);
	\coordinate (D) at (0.2,1);
 	\coordinate (E) at (1.8,1);
 	\draw[double,->] (D) -- node[midway, above]{reduced}node[midway,below]{graph}(E);
	}&
{ 	
 	\coordinate (O) at (0,0);
	\coordinate (A) at (3,0);
	\coordinate (B) at (3,2);
	\coordinate (C) at (0,2);
	\draw (A)--(B)--(C)--(O);
	\draw (O)--node[midway,below]{\begin{tabular}{c}
		reduced LQG \\ phase space
	\end{tabular}}(A);
	
	\coordinate (l) at (0,1);
	\coordinate (r) at (3,1);
	\coordinate (u) at (1.5,2);
	\coordinate (d) at (1.5,0);
	
	\coordinate (CC) at (1.5,1);
		
	\draw[thick] (CC) --node[midway,above]{$A$}(r); 
	\draw[thick] (CC) --node[midway,left]{$B$}(u);
	\draw[thick] (CC) --(l);
	\draw[thick] (CC) --(d);

}&
{5}\\
 	 {
	\coordinate (O) at (0,0);
	\coordinate (A) at (3*1.5,0);
	\coordinate (B) at (3*1.5,2*1.5);
	\coordinate (C) at (0,2*1.5);
	\draw (A)--(B)--(C)--(O);
	\draw (O)--node[midway,below]{\begin{tabular}{c}
		Fock-Rosly phase space
	\end{tabular}} (A);
	
	\coordinate (a1) at (3.8/3*1.5,1.3/3*1.5);
	\coordinate (b1) at (5.2/3*1.5,1.3/3*1.5);
	\coordinate (c1) at (3.8/3*1.5,4.7/3*1.5);
	\coordinate (d1) at (5.2/3*1.5,4.7/3*1.5);
	
	\coordinate (a2) at (2.5/3*1.5,2.3/3*1.5);
	\coordinate (b2) at (6.5/3*1.5,2.3/3*1.5);
	\coordinate (c2) at (2.5/3*1.5,3.7/3*1.5);
	\coordinate (d2) at (6.5/3*1.5,3.7/3*1.5);
	
	\coordinate (ld) at (0,2.3/3*1.5);
    \coordinate (lu) at (0,3.7/3*1.5);
    \coordinate (rd) at (9/3*1.5,2.3/3*1.5);
    \coordinate (ru) at (9/3*1.5,3.7/3*1.5);
    \coordinate (dl) at (3.8/3*1.5,0);
    \coordinate (dr) at (5.2/3*1.5,0);
    \coordinate (ul) at (3.8/3*1.5,6/3*1.5);
    \coordinate (ur) at (5.2/3*1.5,6/3*1.5);
	
	\draw[thick] (ld) --(a2); 
	\draw[thick] (b2)  --node[pos=0.7,above,scale=.8]{$\Ut$}(rd); 
	\draw[thick] (lu) --(c2); 
	\draw[thick] (d2) --node[pos=0.7,above,scale=.8]{$U$}(ru);
	 
	\draw[thick] (dl) -- (a1); 
	\draw[thick] (dr)  --(b1); 
	\draw[thick] (c1) -- node[pos=0.7,left,scale=.8]{$V$}(ul); 
	\draw[thick] (d1)  --node[pos=0.7,left,scale=.8]{$\Vt$}(ur) ;
	
	\draw[thick] (a2) --node[midway,right,scale=.8]{$\Lt$}(c2); 
	\draw[thick] (b2) --node[midway,left,scale=.8]{$L$}(d2); 
	\draw[thick] (b1) --node[midway,above,scale=.8]{$\Mt$}(a1); 
	\draw[thick] (d1) --node[midway,below,scale=.8]{$M$}(c1);
	
	\draw[thick] (d2) --node[midway,above,scale=.8]{$P$}(d1); 
	\draw[thick] (c1) --node[midway,above,scale=.8]{$Q$}(c2); 
	\draw[thick] (b1) --node[midway,below,scale=.8]{$S$}(b2); 
	\draw[thick] (a2) --node[midway,below,scale=.8]{$T$}(a1); 
 	 }
 	 &{2}&{3}&{4}&
 {
  	\coordinate (C) at (1.5,1);
	\coordinate (O) at (1.5,1+0.6);
	\draw (O) node{Goldman bracket: };
 \draw (C) node{$\{\tr A,\tr B\}=2\tr (A_1B_2r_s)$};

 }\\
 {1}&
 {
  	\coordinate (O) at (0,0);
	\coordinate (A) at (3,0);
	\coordinate (B) at (3,2);
	\coordinate (C) at (0,2);
	\draw (A)--(B)--(C)--(O);
	\draw (O)--node[midway,below]{\begin{tabular}{c}
		Fock-Rosly \\ phase space
	\end{tabular}}(A);
	
	\coordinate (a) at (2.3/2,1.3/2);
	\coordinate (b) at (3.7/2,1.3/2);
	\coordinate (c) at (2.3/2,2.7/2);
	\coordinate (d) at (3.7/2,2.7/2);
	
	\coordinate (ld) at (0,1.3/2);
    \coordinate (lu) at (0,2.7/2);
    \coordinate (rd) at (6/2,1.3/2);
    \coordinate (ru) at (6/2,2.7/2);
    \coordinate (dl) at (2.3/2,0);
    \coordinate (dr) at (3.7/2,0);
    \coordinate (ul) at (2.3/2,4/2);
    \coordinate (ur) at (3.7/2,4/2);
	
	\draw[thick] (ld) --(a); 
	\draw[thick] (b) --node[pos=0.7,above,scale=.8]{$\Ut$}(rd); 
	\draw[thick] (lu) --(c); 
	\draw[thick] (d) --node[pos=0.7,above,scale=.8]{$U$}(ru);
	 
	\draw[thick] (dl) -- (a); 
	\draw[thick] (dr)  --(b); 
	\draw[thick] (c) --node[pos=0.7,left,scale=.8]{$V$} (ul); 
	\draw[thick] (d)  --node[pos=0.7,left,scale=.8]{$\Vt$}(ur) ;
	
	\draw[thick] (a) --node[midway,left,scale=.8]{$\Lt$}(c); 
	\draw[thick] (b) --node[midway,right,scale=.8]{$L$}(d); 

	\draw[thick] (b) --node[midway,below,scale=.8]{$\Mt$}(a); 
	\draw[thick] (d) --node[midway,below,scale=.8]{$M$}(c);

 }&
 {
  	\coordinate (O) at (0,0);
	\coordinate (A) at (2,0);
	\coordinate (B) at (2,2);
	\coordinate (C) at (0,2);
	\coordinate (D) at (0.2,1);
 	\coordinate (E) at (	1.8,1);
 	\draw[double,->] (D) -- node[midway, above]{$\text{reduced}$}node[midway,below]{$\text{graph}$}(E);
 	}&
 {
  	\coordinate (O) at (0,0);
	\coordinate (A) at (3,0);
	\coordinate (B) at (3,2);
	\coordinate (C) at (0,2);
	\draw (A)--(B)--(C)--(O);
	\draw (O)--node[midway,below]{\begin{tabular}{c}
		reduced Fock-Rosly \\ phase space
	\end{tabular}}(A);
		
	\coordinate (l) at (0,1);
	\coordinate (r) at (3,1);
	\coordinate (u) at (1.5,2);
	\coordinate (d) at (1.5,0);
	
	\coordinate (CC) at (1.5,1);
		
	\draw[thick] (CC) --node[midway,above]{$A$}(r); 
	\draw[thick] (CC) --node[midway,left]{$B$}(u);
	\draw[thick] (CC) --(l);
	\draw[thick] (CC) --(d);
 }&
 {5}\\
 };
 \coordinate (OO) at (6.5,1.5);
 \coordinate (AA) at (4.5,4);
 \coordinate (CC) at (6.5,-0.5);
 \coordinate (BB) at (4.5,-3);
 \draw[double,->] (AA) -- (OO);
 \draw[double,->] (BB) -- (CC);
 
 \coordinate (DD) at (-6.2,2.3);
 \coordinate (EE) at (-4,4);
 \draw[double,->] (DD) -- node[midway, sloped, above]{loop gravity}node[midway, sloped, below]{gauge fix}(EE);
 \coordinate (FF) at (-6.2,-2.1);
 \coordinate (GG) at (-4,-3);
 \draw[double,->] (FF) -- node[midway, sloped, above]{Fock-Rosly}node[midway, sloped, below]{gauge fix}(GG);

\end{tikzpicture}
\caption{Different choices of gauge fixing from the original Fock-Rosly phase space result in different reduced phase spaces, but with the same Goldman brackets.
Capital letters represent $\SL(2,\C)$ holonomies while $\{\ell,\lt,m,\mt\}\in \SB(2,\C)$ and $\{u,\ut,v,\vt\}\in\SU(2)$.}
\label{fig:relation}
\end{figure}

Let us stress that the partial gauge fixing introduced here mapping the Fock-Rosly phase space to the $q$-deformed loop gravity phase space is very different from the gauge fixing usually done in the Fock-Rosly approach to go from a refined graph to a coarse-grained graph (subgraph of the original graph) by simply setting all the extra $\SL(2,\C)$ group elements to the identity. These different gauge fixing produces different intermediate Poisson brackets, which nevertheless all lead to the same Goldman brackets on the $\SL(2,\C)$ gauge-invariant variables.
The hierarchy of fat and fatter graphs, with their  different Poisson brackets depending on the different choices of (partial) gauge fixing,  is illustrated in fig.\ref{fig:relation}. 

%
%


\section*{Conclusion}
\label{sec:conclusion}

In this present paper, we studied the loop gravity phase space applied to the 2-torus. That is,  our kinematical phase space was constructed for a graph embedded on the torus.  
The cases of a vanishing cosmological constant and of a negative cosmological constant can be treated in a similar way. Indeed, the key step is to recognize that the standard loop gravity phase for a zero cosmological constant can be viewed as the Heisenberg double $\ISU(2)$ where the Poisson brackets are written in terms of a classical $r$-matrix. This phase space is constrained  by the closure constraint (Gauss law) and the $\SU(2)$ flatness constraint that generate Poisson-Lie group symmetries. 

Using this compact formalism, we showed that the standard loop gravity phase space can be written in terms of Poincar\'e holonomies constrained by a Poincar\'e flatness. From this reformulation, the definition of gauge-invariant observables of the theory is straightforward. This already points out toward a connection between the loop gravity phase space and the classical analogue of the combinatorial quantization formulation of Chern-Simons, the so-called Fock-Rosly phase space. Indeed, for a zero cosmological constant, the Chern-Simons gauge group is $\ISU(2)$ and the Fock-Rosly phase space variables are Poincar\'e holonomies living on edges of graphs that are constrained by Poincar\'e flatness on the faces of the graphs.  

This equivalence between the loop gravity phase space and the Fock-Rosly phase space is made explicit in this paper in the case of  3D Euclidean gravity with a negative cosmological constant. For this case, the loop gravity phase space is seen as a deformation of the standard loop gravity phase space. More precisely, the Heisenberg double $\ISU(2)$ is $q$-deformed into the Heisenberg double $\SL(2,\C) \sim \SU(2)\bowtie\SB(2,\C)$ where the deformation $q$, related to the cosmological constant, is introduced to curve the momentum space from $\R^3$ to $\SB(2,\C)$. The constraints are now $\SB(2, \C)$ flatness as well as $\SU(2)$ flatness generating Poisson-Lie group symmetries. To visualize the phase space variables as well as the constraints, it is natural to work with fat graphs, where each edge of the initial graph has been fattened into a ribbon. A ribbon constraint is associated to each ribbon and these constraints are solved by requiring that the left and right Iwasawa decompositions of the $\SL(2, \C)$ group element of each edge are equal.  

In a similar way as in the flat case, we showed that this formulation allows to write the $q$-deformed loop gravity phase space  in terms of $\SL(2,\C)$ holonomies living on the initial graph and satisfying a $\SL(2,\C)$ flatness constraint. Again, this sounds very similar to what would be a Fock-Rosly phase space for $\SL(2,\C)$ Chern-Simons theory (i.e. Chern-Simons describing 3D Euclidean gravity with a negative cosmological constant). However, the brackets at this stage are not the same, although Goldman brackets are recovered going to the reduced phase space in both approaches. The equivalence between the two phase spaces is explicitly shown by going to a ``fatter graph'' (see previous section for an explicit definition). Then, the $q$-deformed loop gravity phase space and its constraints are regained by a partial and asymmetric gauge fixing of the (extended - i.e on the ``fatter" graph) Fock-Rosly phase space. It is worth noticing  showing this equivalence requires working with the Fock-Rosly formalism defined on a general graph and not on a flower graph as it is very often the case.

This paper only focused on the simple topology of a 2-torus, i.e our discrete kinematical phase spaces were defined through graphs embedded on the torus, which is the simplest non-trivial case. But we expect that  our results and in particular the equivalence between the $q$-deformed loop gravity phase space and the Fock-Rosly phase space naturally generalize to manifolds of topology $\cM \sim \R \times \Sigma$, with $\Sigma$ a surface of general genus $g$ (and with punctures).
We leave this full analysis for future work. This would conclude the convergence of the various approaches to the quantization of 3D gravity. 
\section*{Acknowledgements}

The authors would like to thank Florian Girelli for helpful discussions.
This research was supported in part by Perimeter Institute for Theoretical
Physics. Research at Perimeter Institute is supported by the Government
of Canada through the Department of Innovation, Science and Economic
Development Canada  and by the Province of Ontario through the Ministry
of Research, Innovation and Science. QP is supported by a NSERC Discovery grant awarded to MD.

\bibliographystyle{bib-style}
\bibliography{qLQGonTorus}

\providecommand{\href}[2]{#2}\begingroup\raggedright\begin{thebibliography}{10}

\bibitem{DePietri:1998hnx}
R.~De~Pietri and L.~Freidel, ``{so(4) Plebanski action and relativistic spin
  foam model},'' Class. Quant. Grav. {\bf 16} (1999) 2187--2196,
\href{http://arXiv.org/abs/gr-qc/9804071}{{\texttt{arXiv:gr-qc/9804071}}}.

\bibitem{Barrett:1997gw}
J.~W. Barrett and L.~Crane, ``{Relativistic spin networks and quantum
  gravity},'' J. Math. Phys. {\bf 39} (1998) 3296--3302,
\href{http://arXiv.org/abs/gr-qc/9709028}{{\texttt{arXiv:gr-qc/9709028}}}.

\bibitem{Barrett:1999qw}
J.~W. Barrett and L.~Crane, ``{A Lorentzian signature model for quantum general
  relativity},'' Class. Quant. Grav. {\bf 17} (2000) 3101--3118,
\href{http://arXiv.org/abs/gr-qc/9904025}{{\texttt{arXiv:gr-qc/9904025}}}.

\bibitem{Smolin:2003qu}
L.~Smolin and A.~Starodubtsev, ``{General relativity with a topological phase:
  An Action principle},''
\href{http://arXiv.org/abs/hep-th/0311163}{{\texttt{arXiv:hep-th/0311163}}}.

\bibitem{Freidel:2005ak}
L.~Freidel and A.~Starodubtsev, ``{Quantum gravity in terms of topological
  observables},''
\href{http://arXiv.org/abs/hep-th/0501191}{{\texttt{arXiv:hep-th/0501191}}}.

\bibitem{Witten:1988hc}
E.~Witten, ``{(2+1)-Dimensional Gravity as an Exactly Soluble System},'' Nucl.
  Phys. {\bf B311} (1988)
46.

\bibitem{Carlip:2003}
S.~{Carlip}, {\em {Quantum Gravity in 2+1 Dimensions}}.
\newblock Dec., 2003.

\bibitem{Alekseev:1994pa}
A.~{\relax Yu}. Alekseev, H.~Grosse, and V.~Schomerus, ``{Combinatorial
  quantization of the Hamiltonian Chern-Simons theory},'' Commun. Math. Phys.
  {\bf 172} (1995) 317--358,
\href{http://arXiv.org/abs/hep-th/9403066}{{\texttt{arXiv:hep-th/9403066}}}.

\bibitem{Alekseev:1994au}
A.~{\relax Yu}. Alekseev, H.~Grosse, and V.~Schomerus, ``{Combinatorial
  quantization of the Hamiltonian Chern-Simons theory. 2.},'' Commun. Math.
  Phys. {\bf 174} (1995) 561--604,
\href{http://arXiv.org/abs/hep-th/9408097}{{\texttt{arXiv:hep-th/9408097}}}.

\bibitem{Fock:1998nu}
V.~V. Fock and A.~A. Rosly, ``{Poisson structure on moduli of flat connections
  on Riemann surfaces and r matrix},'' Am. Math. Soc. Transl. {\bf 191} (1999)
  67--86,
\href{http://arXiv.org/abs/math/9802054}{{\texttt{arXiv:math/9802054}}}.

\bibitem{Buffenoir:2002tx}
E.~Buffenoir, K.~Noui, and P.~Roche, ``{Hamiltonian quantization of
  Chern-Simons theory with SL(2,C) group},'' Class. Quant. Grav. {\bf 19}
  (2002) 4953,
\href{http://arXiv.org/abs/hep-th/0202121}{{\texttt{arXiv:hep-th/0202121}}}.

\bibitem{PR}
G.~Ponzano and T.~Regge, ``{Semiclassical Limit of Racah Coefficients},'' in
  {\em Spectroscopic and Group Theoretical Methods in Physics}, F.~Bloch, ed.,
  pp.~1--58.
\newblock North-Holland Publ. Co., Amsterdam, Netherlands, 1968.

\bibitem{Freidel:2004vi}
L.~Freidel and D.~Louapre, ``{Ponzano-Regge model revisited I: Gauge fixing,
  observables and interacting spinning particles},'' Class. Quant. Grav. {\bf
  21} (2004) 5685--5726,
\href{http://arXiv.org/abs/hep-th/0401076}{{\texttt{arXiv:hep-th/0401076}}}.

\bibitem{Freidel:2005bb}
L.~Freidel and E.~R. Livine, ``{Ponzano-Regge model revisited III: Feynman
  diagrams and effective field theory},'' Class. Quant. Grav. {\bf 23} (2006)
  2021--2062,
\href{http://arXiv.org/abs/hep-th/0502106}{{\texttt{arXiv:hep-th/0502106}}}.

\bibitem{Barrett:2008wh}
J.~W. Barrett and I.~Naish-Guzman, ``{The Ponzano-Regge model},'' Class. Quant.
  Grav. {\bf 26} (2009) 155014,
\href{http://arXiv.org/abs/0803.3319}{{\texttt{arXiv:0803.3319}}}.

\bibitem{Rovelli:2007quantum}
C.~Rovelli, {\em Quantum gravity}.
\newblock Cambridge university press, 2007.

\bibitem{Turaev:1992state}
V.~G. Turaev and O.~Y. Viro, ``State sum invariants of 3-manifolds and quantum
  6j-symbols,'' Topology {\bf 31} (1992), no.~4, 865--902.

\bibitem{Freidel:2004ponzano}
L.~Freidel and D.~Louapre, ``Ponzano-Regge model revisited II: equivalence with
  Chern-Simons,'' arXiv preprint gr-qc/0410141 (2004).

\bibitem{Meusburger:2010hilbert}
C.~Meusburger, K.~Noui, {\em et al.}, ``The Hilbert space of 3d gravity:
  quantum group symmetries and observables,'' Advances in Theoretical and
  Mathematical Physics {\bf 14} (2010), no.~6, 1651--1715.

\bibitem{Meusburger:2003hc}
C.~Meusburger and B.~J. Schroers, ``{The quantisation of Poisson structures
  arising inChern-Simons theory with gauge group $G \ltimes \mathfrak{g}^*$},''
  Adv. Theor. Math. Phys. {\bf 7} (2003), no.~6, 1003--1043,
\href{http://arXiv.org/abs/hep-th/0310218}{{\texttt{arXiv:hep-th/0310218}}}.

\bibitem{Roberts:1995skein}
J.~Roberts, ``Skein theory and Turaev-Viro invariants,'' Topology {\bf 34}
  (1995), no.~4, 771--787.

\bibitem{Noui:2011im}
K.~Noui, A.~Perez, and D.~Pranzetti, ``{Canonical quantization of
  non-commutative holonomies in 2+1 loop quantum gravity},'' JHEP {\bf 10}
  (2011) 036,
\href{http://arXiv.org/abs/1105.0439}{{\texttt{arXiv:1105.0439}}}.

\bibitem{Pranzetti:2014xva}
D.~Pranzetti, ``{Turaev-Viro amplitudes from 2+1 Loop Quantum Gravity},'' Phys.
  Rev. {\bf D89} (2014), no.~8, 084058,
\href{http://arXiv.org/abs/1402.2384}{{\texttt{arXiv:1402.2384}}}.

\bibitem{Bonzom:2014wva}
V.~Bonzom, M.~Dupuis, F.~Girelli, and E.~R. Livine, ``{Deformed phase space for
  3d loop gravity and hyperbolic discrete geometries},''
\href{http://arXiv.org/abs/1402.2323}{{\texttt{arXiv:1402.2323}}}.

\bibitem{Bonzom:2014bua}
V.~Bonzom, M.~Dupuis, and F.~Girelli, ``{Towards the Turaev-Viro amplitudes
  from a Hamiltonian constraint},'' Phys. Rev. {\bf D90} (2014), no.~10,
  104038,
\href{http://arXiv.org/abs/1403.7121}{{\texttt{arXiv:1403.7121}}}.

\bibitem{Dupuis:2013lka}
M.~Dupuis and F.~Girelli, ``{Observables in Loop Quantum Gravity with a
  cosmological constant},'' Phys. Rev. {\bf D90} (2014), no.~10, 104037,
\href{http://arXiv.org/abs/1311.6841}{{\texttt{arXiv:1311.6841}}}.

\bibitem{Goldman:1986invariant}
W.~M. Goldman, ``Invariant functions on Lie groups and Hamiltonian flows of
  surface group representations,'' Inventiones mathematicae {\bf 85} (1986),
  no.~2, 263--302.

\bibitem{Dupuis:2017otn}
M.~Dupuis, L.~Freidel, and F.~Girelli, ``{Discretization of 3d gravity in
  different polarizations},'' Phys. Rev. {\bf D96} (2017), no.~8, 086017,
\href{http://arXiv.org/abs/1701.02439}{{\texttt{arXiv:1701.02439}}}.

\bibitem{Freidel:2018pbr}
L.~Freidel, F.~Girelli, and B.~Shoshany, ``{2+1D Loop Quantum Gravity on the
  Edge},'' Phys. Rev. {\bf D99} (2019), no.~4, 046003,
\href{http://arXiv.org/abs/1811.04360}{{\texttt{arXiv:1811.04360}}}.

\bibitem{Freidel:2010aq}
L.~Freidel and S.~Speziale, ``{Twisted geometries: A geometric parametrisation
  of SU(2) phase space},'' Phys. Rev. {\bf D82} (2010) 084040,
\href{http://arXiv.org/abs/1001.2748}{{\texttt{arXiv:1001.2748}}}.

\bibitem{Dupuis:2012yw}
M.~Dupuis, J.~P. Ryan, and S.~Speziale, ``{Discrete gravity models and Loop
  Quantum Gravity: a short review},'' SIGMA {\bf 8} (2012) 052,
\href{http://arXiv.org/abs/1204.5394}{{\texttt{arXiv:1204.5394}}}.

\bibitem{Freidel:2013bfa}
L.~Freidel and J.~Ziprick, ``{Spinning geometry = Twisted geometry},'' Class.
  Quant. Grav. {\bf 31} (2014), no.~4, 045007,
\href{http://arXiv.org/abs/1308.0040}{{\texttt{arXiv:1308.0040}}}.

\bibitem{Freidel:2018pvm}
L.~Freidel and E.~R. Livine, ``{Bubble networks: framed discrete geometry for
  quantum gravity},'' Gen. Rel. Grav. {\bf 51} (2019), no.~1, 9,
\href{http://arXiv.org/abs/1810.09364}{{\texttt{arXiv:1810.09364}}}.

\bibitem{Bonzom:2011hm}
V.~Bonzom and L.~Freidel, ``{The Hamiltonian constraint in 3d Riemannian loop
  quantum gravity},'' Class. Quant. Grav. {\bf 28} (2011) 195006,
\href{http://arXiv.org/abs/1101.3524}{{\texttt{arXiv:1101.3524}}}.

\bibitem{Bonzom:2011nv}
V.~Bonzom and E.~R. Livine, ``{A New Hamiltonian for the Topological BF phase
  with spinor networks},'' J. Math. Phys. {\bf 53} (2012) 072201,
\href{http://arXiv.org/abs/1110.3272}{{\texttt{arXiv:1110.3272}}}.

\bibitem{Bonzom:2013tna}
V.~Bonzom and B.~Dittrich, ``{Dirac's discrete hypersurface deformation
  algebras},'' Class. Quant. Grav. {\bf 30} (2013) 205013,
\href{http://arXiv.org/abs/1304.5983}{{\texttt{arXiv:1304.5983}}}.

\bibitem{Dupuis:2014fya}
M.~Dupuis, F.~Girelli, and E.~R. Livine, ``{Deformed Spinor Networks for Loop
  Gravity: Towards Hyperbolic Twisted Geometries},'' Gen. Rel. Grav. {\bf 46}
  (2014), no.~11, 1802,
\href{http://arXiv.org/abs/1403.7482}{{\texttt{arXiv:1403.7482}}}.

\bibitem{Ahluwalia:1993rq}
K.~S. Ahluwalia, ``{Fundamentals of poisson lie groups with application to the
  classical double},''
\href{http://arXiv.org/abs/hep-th/9310068}{{\texttt{arXiv:hep-th/9310068}}}.

\bibitem{Kosmann:1997lie}
Y.~Kosmann-Schwarzbach, ``Lie bialgebras, Poisson Lie groups and dressing
  transformations,'' in {\em Integrability of nonlinear systems}, pp.~104--170.
\newblock Springer, 1997.

\bibitem{Dupuis:2013quantum}
M.~Dupuis and F.~Girelli, ``Quantum hyperbolic geometry in loop quantum gravity
  with cosmological constant,'' Physical Review D {\bf 87} (2013), no.~12,
  121502.

\end{thebibliography}\endgroup

\end{document}